\documentclass[useAMS,usenatbib]{mn2e}
\usepackage[usenames,dvipsnames]{color}
\usepackage{multicol}
\usepackage{multirow}
\usepackage{natbib}
\usepackage{float}
\usepackage{flafter}
\usepackage{amsmath}
\usepackage{graphicx}
\usepackage{txfonts}
\usepackage{psfig}
\usepackage{subfigure}
\usepackage{natbib}
\usepackage{tikz}
\usepackage{rotating}

\newcommand{\myemail}{walch@ph1.uni-koeln.de}

\title[SILCC: Evolution of the SN-driven ISM]
{The SILCC (SImulating the LifeCycle of molecular Clouds) project: \\I.~Chemical evolution of the supernova-driven ISM}
\author[Walch et al.]
{S. Walch$^{1,2}$\thanks{\myemail}, 
P.~Girichidis$^{2}$, T.~Naab$^{2}$, A.~Gatto$^{2}$, S.C.O.~Glover$^{3}$, R.~W\"unsch$^{4}$, \\ \\
{\LARGE \rm R.S.~Klessen$^{3,7,8}$, P.~C.~Clark$^{5}$, T.~Peters$^{2,6}$, C.~Baczynski$^{3}$}\\ \\
$^{1}$I. Physikalisches Institut, Universit\"at zu K\"oln, Z\"ulpicher Str. 77, 50937 K\"oln, Germany\\
$^{2}$Max-Planck-Institute for Astrophysics, Karl-Schwarzschild-Str. 1, 85741 Garching, Germany \\
$^{3}$Zentrum f\"ur Astronomie der Universit\"at Heidelberg, Institut f\"ur Theoretische Astrophysik, Albert-Ueberle-Str. 2, 69120 Heidelberg, Germany\\
$^{4}$Astronomical Institute, Academy of Sciences of the Czech Republic, Bocni II 1401, 141 31 Prague, Czech Republic\\
$^{5}$School of Physics \& Astronomy, Cardiff University, 5 The Parade, Cardiff CF24 3AA, Wales, UK \\
$^{6}$Institut f\"{u}r Computergest\"{u}tzte Wissenschaften, Universit\"{a}t Z\"{u}rich, Winterthurerstrasse 190, CH-8057 Z\"{u}rich, Switzerland. \\
$^{7}$Department of Astronomy and Astrophysics, University of California, 1156 High Street, Santa Cruz, CA 95064, USA\\
$^{8}$Kavli Institute for Particle Astrophysics and Cosmology, Stanford University, SLAC National Accelerator Laboratory, Menlo Park, CA 94025, USA. }

\begin{document}

\date{Accepted . Received 2014 December 8th; in original form }

\pagerange{\pageref{firstpage}--\pageref{lastpage}} \pubyear{2014}

\maketitle
  
\label{firstpage}
\begin{abstract}
The SILCC project (SImulating the Life-Cycle of molecular Clouds) aims at a more self-consistent understanding of the interstellar medium (ISM) on small scales and its link to galaxy evolution. We simulate the evolution of the multi-phase ISM in a (500 pc)$^2 \times \pm$ 5 kpc region of a galactic disc, with a gas surface density of  $\Sigma_{_{\rm GAS}} =  10 \;{\rm M}_\odot/{\rm pc}^2$. The {\sc Flash} 4.1 simulations include an external potential, self-gravity, magnetic fields, heating and radiative cooling, time-dependent chemistry of H$_2$ and CO considering (self-) shielding, and supernova (SN) feedback. We explore SN explosions at different (fixed) rates in high-density regions ({\it peak}), in random locations ({\it random}), in a combination of both ({\it mixed}), or clustered in space and time ({\it clustered}). 
Only {\it random} or {\it clustered} models with self-gravity (which evolve similarly) are in agreement with observations. Molecular hydrogen forms in dense filaments and clumps and contributes 20\% - 40\% to the total mass, whereas most of the mass (55\% - 75\%) is in atomic hydrogen. The ionised gas contributes $<$10\%. For high SN rates (0.5 dex above Kennicutt-Schmidt) as well as for {\it peak} and {\it mixed} driving the formation of H$_2$ is strongly suppressed. Also without self-gravity the H$_2$ fraction is significantly lower ($\sim$ 5\%). Most of the volume is filled with hot gas ($\sim$90\% within $\pm$2 kpc). Only for {\it random} or {\it clustered} driving, a vertically expanding warm component of atomic hydrogen indicates a fountain flow. Magnetic fields have little impact on the final disc structure. However, they affect dense gas ($n\gtrsim 10\;{\rm cm}^{-3}$) and delay H$_2$ formation. We highlight that individual chemical species, in particular atomic hydrogen, populate different ISM phases and cannot be accurately accounted for by simple temperature-/density-based phase cut-offs.    

\end{abstract}


\section{Introduction}
Within a typical disc galaxy, gas is continually cycled between several very different phases. Warm atomic gas cools and undergoes gravitational collapse, forming cold, dense molecular clouds. Stars form in many of these clouds and exert strong feedback on the surrounding gas, in the form of UV radiation \citep{Dale2005, Dale2012, Walch2012a, Walch2013}, stellar winds \citep{Toala2011, Rogers2013}, and supernova explosions \citep{Hill2012, Kim2014, Martizzi2014, Walch2014, Gatto2014}. This feedback helps to disperse many of the clouds, returning gas to the warm atomic phase, or heating it so much that it reaches a highly diffuse, ionized phase. Other clouds are destroyed by large-scale dynamical processes in the disk, such as the strong shear flows around spiral arms \citep[see e.g.][]{Dobbs2013}. Dynamically significant magnetic fields help to guide the gas flows that form and disperse the clouds \citep[e.g.][]{deAvillez2004, deAvillez2005, Heitsch2009, Banerjee2009}, and cosmic rays can also play an important role, as they represent a significant fraction of the local energy density in the ISM \citep[e.g.][]{Breitschwerdt2008, Dorfi2012, Hanasz2013, Girichidis2014CR}.

Understanding how gas is cycled between the different phases in the interstellar medium (ISM) and how molecular clouds are formed and destroyed is important for understanding how star formation is regulated in galaxies. The interplay between feedback from massive stars and the surrounding ISM also governs the launching of galactic winds, which play an important role for regulating spiral galaxy evolution on larger spatial and temporal scales at all cosmic epochs \citep{Oppenheimer2010, Aumer2013, Stinson2013, Marinacci2014, Hirschmann2013, Brook2014, Hopkins2014, Murante2014, Ubler2014}. Substantial effort has already been devoted to the study of stellar feedback and the regulation of star formation on galactic scales (see e.g. \citealt{Stinson2010}, \citealt{Hopkins2011}, \citealt{Agertz2012}, and \citealt{Genel2014} for a few recent examples), but much of this work considers only large scales and fails to resolve the small-scale processes responsible for driving the large-scale flows. 

A better understanding of the ISM is therefore of fundamental astrophysical importance. However, the ISM is governed by a complex network of many interacting physical processes. Any of these processes alone can affect the properties of the ISM significantly and their mutual interaction may well be non-linear. For a comprehensive model of the ISM, it is therefore important to consider all of the above processes (heating, cooling, gravity, chemistry, magnetic fields, etc.), but for a concise understanding it is necessary to first investigate their impact separately. Within the framework of SILCC (SImulating the Life-Cycle of molecular Clouds\footnote{For movies of the simulations and download of selected {\sc Flash} data see the SILCC website: {\bf www.astro.uni-koeln.de/silcc}}), we have started this enterprise as a collaboration in which we aim to understand the full life-cycle of molecular clouds using state-of-the-art numerical simulations. In this first SILCC paper, we introduce the elements of our numerical model, and use it to study the effect of the spatial and temporal distribution of supernova (SN) explosions on the multi-phase structure of the ISM and the formation of molecular gas.
 
Specifically, we use the adaptive mesh refinement (AMR) code {\sc Flash} 4.1 to model the multi-phase ISM in a representative portion of a disc galaxy in three-dimensional, magnetohydrodynamic (MHD) simulations, including
\begin{itemize}
\item an external galactic potential,
\item self-gravity,
\item radiative heating and cooling coupled to
\item a chemical network to follow the formation of H$_{2}$ and CO,
\item diffuse heating and its attenuation by dust shielding, and
\item feedback from supernova explosions of massive stars.
\end{itemize}
This setup allows us to self-consistently follow the formation and evolution of molecular clouds in disc galaxies. Since much of the gas in the ISM is highly turbulent and out of chemical equilibrium, it is necessary to follow molecule formation on-the-fly, even though this process has only a minor impact on the gas dynamics (\citeauthor{Glover2012a}~2012a).

In this paper, we investigate the specific location and clustering properties of the supernovae (SNe) on the structure of the multi-phase ISM and in particular on the evolution of the molecular gas phase. We study four different models for SN driving: (1) SNe explode at random positions, modulo the constraint that the positions form a Gaussian distribution in the vertical direction about the disc mid-plane; (2) SNe explode within forming molecular clouds, i.e.\ at the peaks in the density distribution; (3) half of the SNe explode randomly, and half of them explode within dense clouds (mixed driving); or (4) the SNe explode in clusters and are therefore correlated in space and time \citep[e.g.][]{MacLow1988, Matzner2002}. These different driving mechanisms have been employed in previous studies (see below), but a conclusive comparison between them has not yet been carried out.

Previously, many groups have investigated the development of the multi-phase ISM structure in a representative portion (with similar size to our setup) of a stratified galactic disc. In particular, \citet{Joung2006}, \citet{Hill2012}, \citeauthor{Gent2013a} (2013a), \citeauthor{Gent+13b} (2013b), and \citet{Creasey2013} have carried out simulations of stratified discs with the {\sc Flash} code (although older releases of it), employing mostly randomly placed or clustered SNe. They ran models with and without self-gravity (although these are not compared in the way presented here) as well as with and without magnetic fields and found that the magnetic field is generally unimportant for the global disc structure. Recently,  \citet{Hennebelle2014} computed models of low surface density discs with $\Sigma_{\rm GAS} \sim 3 \; {\rm M_{\odot} / pc^{2}}$ and galactic SN rates with {\sc Ramses} 
\citep{Teyssier2002}. They include magnetic fields and self-gravity, which leads to the formation of sink particles, and show that a weak magnetic field (2.5 $\mu$G) reduces the star formation rate by a factor of $\sim 2$.

The setup used in the work of \citet{Kim2011}, \citet{Shetty2012}, and \citet{Kim2013}, carried out with the {\sc Athena} MHD code  \citep{Stone2008}, is different since they correlate SN explosions with dynamically forming, local density maxima, and include shearing box boundary conditions. The SNe deposit a fixed amount of momentum ($3 \times 10^{5} \: {\rm M_{\odot}  km / s}$ per event) into the surrounding gas, but no thermal energy is added, and thus there is no hot gas present in these simulations. With this method, the star formation rate within the discs can be regulated to the observed level. The role of shear -- which we do not include at the moment -- is potentially important, in particular for gas-rich disc models  \citep{Shetty2012}, which resemble the conditions within the Galactic Center or in high-redshift, star-forming galaxies \citep{Genzel2011, FoersterSchreiber2011, Fisher2013}. In these environments, shear could limit the size of the cold clouds that form. Models of full galactic discs \citep{Dobbs2012, Bonnell2013} even describe shear as one of the main drivers of ISM turbulence (e.g.\ \citealt{Piontek2005}; \citealt{Kevlahan2009}; and see \citealt{Agertz2009} for a discussion), although the main role of the spiral arm structure to form and cause collisions between molecular clouds is also emphasised. A detailed investigation of the relative contribution of shear will be left to future study.

Most of the models discussed above use highly simplified cooling functions and do not follow the formation of molecular gas. A few attempts have been made to account for the non-equilibrium chemistry of H$_2$ within galactic-scale simulations of the ISM \citep[see e.g.][]{Dobbs2008, Dobbs2011, Gnedin2009, Christensen2012}, and more recently a couple of studies have also treated CO formation \citep{Smith2014, Pettitt2014}. However, the spatial and mass resolution in most of these studies is poor (with the exception of the recent simulations by \citet{Smith2014}, which have a mass resolution of only $4 \: {\rm M_{\odot}}$, but these simulations did not account for either stellar feedback or for the self-gravity of the gas).

In this paper, we improve on this earlier work by combining high-resolution AMR simulations that properly account for the effects of magnetic fields, SN feedback and self-gravity with a non-equilibrium treatment of the H$_2$ and CO chemistry of the gas. The plan of this paper is the following. The numerical method is described in section~\ref{SEC_NM}, and the simulation setup, the SN positioning, and the adopted SN rates are outlined in section~\ref{SEC_Setup}. In section~\ref{SEC_FIDUCIAL}, we discuss the morphologies and chemical evolution of simulations that use the same SN rate but different physics or SN positioning. The influence of the SN rate is discussed in section~\ref{SEC_SFR}, and section~\ref{SEC_VERTICAL} shows the vertical distributions and the evolution of the disc in the $z$-direction. In section~\ref{SEC_PHASES}, we show the difference between temperature phases and the chemical composition of the gas. We summarise our findings with respect to the total gas mass fractions in the different chemical species and the average volume filling fractions of the different temperature phases in section~\ref{SEC_SUMMARY}, and give our conclusions in section~\ref{SEC_CONCLUSIONS}.

\section{Numerical Method} \label{SEC_NM}
We use the fully three-dimensional, Adaptive Mesh Refinement (AMR) code Flash 4.1 \citep{Fryxell2000, Dubey2009, Dubey2012} to simulate representative pieces of galactic discs with dimensions 500 pc x 500 pc x $\pm$ 5 kpc. {\sc Flash} is parallelised via domain decomposition under the Message Passing Interface (MPI). It further employs a finite volume scheme, in which the physical variables are represented as zone averages. The AMR is handled by the PARAMESH library. 

\subsection{Magneto-Hydrodynamics}\label{SEC_MHD}

Using {\sc Flash} 4.1, we model the evolution of the gas via solving the ideal MHD equations. For all simulations (magnetic and non-magnetic), we use a directionally split, finite-difference scheme that is based on the 5-wave Bouchut MHD solver HLL5R \citep{Bouchut2007, Klingenberg2007, Bouchut2010, Waagan2009, Waagan2011}, which is stable and preserves positivity for high Mach number flows. The nonlinear flux of quantities between zones is obtained by solving a Riemann problem at each zone boundary in alternating one-dimensional sweeps through the grid (MUSCL scheme, VanLeer 1979). 
With (self-)gravity (see section \ref{SEC_GRAV}), the equations read:

\begin{eqnarray}
\frac{\partial \rho}{\partial t} + \bf{\bigtriangledown} \cdot \left(\rho \bf{v} \right) &=& 0, \\
\frac{\partial \rho {\bf v}}{\partial t} + \bigtriangledown \cdot \left [ \rho {\bf v v}^{\rm T} + \left(P + \frac{\|{\bf B} \|^2}{8 \pi} \right) {\bf I} - \frac{{\bf B B}^{\rm T}}{4 \pi}\right ] &=& \rho {\bf g} + \dot{q}_\mathrm{inj},\\
\frac{\partial {\bf B}}{\partial t} - \bigtriangledown \times  \left ( {\bf v} \times {\bf B} \right ) &=& 0.
\end{eqnarray}
Here, $t$ is the time, $\rho$ is the mass density, $\bf{v}$ is the gas velocity, $\bf{B}$ is the magnetic field, and $P = (\gamma -1) u$ is the thermal pressure, where $u $ is the internal energy per unit volume. We assume an ideal gas with $\gamma=5/3$. {\bf I} is the $3 \times 3$ identity matrix, and ${\bf B B}^{\rm T}$ is the outer product of $\bf{B}$ with itself. The divergence-free constraint applies ($\bf{\bigtriangledown} \cdot {\bf B} = 0$). The momentum input from SNe per time step is $\dot{q}_\mathrm{inj}$. 
The gravitational acceleration ${\bf g}$ combines the contributions from the self-gravity of the gas (${\bf g}_\mathrm{sg}$; see section \ref{SEC_GRAV}), and from the external gravitational potential (${\bf g}_\mathrm{ext}$; see section \ref{SEC_GRAV_EXT}), which is provided by the stellar distribution within the galactic disc:
\begin{equation}
{\bf g} = {\bf g}_\mathrm{sg} + g_\mathrm{ext}(z)\hat{z}.
\end{equation}

We modify the energy equation to include heating and cooling effects as well as the energy input by supernovae:
\begin{eqnarray}\label{energy_eq}
\frac{\partial E}{\partial t} + 
\bigtriangledown \cdot \left [ \left ( E + \frac{\|{\bf B} \|^2}{8 \pi} + \frac{P}{\rho} \right) {\bf v} - \frac{\left( {\bf B} \cdot {\bf v}\right) {\bf B}}{4 \pi} \right ] &=&\nonumber \\
  {\bf v} \cdot {\bf g} + \dot{u}_\mathrm{chem} + \dot{u}_{\rm inj} & &
 \end{eqnarray}
where the energy density (in erg per unit volume) is
\begin{equation}
E = u+ \frac{\rho \|{\bf v} \|^2}{2} +\frac{\|{\bf B} \|^2}{8 \pi}.
\end{equation}
${\dot{u}_\mathrm{chem}}$ is the net rate of change in internal energy due to diffuse heating and radiative cooling, both of which are computed through the chemical network (as described in section \ref{SEC_CHEM_MODEL}). $\dot{u}_{\rm inj}$ is the thermal energy input (per unit volume) from supernovae. The detailed implementation of SN feedback is described in section \ref{SEC_SN}.

In the case of fast advection, where the kinetic energy dominates the internal energy by more than four orders of magnitude (parameter {\it eint\_switch} $=10^{-4}$), we separately solve for the internal energy, $u$, to avoid truncation errors:
\begin{equation}
\frac{\partial u}{\partial t} + \bigtriangledown \cdot \left[ \left(u + \frac{P}{\rho} \right) {\bf v}\right] = {\bf v} \cdot \bigtriangledown \left(\frac{P}{\rho}\right)
\end{equation}
Then the total energy is recomputed using the velocities from the momentum equation as well as the new internal energy.
 


\subsection{Chemistry and Cooling}\label{SEC_CHEM_MODEL}
We follow the chemical evolution of the gas in our simulations using a simplified chemical network that tracks the ionisation fraction of the gas, and the formation and destruction of  H$_{2}$ and CO. We do not assume chemical equilibrium and therefore have to solve a continuity equation of the form \citep[see e.g.][]{Glover2007,Micic2012}
\begin{equation}
\frac{\partial \rho_i}{\partial t} +\bigtriangledown \cdot \left( \rho_i {\bf v} \right )= C_i\left(\rho,T,\ldots \right ) - D_i\left(\rho,T, \ldots \right)
\end{equation}
for every chemical species $i$ included in our network. The terms $C_i$ and $D_i$ in this equation represent the creation and destruction of species $i$ due to chemical reactions. These terms generally depend on the density and temperature of the gas, and also on the abundances of the other chemical species. Therefore, we have to solve a set of coupled partial differential equations for the mass densities of the different chemical species. In practice, we can make the problem substantially easier to handle by operator splitting the chemical source and sink terms from the advection terms. With this approach, the continuity equations that we have to solve during the advection step simplify to
\begin{equation}
\frac{\partial \rho_i}{\partial t} +\bigtriangledown \cdot \left( \rho_i {\bf v} \right ) = 0.
\end{equation}
These equations describe the evolution of a set of scalar tracer fields, and can be handled using the standard {\sc Flash} infrastructure for such fields.  Changes in the chemical composition of the gas resulting from chemical reactions are then computed in a separate chemistry step, during which we solve the following set of coupled ordinary differential equations (ODEs):
\begin{equation}
\frac{{\rm d} \rho_i}{{\rm d}t} = C_i\left(\rho,T,\ldots \right ) - D_i\left(\rho,T, \ldots \right).
\end{equation}
Because the radiative cooling rate can often depend sensitively on the chemical abundances, we also operator split the {\sc Flash} energy equation, solving for the rate of change in $E$ due to radiative heating and cooling, $\dot{u}_\mathrm{chem}$, separately from the other terms in Equation~\ref{energy_eq}. The resulting ODE describing $\dot{u}_\mathrm{chem}$ is solved simultaneously with the chemical ODEs, using the implicit solver {\sc Dvode} \citep{Brown1989}. A similar basic approach was used in the {\sc Zeus-mp} MHD code by \citet{Glover2010} and in an earlier version of the {\sc Flash} code by \citet{Walch2011a} and \citet{Micic2012,Micic2013}.

If the chemistry or cooling time-steps are much shorter than the hydrodynamical time-step, then sub-cycling is used to treat the cooling and chemistry, thereby avoiding the need to constrain the global time-step. We limit the maximum size of the time-step taken within the ODE solver to be
\begin{equation}
\Delta t_{\rm max} = {\rm min} \left(\Delta t_{\rm cool}, \Delta t_{\rm chem} \right),
\end{equation}
where $\Delta t_{\rm cool}$ and $\Delta t_{\rm chem}$ are the cooling time-step and the chemical time-step, respectively, which are estimated using the current values of $\dot{u}_\mathrm{chem}$, $C_{i}$ and $D_{i}$.
 
The chemical network used in our simulations is based on the network for hydrogen chemistry presented in \citet{Glover2007,Glover2007b}, supplemented with the simplified model for CO formation introduced by \citet{Nelson1997}. We model the evolution of seven chemical species: free electrons, H$^{+}$, H, H$_{2}$, C$^{+}$, O and CO. The fractional abundances of these species are constrained by several different conservation laws. We conserve total charge, allowing us to write the free electron abundance as
\begin{equation}
x_{\rm e} = x_{\rm H^{+}} + x_{\rm C^{+}} + x_{\rm Si^{+}},
\end{equation}
where $x_{i}$ denotes the fractional abundance of species $i$, relative to the total abundance of hydrogen nuclei in all forms. Note that although we include a contribution from ionised silicon in our expression for $x_{\rm e}$, we do not track the chemical evolution of silicon, instead simply assuming that it remains singly ionised throughout the gas. In addition to conserving charge, we also conserve the total abundances of hydrogen, carbon and oxygen, allowing us to derive the fractional abundances of atomic hydrogen, ionised carbon and atomic oxygen from the following expressions:
\begin{eqnarray}
x_{\rm H} & = & 1 - 2 x_{\rm H_{2}} - x_{\rm H^{+}}, \\
x_{\rm C^{+}} & = & x_{\rm C, tot} - x_{\rm CO}, \\
x_{\rm O} & = & x_{\rm O, tot} - x_{\rm CO},
\end{eqnarray}
where $x_{\rm C, tot}$ and $x_{\rm O, tot}$ are the total fractional abundances of carbon and oxygen (in all forms) relative to hydrogen. Using these relationships allows us to reduce the number of chemical species for which we need to solve the full chemical rate equations from seven to three: H$^{+}$, H$_{2}$ and CO.

In our current set of simulations, we assume a constant gas-phase metallicity ${\rm Z = Z_{\odot}}$ and a constant dust-to-gas ratio of 0.01. We take the total gas-phase carbon, oxygen and silicon abundances to be $x_{_{\rm C,tot}}=1.41 \times 10^{-4}$, $x_{_{\rm O, tot}}= 3.16 \times 10^{-4}$ and $x_{_{\rm Si, tot}}= 1.5 \times 10^{-5}$ \citep{Sembach2000}.

\subsubsection{Hydrogen chemistry}
The full set of chemical reactions that make up our implementation of the non-equilibrium hydrogen chemistry are given in Table 1 of \citet{Micic2012}. We include the standard processes governing the formation and destruction of ionised hydrogen (collisional ionisation, ionisation by cosmic rays and X-rays, radiative recombination, etc.) plus a simplified treatment of H$_{2}$ formation and destruction. H$_{2}$ is assumed to form only on the surface of dust grains, following the prescription given in \citet{Hollenbach1989}, as this dominates over gas-phase formation via the H$^{-}$ or H$_{2}^{+}$ ions at metallicities close to solar \citep{Glover2003}. Some H$_{2}$ is destroyed by cosmic ray ionisation and by collisional dissociation in hot gas, but in most regions, the dominant destruction process is photo-dissociation by the interstellar radiation field (ISRF). In our treatment, we compute the H$_{2}$ photo-dissociation rate using the following expression:
\begin{equation}
R_{\rm pd, H_{2}} = R_{\rm pd, H_{2}, thin} f_{\rm dust, H_{2}} f_{\rm shield, H_{2}}.
\end{equation}
Here, $R_{\rm pd, H_{2}, thin} = 3.3 \times 10^{-11} G_{0} \: {\rm s^{-1}}$ is the photo-dissociation rate in optically thin gas,  taken from \citet{DraineBertoldi1996}, $G_{0}$ is the strength of the interstellar radiation field in units of the \citet{Habing1968} field, $f_{\rm dust, H_{2}}$ is a factor accounting for the effects of dust extinction and $f_{\rm shield, H_{2}}$ is a factor accounting for H$_{2}$ self-shielding. Along any particular line of sight, $f_{\rm dust, H_{2}}$ is related to the visual extinction $A_{\rm V}$ by
\begin{equation}
f_{\rm dust, H_{2}} = \exp{\left( -3.5 A_{\rm V}\right)}. \label{EQ_fdust}
\end{equation}
In the diffuse ISM, $A_{\rm V}$ is related to the total hydrogen column density $N_{\rm H, tot}$ by \citep{Bohlin1978}
\begin{equation}
A_{\rm V} = \frac{N_{\rm H,tot}}{1.87 \times 10^{21} \; {\rm cm}^{-2}}, \label{AV_Ntot}
\end{equation}
where $N_{\rm H,tot} = N_{\rm H^{+}} + N_{\rm H} + 2 N_{\rm H_2}$. In dense clouds, grain coagulation leads to a slightly different relationship between $A_{\rm V}$ and $N_{\rm H, tot}$ \citep[see e.g.][]{Foster2013}, but we neglect this complication here.

To determine $f_{\rm shield, H_{2}}$ for a particular line of sight, we use the following self-shielding function from \citet{DraineBertoldi1996}:
\begin{eqnarray}
f_{\rm shield, H_{2}} & = & \frac{0.965}{\left(1+x/b_5 \right)^2} +  \frac{0.035}{\sqrt{1+x}} \nonumber \\
& \times & \exp \left( -8.5 \times 10^{-4} \sqrt{1+x} \, \right)
\end{eqnarray}
where $x=N_{\rm H_2}/\left(5 \times 10^{14} \;{\rm cm}^{-2} \right)$, $b_5 = b/\left(10^5\;{\rm cm\; s}^{-1} \right)$, and $b$ is the Doppler broadening parameter, which is related to the gas temperature by $b^{2} = k_\mathrm{B}T / m_{\rm H}$. We compute the values of $f_{\rm dust, H_{2}}$ and $f_{\rm shield, H_{2}}$ for each grid cell\footnote{We use the local temperature to estimate $b$, assuming that this is reasonably representative of the bulk temperature of H$_2$ along the line of sight. Note also, that $f_{\rm shield, H_{2}}$ is only weakly dependent on $b$. } by averaging over multiple lines of sight using the TreeCol algorithm \citep{Clark2012}, which we describe in more detail in Section~\ref{SEC_TREECOL}.

\subsubsection{Carbon chemistry}
To model the transition from C$^{+}$ to CO, we use a highly simplified treatment first introduced by \citeauthor{Nelson1997}~(1997; hereafter, NL97). Their approach is based on the assumption that the rate limiting step in the formation of CO is the radiative association of C$^{+}$ with H$_{2}$ to form the CH$_{2}^{+}$ molecular ion. Once formed, this ion is assumed to have only two possible fates: either it reacts with atomic oxygen to produce CO, or it is photodissociated by the interstellar radiation field, returning the carbon to the gas as C$^{+}$.  This model tends to overproduce CO somewhat at intermediate densities and extinctions compared to more detailed chemical models (see e.g.\ the detailed comparison in \citeauthor{Glover2012b}~2012b). However, we do not expect this to significantly affect the spatial distribution of CO or the temperature structure of the gas on the scales resolved in the set of simulations presented in this paper. 

The main process responsible for destroying CO is photodissociation by the ISRF. We adopt a photodissociation rate given by
\begin{equation}
R_{\rm pd, CO} = R_{\rm pd, CO, thin} f_\mathrm{dust, CO} f_{\rm shield, CO} \; {\rm s}^{-1}. \label{EQ_CO}
\end{equation}
Here, $R_{\rm pd, CO, thin} = 2.1 \times 10^{-10} (G_{0} / 1.7) \: {\rm s^{-1}}$ is the photo-dissociation rate of CO in optically thin gas, taken from 
\citet{vdb88}, $f_{\rm shield, CO}$ is a shielding factor quantifying the effects of CO self-shielding and the shielding of CO by the Lyman-Werner bands of H$_{2}$, and $f_{\rm dust}$ accounts for the effects of dust absorption. To compute $f_{\rm shield, CO}$ for a particular line of sight, we first compute the H$_{2}$ and CO column densities along that line of sight, and then convert these into a value for $f_{\rm shield, CO}$ using data from \citet{Lee96}. For the dust absorption term $f_{\rm dust, CO}$, we use the expression \citep{vdb88}:
\begin{equation}
f_{\rm dust, CO} = \exp{\left( -2.5 A_{\rm V}\right)}.
\end{equation}
using the same relationship as before between the visual extinction $A_{\rm V}$ and the total hydrogen column density $N_{\rm H, tot}$. 
To compute $f_{\rm dust, CO}$ and $f_{\rm shield, CO}$ for each grid cell, we once again average their values over multiple lines of
sight using the TreeCol algorithm (see Section~\ref{SEC_TREECOL}).

\subsubsection{Radiative cooling}
To model the effects of radiative cooling, we use the detailed atomic and molecular cooling function outlined in \citet{Glover2010} and updated in \citeauthor{Glover2012b}~(2012b). This includes contributions from the fine structure lines of C$^{+}$, O and Si$^{+}$, the rotational and vibrational lines of H$_{2}$ and CO, the electronic lines of atomic hydrogen (i.e.\ Lyman-$\alpha$ cooling), and also the transfer of energy from the gas to the dust, although the latter effect is generally unimportant at the densities reached in these simulations. Since we do not explicitly track the ionisation state of helium, or any ionisation states of the metals beyond the singly ionised state of carbon, we cannot compute the contribution of these to the high temperature cooling rate in a completely self-consistent fashion. Instead, we assume that at $T > 10^{4} \: {\rm K}$, the helium and the metals are in collisional ionisation equilibrium, and determine their contribution to the cooling rate using the ion-by-ion cooling rates given in \citet{gf12}. Note that at no point do we assume that hydrogen is in collisional ionisation equilibrium: its contribution to the cooling rate is determined self-consistently at all temperatures, allowing us to properly model the non-equilibrium effects that can strongly affect the cooling rate at temperatures close to $10^{4} \: {\rm K}$ \citep[see e.g.][]{Kaf73,Sutherland1993,Micic2013,rich14}.

\subsubsection{Radiative heating}
We account for the radiative heating of the gas by cosmic rays, soft X-rays, and photoelectric emission from small grains and polycyclic aromatic hydrocarbons (PAHs). We adopt a cosmic ray ionisation rate for neutral atomic hydrogen given by $\zeta_{\rm H} = 3 \times 10^{-17} \: {\rm s^{-1}}$ and a rate twice this size for the cosmic ray ionisation of H$_{2}$. The heating rate per unit volume due to cosmic ray ionisation is then given by $\Gamma_{\rm cr} = 20 \, \zeta_{\rm H} n \: {\rm eV \, s^{-1} \, cm^{-3}} = 3.2 \times 10^{-11} \zeta_{\rm H} n \: {\rm erg \, s^{-1} \, cm^{-3}}$ \citep{gl78}. To account for X-ray heating and ionisation, we use the prescription given in Appendix A of \citet{Wolfire1995}, assuming a fixed absorbing column of hydrogen ${\rm N}_\mathrm{H} = 10^{20} \: {\rm cm^{-2}}$. Finally, we model the effects of photoelectric heating using a heating rate given by
\citep{bt94,bergin04}
\begin{equation}
\Gamma_{\rm pe} = 1.3 \times 10^{-24} \epsilon G_{\rm eff} n  \: {\rm erg} \: {\rm s^{-1}} \: {\rm cm^{-3}},
\end{equation}
where $G_{\rm eff} = G_{0} \exp{\left( -2.5 A_{\rm V}\right)}$, and where the photoelectric heating efficiency $\epsilon$ is given by \citep{bt94,wolf03}
\begin{equation}
\epsilon = \frac{0.049}{1 + (\psi / 963)^{0.73}} + \frac{0.037 (T / 10000)^{0.7}}{1 + (\psi / 2500)},
\end{equation}
with
\begin{equation}
\psi = \frac{G_{\rm eff} T^{1/2}}{n_{\rm e}}.
\end{equation}
To compute $A_{\rm V}$, we use the same approach as described above in our discussion of the carbon chemistry.

\subsubsection{Chemical heating and cooling}
We account for changes in the thermal energy of the gas due to several different chemical processes: cooling from H$^{+}$ recombination, the collisional ionisation of H and the collisional dissociation of H$_{2}$, and heating from H$_{2}$ formation, H$_{2}$ photodissociation, and UV pumping of vibrationally excited states of H$_{2}$ by the ISRF. Full details of our treatment of these processes can be found in \citet{Glover2007,Glover2007b} and \citeauthor{Glover2012b}~(2012b).

\subsubsection{Dust}
We self-consistently determine the dust temperature in every grid cell by assuming that the dust is in thermal equilibrium and solving for the dust temperature $T_{\rm d}$ for which the heating of the dust grains by the ISRF and by collisions with the gas is balanced by their thermal emission. We assume that the dust-to-gas ratio is 0.01, and that the dust has a size distribution typical for the Milky Way. For the dust opacities, we use values from 
\citet{Mathis1983} for wavelengths shorter than 1 $\mu$m and from \citet{Ossenkopf1994} for longer wavelengths. To treat the attenuation of the ISRF by dust absorption, which reduces the heating rate of the dust in dense regions of the ISM, we use the column density dependent attenuation factor $\chi(N_{\rm H})$ computed by \citeauthor{Glover2012b}~(2012b), together with the values of $N_{\rm H}$ provided by the TreeCol algorithm (see below).

\subsection{Modeling molecular self-shielding and dust shielding using TreeCol}\label{SEC_TREECOL}
In order to model the attenuation of the ISRF by H$_2$ self-shielding, CO self-shielding, the shielding of CO by H$_{2}$, and absorption by dust, we need a fast but accurate method for computing the column densities of hydrogen nuclei, H$_{2}$ and CO along many different sight-lines in our simulations. We do this by means of the {\sc TreeCol} algorithm \citep{Clark2012}. This algorithm uses information stored in the oct-tree structure that {\sc Flash} also uses to compute the effects of self-gravity (see section~\ref{SEC_GRAV}) to compute $4\pi$ steradian maps of the dust extinction, and H$_{2}$ and CO column density distributions surrounding each grid cell in the computational domain. These maps are discretised onto $N_{\rm pix}$ equal-area pixels using the  {\sc healpix} algorithm \citep{Gorski2011}. We can then compute values of $f_{\rm dust, H_{2}}$, $f_{\rm dust, CO}$, $\chi$, etc.\ for each pixel, as outlined above, and finally can determine a single mean value for each cell simply by taking the arithmetic mean of the values for the individual pixels. In the simulations presented here, we use $N_{\rm pix} = 48$ pixels. We have explored the effects of increasing $N_{\rm pix}$, but find that at our current level of hydrodynamical resolution, and with the employed tree opening angle, it makes little difference to the outcome of the simulations. 

The version of {\sc TreeCol} described in \citet{Clark2012} includes all of the gas between the grid cell of interest and the edge of the simulation volume in its calculation of the column densities. However, we cannot use the same approach in our present simulations, as we expect that the typical separation between individual UV sources at the given disc surface density will be much less than 500~pc. We therefore use a similar strategy to that used by \citet{Smith2014}: we define a shielding length $L_{\rm sh} = 50 \: {\rm pc}$ and only include gas located at $L \leq L_{\rm sh}$ in our calculation of the column densities, where we allow the search radius to extend across periodic boundaries. In practice, we do not expect our results to be particularly sensitive to moderate variations in $L_{\rm sh}$.

In addition, we have improved and optimised the algorithm of \citet{Clark2012} significantly, leading to a speedup of more than a factor of 10 with respect to the original implementation. Details of the {\sc Flash} implementation of {\sc TreeCol} are presented in \citet{Wunsch2014}.

In the runs that we perform with our standard supernova rate of $15 \: {\rm Myr}^{-1}$, we use the estimate of \citet{Draine1978} for the strength of the UV radiation field, and set $G_{0} = 1.7$. For the runs with different supernova rates, we assume that $G_{0}$ scales linearly with the supernova rate, since the main contributions to the strength of the UV radiation field comes from the same massive stars that will ultimately explode as supernovae. For simplicity, in this first set of runs we take $G_{0}$ to be spatially constant. 


\subsection{Self-gravity} \label{SEC_GRAV}
We solve Poisson's equation
\begin{equation}
\Delta \Phi = 4 \pi G \rho \label{EQ_Poisson}
\end{equation}
with a tree-based algorithm developed by R. W\"unsch, which is part of the {\sc Flash} 4.1 release. It is a Barnes-Hut type octal-spatial tree \citep{Barnes1986} described in detail in \citet{Wunsch2014}. 
However, our implementation employs several modifications to this algorithm, which we briefly describe here. 

We use the tree to calculate the gravitational potential and also use the information stored in it to model the attenuation of the ISRF, as described in section \ref{SEC_TREECOL} above. Since both the calculation of the gravitational potential and that of the attenuation of the ISRF share the same tree walk, the overhead costs of the latter are minimal. Moreover, the AMR grid used in our simulations is already organised into an octal tree allowing a further increase of the code efficiency by making use of the existing data structures.

Our implementation of the tree uses only monopole moments, i.e.\ it stores total masses, masses of $\mathrm{H}_2$ and $\mathrm{CO}$ components and mass centre positions of the tree nodes. Therefore, to reach a specified accuracy when computing the gravitational potential, a larger number of tree nodes have to be taken into account than in the case of using quadrupole or higher order moments. However, the oct-tree with monopole moments can be constructed in an extremely memory efficient way and consequently, computational costs are spared due to efficient cache usage and faster network communication \citep{Springel2005, Wetzstein2009}.

In this work, a {\sc Gadget2}-style multipole acceptance criterion (MAC; see \citealt{Springel2005}) is used to decide whether a contribution of a given node to the gravitational potential or to the column density is accurate enough, or whether children of the node should be opened. The node is accepted if  
\begin{equation}
\label{eq:MAC:APE}
D^{4} > \frac{GMh^2}{\Delta a_\mathrm{p}}
\end{equation}
where $D$ is the distance between the cell where the potential and the column density are calculated (target cell) and the mass centre of the tree node, $h$ is the node size (edge of the corresponding cube), $M$ is the node mass, $G$ is the gravitational constant and $\Delta a_\mathrm{p}$ is the estimated error in the gravitational acceleration produced by the contribution from the node. Controlling the code accuracy by setting $\Delta a_\mathrm{p}$ is not as safe as criteria introduced by \citet{SalmonWarren1994} that allow control of the total error in gravitational acceleration, resulting from contributions of all nodes. However, \citet{Springel2005} shows that the worst-case errors, resulting from situations when the target cell is inside the contributing node, can be avoided by introducing an additional criterion that does not allow the target cell lie within a volume of the node increased by a certain factor, $\eta_\mathrm{SB}$. Here we follow this approach, setting $\eta_\mathrm{SB} = 1.2$.

Communicating the whole tree to all processors would lead to prohibitively high bandwidth and memory requirements. Therefore, only the top part of the tree, between the root node and the nodes, which correspond to leaf blocks of the AMR grid, is communicated globally. Lower tree levels, between AMR leaf blocks and individual grid cells, are communicated only to processors where they are needed during the tree walk. Parts of the tree that need to be communicated are determined by traversing the tree for block centres only and evaluating the MAC. After that, the main tree walk proceeds by traversing the tree in a simple 'depth-first' way, running independently on each processor for all local grid cells. This part of the algorithm typically uses the majority of computational time and since it does not include any communication the code reaches almost linear scalability up to at least 2048 processors \citep{Wunsch2014}.

During the tree walk, the gravitational potential is calculated by the modified Ewald method \citep{Ewald1921,Klessen1997}. In order to take into account contributions of an infinite number of periodic copies of each tree node efficiently, the node contribution to the gravitational potential is split into a short--range and a long--range part ($GM/D = GM \mathrm{erf}{(\alpha D)}/D + GM \mathrm{erfc}{(\alpha D)}/D$) with $\alpha$ being a number between $0$ and $1$. The Ewald method utilises the fact that the sum over all periodic node copies of the first term converges rapidly in physical space, while the sum of the second term converges rapidly in Fourier space. Since in our setup, the
periodic node copies exist only in two directions (we use mixed boundary conditions which are periodic in $x$ and $y$, but not in the $z$-direction), the standard equations of the Ewald method were modified by calculating analytically the limit of the periodic box size in the third (isolated) direction going to infinity, taking into account an increasing number of wave-numbers in that direction. The appropriate equations are given in \citet{Wunsch2014}.


\section{Simulation Setup}\label{SEC_Setup}

\subsection{Gas distribution}
We choose a box size of 500 pc x 500 pc x $\pm$ 5 kpc.
Our initial conditions are motivated by a typical galactic disc at low redshift with solar neighbourhood properties. In particular, we study a gas surface density of $\Sigma_{_{\rm GAS}} = 10\; {\rm M}_\odot/{\rm pc}^2$. We choose a SN rate in agreement with the KS relation for this surface density (see section 3.3) and adjust the strength of the interstellar radiation field (ISRF) accordingly. The main parameters are explained below and summarised in Table \ref{table1}.

At $t=0$, we use a Gaussian profile to model the gaseous disc density distribution in the vertical direction:
\begin{equation}
\rho(z) = \rho_0 \exp{\left[ - \left(\frac{z}{h_z} \right)^2 \right ]}
\end{equation}
Thus, two parameters constrain the initial gas density profile, the scale height of the gas disc, $h_z=60$ pc, and the surface density, $\Sigma_{_{\rm GAS}}$, or equivalently, the midplane density $\rho_0= 9 \times 10^{-24} \;{\rm g\; cm}^{-3}$.
The Gaussian distribution is cut off at the height at which the disc density is equal to the uniform background density of $\rho_b =  10^{-27} \;\textrm{g cm}^{-3}$, i.e. at $|z|\approx240$ pc. The resulting total gas mass in the disc is $M_\mathrm{disc}=2.5\times 10^6\;{\rm M}_\odot$. 
The initial temperature within the gas disc is set to $5000\;$K.
 In magnetic runs we initialise the field in $x$-direction as 
\begin{equation}
B_x(z) = B_{x,0} \sqrt{\rho(z)/\rho_0}
\end{equation}
where the magnetic field in the mid plane is initialised to $B_{x,0}=3\;\mu$G.
The fiducial resolution close to the disc midplane is 3.9 pc corresponding to 128 cells in $x$- and $y$-direction. In $z$-direction we reduce the resolution by a factor of 2 above $|z|=2$ kpc.

\subsection{External Gravitational Potential}\label{SEC_GRAV_EXT}
In the MHD equations, both the gravitational acceleration due to stars and due to the gas self-gravity are included as source terms.
The gravitational acceleration due to the stellar component in the galactic disc, $g_\mathrm{ext}(z)$, is modelled with an external potential. We use an isothermal sheet, originally proposed by \citet{Spitzer1942}, in which the distribution function of stars is Maxwellian. The vertical density distribution then has the functional form
\begin{equation}
 \rho_\star(R,z) = \rho_\star(R,0)\mathrm{sech}^2 (z/2z_\mathrm{d}),
\end{equation}
where $R$ is the radial distance to the galactic centre, $z$ is the height above the disc, and $z_\mathrm{d}$ is the vertical scale height of the stellar disc. The midplane density $\rho_\star(R,0)$ is related to the surface density $\Sigma_\star(R)$,
\begin{equation}
 \rho_\star(R,0) = \frac{\Sigma_\star(R)}{4z_\mathrm{d}}.
\end{equation}
For our setups we choose a stellar surface density of $\Sigma_\star=30\,{\rm M}_\odot/\mathrm{pc}^{2}$ and a vertical scale height of $z_\mathrm{d}=100\,\mathrm{pc}$.

We solve Poisson's equation to compute the gravitational potential of the stellar disc from 
\begin{equation}
\bigtriangleup \Phi_{\star} = 4 \pi G \rho_{\star}.
\end{equation}
The gravitational acceleration caused by the stars is then found from
\begin{equation}
g_\mathrm{ext}(z) = - \frac{\partial \Phi_{\star}}{\partial z}.
\end{equation}


\begin{table*} 
\begin{center}

\begin{tabular}{clcccl}
 &   & SN rate   & $f_{_{\rm RAND}} $  & $G_0$ & \\
symbol   &  sim name       & [Myr$^{-1}$]&  & $[{\rm Habing\; field}]$&comment\\
\hline

\tikz\draw[red!20!yellow,fill=red!20!yellow] (0,0) circle (0.9ex); & S10-KS-rand-nsg & 15   & 1.0 & 1.70 & no self-gravity, random SNe \\

\hline
\tikz\draw[black,fill=black] (0,0)--(0.15,0.2) -- (0.3,0)-- (0,0); & S10-lowSN-rand  & 5   &  1.0 & 0.56 & random SNe\\
\tikz\draw[green,fill=green] (0,0)--(0.15,0.2) -- (0.3,0)-- (0,0); & S10-lowSN-peak  & 5  &  0.0 & 0.56 & peak SNe\\
\tikz\draw[blue!30!red,fill=blue!30!red] (0,0)--(0.15,0.2) -- (0.3,0)-- (0,0); & S10-lowSN-mix  & 5   &  0.5 & 0.56 & mixed SNe \\
\tikz\draw[black,fill=black] (0,0) circle (0.9ex); & S10-KS-rand  & 15   &  1.0 & 1.70 & random SNe \\
\tikz\draw[green,fill=green] (0,0) circle (0.9ex);  & S10-KS-peak  & 15   &  0.0 & 1.70 & peak SNe \\
\tikz\draw[blue!30!red,fill=blue!30!red] (0,0) circle (0.9ex);  & S10-KS-mix  & 15  &  0.5 & 1.70 &  mixed SNe\\
\tikz\draw[black,fill=black] (0,0)--(0.2,0.) -- (0.2,0.2)-- (0,0.2); & S10-highSN-rand & 45    & 1.0 & 5.10 & random SNe \\
\tikz\draw[green,fill=green] (0,0)--(0.2,0.) -- (0.2,0.2)-- (0,0.2); & S10-highSN-peak & 45    & 0.0 & 5.10 & peak SNe \\
\tikz\draw[blue!30!red,fill=blue!30!red] (0,0)--(0.2,0.) -- (0.2,0.2)-- (0,0.2); & S10-highSN-mix  & 45   & 0.5 & 5.10 & mixed SNe \\
\hline
\tikz\draw[blue,fill=blue] (0,0) circle (0.9ex);  & S10-KS-clus & 15   & 1.0 & 1.70 & clustered SNe \\
\hline

\tikz\draw[red,fill=red] (0,0) circle (0.9ex); & S10-KS-clus-mag3 & 15 & 1.0 & 1.70 &  $B_0=3 \;\mu$G, clustered SNe\\
\hline


\end{tabular}
\end{center}

\caption{List of simulation properties. Column 1 gives the symbol and colour used to denote the run in future plots, and column 2 gives the name of the run. In column 3, we list the supernova rate per Megayear. The SNe are distributed with a certain fraction of random locations, $f_{_{\rm RAND}}$, which is listed in column 4. $f_{_{\rm RAND}} =1.0$ corresponds to 100\% random driving, whereas $f_{_{\rm RAND}}=0.0$ corresponds to 100\% peak driving. Column 5 gives the scaling for the background UV field following the notation of \citet{Habing1968}. In column 6, we highlight specific properties.}
\label{table1}
\end{table*}

\subsection{Supernova treatment}
\subsubsection{Supernova feedback and positioning}\label{SEC_SN}
We model a supernova (SN) event by injecting thermal energy into a spherical injection region with radius $R_{\rm inj}$. The radius $R_{\rm inj}$ is adjusted at runtime, such that it contains a given gas mass, $M_{\rm inj}$. This means that $R_{\rm inj} \ge 4\; \Delta x$ can increase freely, where $\Delta x$ is the cell size on the highest refinement level. 

For a single explosion, we add a thermal energy of $E_{_{\rm SN}} = 10^{51}$ erg. 
We typically use $M_{\rm inj} = 800 \;{\rm M}_\odot$ for single SNe, which roughly corresponds to a star formation efficiency of $\sim$8\% for a Chabrier initial stellar mass function \citep[IMF; ][]{Chabrier2001}.  
In a low density environment ($n \lesssim 10\;{\rm cm}^{-3}$) this implementation results in a typical temperature of $\gtrsim 5\times 10^6\;{\rm K}$ within the injection regions with $R_{\rm inj} \gtrsim 15.6\;{\rm pc}$ (which corresponds to a minimum of 4 cells in radius), $M_{\rm inj} \approx 800 \;{\rm M}_\odot$ and $E_{\rm inj} = 10^{51} \;{\rm erg}$. This temperature is well above the local minimum in the cooling curve at high temperatures at a few $\times 10^5$ K \citep[see e.g.][]{Dalgarno1972}. The (momentum-generating) Sedov-Taylor phase of supernovae which explode in such low density environments is therefore resolved and we add $E_{\rm inj}$ in the form of thermal energy (see Eq. \ref{energy_eq}):
\begin{equation}
\dot{u}_{\rm inj}=\rho \frac{E_{\rm inj}}{M_{\rm inj}} \frac{1}{\Delta t}
\end{equation}

In a high density environment \citep[$n \gtrsim 100\;{\rm cm}^{-3}$; see e.g.][]{Walch2014} the condition that $R_{\rm inj} \ge 4\; \Delta x$ can typically not be met with $M_{\rm inj} = 800 \;{\rm M}_\odot$. Instead, the $M_{\rm inj}$ will be higher in high density environments. Therefore, the initial Sedov-Taylor expansion of the blast wave cannot be resolved and we switch to a momentum input scheme as described in \citet{Gatto2014}. In this case we compute the momentum input at the end of the Sedov-Taylor phase for the mean density that is found within the injection region \citep{Blondin1998}. We then distribute this momentum evenly over all cells within the injection region leading to an effective $\dot{q}_\mathrm{inj} >0$ for these cells. In addition, the cells are heated to $10^4$ K. To stabilise the simulations we modify the chemical abundances within the SN injection region, i.e. we reduce the H$_2$ abundance assuming that all H$_2$ is dissociated by the SN, while increasing the H$^+$ abundance accordingly.
At each SN event the global time step can be reduced according to a modified CFL criterion using the maximum of the global sound speed or velocity to determine the global time step.\\

We do not follow the formation of massive stars or star clusters self-consistently but choose a constant SN rate, which is informed by observations (see section \ref{SEC_SNR}). Given a certain SN rate, we generate and tabulate one random sequence of SN positions at given explosion times {\it before} the start of the simulation. This sequence is read in upon start-up. The advantage of this approach is that it enables us to accurately compare different physics models e.g. simulations without self-gravity or with additional magnetic fields, without the bias of having different star formation and therefore SN rates and positions in the respective runs. For simulations with peak driving, the SN rate is still the same. However, each SN explodes at the position of the global density maximum at the given time.\\

\begin{figure}
\begin{center}
\begin{tabular}{r r}
\includegraphics[ width=80mm]{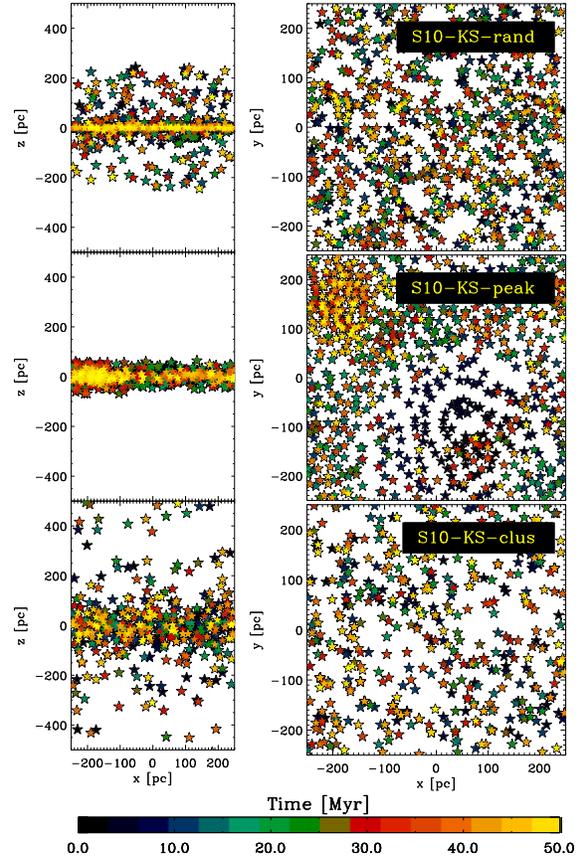} \\
\end{tabular}
\caption{Distribution of SN events within the first 50 Myr (see colour code) for (i) random driving in x-z-projection and in x-y-projection ({\it top panels}); (ii) peak driving ({\it center}); and (iii) clustered driving including Type Ia's ({\it bottom panels}).}
\label{FIG_SNe}
\end{center}
\end{figure}

The SNe are positioned in four different ways (see Fig. \ref{FIG_SNe}), where the different positioning of the SNe allows us to mimic explosions of massive stars that are still embedded within their parental molecular cloud as well as the explosion of massive stars in evolved, gas-poor environments or runaway O-stars, respectively. We use:
\begin{enumerate}
\item {\it Random driving:} the SN distribution is completely random; 
\item {\it Peak driving:} the SNe are positioned on local peaks in the density field; 
\item {\it Mixed driving:} their locations are a mix of random and local peak positions, where $f_{_{\rm RAND}}$ is the ratio of random SNe to the total number of SNe. $f_{_{\rm RAND}} =1.0$ corresponds to purely random SN driving, and $f_{_{\rm RAND}}=0.0$ corresponds to pure peak driving. 
\item {\it Clustered random driving:} here we consider the temporal and spatial correlation of Type II SNe as they stem from massive stars, which are born in associations or clusters. 60\% of the Type II SNe are assumed to explode in clusters, and 40\% are single SNe exploding at random positions. Each cluster has a pre-defined position and a constant lifetime of 40 Myr \citep{Oey1997}, and the total number of SNe per cluster, $N_\mathrm{SNe,clus}$, are drawn from a truncated power-law distribution $f(N_\mathrm{SNe,clus})=N_\mathrm{SNe,clus}^{-2}$ \citep{Clarke2002} with an upper limit of $N_\mathrm{SNe,clus}=7$ and a lower limit of $N_\mathrm{SNe,clus}=40$. Within a given cluster the temporal offset of individual SNe is set to $dt_\mathrm{SN,clus}=40 \;{\rm Myr}/N_\mathrm{SNe,clus}$. 
\end{enumerate}
For all Type II SNe we use a scale height of 50 pc \citep{Tammann1994}. 
Only in case (iv), we assume that 80\% of all SNe are Type II's, and 20\% are Type Ia's, which have a broader vertical distribution with a scale height of 325 pc \citep{Joung2006}. In this case, to keep the overall SN rate constant at the required value, we first determine the type of each explosion (Type Ia, Type II in a cluster, or Type II at a random position), and then choose the current SN position accordingly. If the type is of 'Type II SN in a cluster', we find the cluster with the smallest time difference between the current simulation time and the formation time of the cluster plus $dt_\mathrm{SN,clus}$. If no suitable cluster is available we create a new one. Otherwise, new clusters are created when an existing one has set off all its SNe after 40 Myr.

\subsubsection{Supernova rates}\label{SEC_SNR}

\begin{figure}
\begin{center}
\includegraphics[width=80mm]{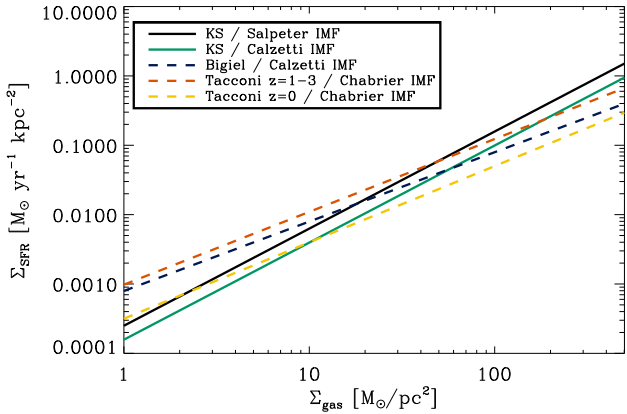}
\caption{A compilation of proposed power-law scaling relations of the SFR surface density, $\Sigma_{_{\rm SFR}}$, vs. total gas surface density, $\Sigma_{_{\rm H_2 + HI}}$ ({\it full lines}), or molecular mass surface density, $\Sigma_{_{\rm H_2}}$ ({\it dashed lines}), respectively. The scaling relations of $\Sigma_{_{\rm SFR}}$ vs. $\Sigma_{_{\rm H_2+HI}}$ follow a power law index of 1.4. The offset depends on the underlying stellar IMF. Here we show two possibilities: (1){\it Black line:} the standard Kennicutt-Schmidt (KS) relation \citep{Kennicutt1998} as derived for a Salpeter IMF \citep{Salpeter1955}, and (2) {\it green line:} the KS relation scaled with the IMF from \citet{Calzetti2007}, which is the standard IMF in STARBURST99 \citep{Leitherer1999}. 
The scaling relations of $\Sigma_{_{\rm SFR}}$ vs. $\Sigma_{_{\rm H_2}}$ are approximately linear. We show the results of (3) {\it blue line:} \citet{Bigiel2008}, who use the same IMF as (2); (4) {\it red line:} \citet{Tacconi2013}, their fit to high redshift galaxies, and (5) {\it yellow line:} \citet{Tacconi2013}, their fit to low redshift galaxies, both of which assume a Chabrier IMF \citep{Chabrier2003}.}
\label{FIG_KS}
\end{center}
\end{figure}

Since we do not follow star formation self-consistently, we have to choose a SN rate for our simulations. For this reason, we compile all popular scaling relations of $\Sigma_{_{\rm SFR}}$ vs. $\Sigma_{_{\rm GAS}}$, which have been inferred from observations, in Fig. \ref{FIG_KS}. These are (i) the Kennicutt-Schmidt relation \citep{Kennicutt1998} with $\Sigma_{_{\rm SFR}} \propto \Sigma_{_{\rm H_2+HI}}^{1.4}$, as well as (ii) newer results which relate the star formation rate surface density to the molecular gas surface density, $\Sigma_{_{\rm H_2}}$, in a galaxy in a linear fashion\footnote{Recently, using Bayesian linear regression, \citet{Shetty2013} have pointed out that the relation might actually be sub-linear. For simplicity, we do not consider this here.} \citep{Bigiel2008, Tacconi2013}. Please note that $\Sigma_{_{\rm GAS}}$ (as plotted on the x-axis in Fig. \ref{FIG_KS}) may therefore be equal to $\Sigma_{_{\rm H_2+HI}}$ or $\Sigma_{_{\rm H_2}}$ depending on which line one 
 refers to. In any case, $\Sigma_{_{\rm H_2}}$ should approach $\Sigma_{_{\rm H_2+HI}}$ for high gas surface densities, where basically all gas is in molecular form. We argue that according to Fig. \ref{FIG_KS} a factor of 3 uncertainty in $\Sigma_{_{\rm SFR}}$ for any given $\Sigma_{_{\rm GAS}}$ can easily be justified.  

We then translate $\Sigma_{_{\rm SFR}}$ into a SN rate by assuming a standard IMF \citep{Chabrier2001}, which implies that approximately one massive star forms for every $100\; {\rm M}_\odot$ of gas that is turned into stars. Now scaling this value to our simulated volume results in a standard (KS) SN rate of 15/Myr for $\Sigma_{_{\rm GAS}}=10\;{\rm M}_\odot/{\rm pc}^2$ 
(see Table \ref{table1}). We test the influence of the SN rate on the resulting ISM distribution by also performing simulations with a 3 times lower and 3 times higher SN rate ({\it lowSN} and {\it highSN} runs). The ISRF is linearly correlated with the star formation rate (see section \ref{SEC_TREECOL}) and therefore it is changed accordingly in the simulations (see Table \ref{table1}).

\section{Fiducial set of runs}\label{SEC_FIDUCIAL}
\begin{figure*}   

\begin{minipage}[b]{0.8\linewidth}
\begin{center}
  \includegraphics[ width=140mm]{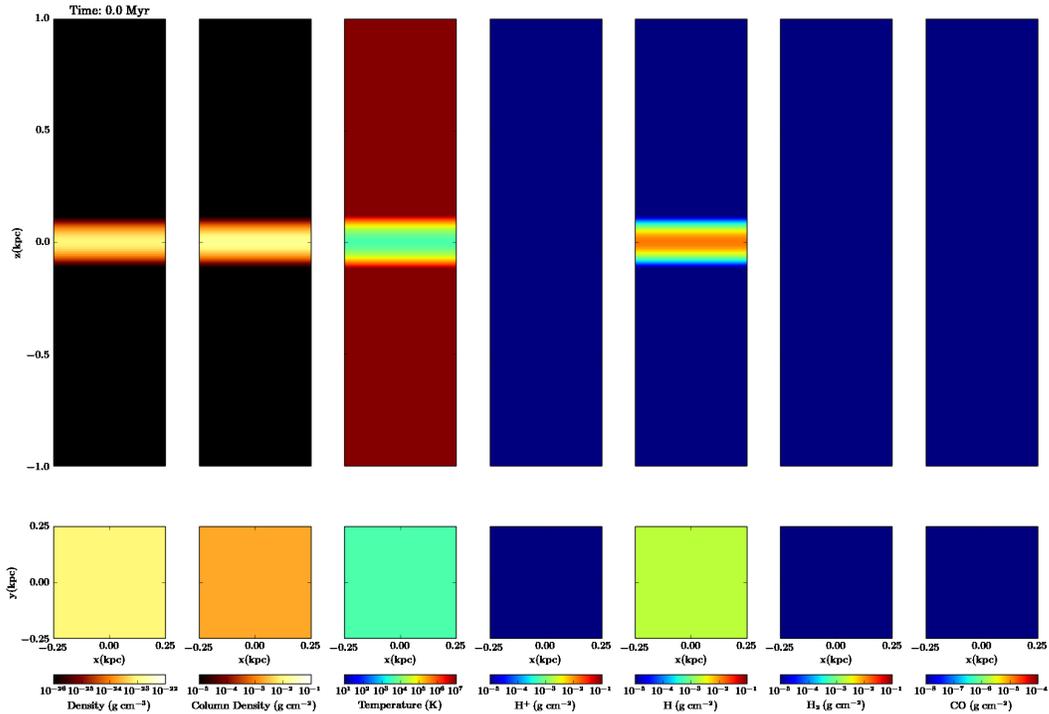} 
\end{center}
 \caption{Initial conditions for all runs around the disc midplane. {\it From left to right:} Gas density (slice), column density (projection), gas temperature (slice), and column densities of H$^{+}$, H and H$_{2}$, and CO. All slices are taken at $y=0$ (top panels), or at $z=0$ (bottom panels), respectively. Accordingly, the column density is projected along the $y$- or $z$-axis.  } \label{FIG_IC}
\end{minipage}
\end{figure*}

In our six fiducial models, we study the evolution of a stratified disc with $\Sigma_{_{\rm GAS}}=10\;{\rm M}_\odot/{\rm pc}^2$, which is driven by randomly placed or clustered SNe at a rate of 15/Myr, which corresponds to the KS value. Initially, the disc is represented by a Gaussian gas density distribution in vertical direction. The disc gas is initialised as atomic hydrogen (see Fig. \ref{FIG_IC}). When starting the simulation, the gas collapses towards the midplane due to the external gravitational potential. At the same time, it is stirred by SN-driven turbulence until a complex, multi-phase ISM emerges. In particular, we compare runs 
\begin{enumerate}
\item[1.]{\it S10-KS-rand-nsg:} without self-gravity and random driving;
\item[2.]{\it S10-KS-rand:} the standard run with random driving;
\item[3.]{\it S10-KS-peak:} using peak driving;
\item[4.]{\it S10-KS-mix:} using mixed driving with $f_{_{\rm RAND}}=0.5$;
\item[5.]{\it S10-KS-clus:} using $\sim$50\% of clustered Type II SNe, $\sim$30\% of random SNe, and 20\% Type Ia SNe with a larger scale height;
\item[6.]{\it S10-KS-clus-mag3:} MHD run with clustered SN driving.
\end{enumerate}
This set of simulations represents a sequence of runs with increasing physical complexity. As an additional branch, the influence of the SN positioning (random/peak/mix) is explored.

We also perform runs with lower and higher SN rates. These are discussed later on (see section \ref{SEC_SFR}). In Table \ref{table1} we list all of the runs and their properties.

\subsection{Example: Evolution of run {\it S10-KS-rand}}
\begin{figure*}   

\begin{minipage}[b]{0.8\linewidth}
\begin{center}
  \includegraphics[width=150mm]{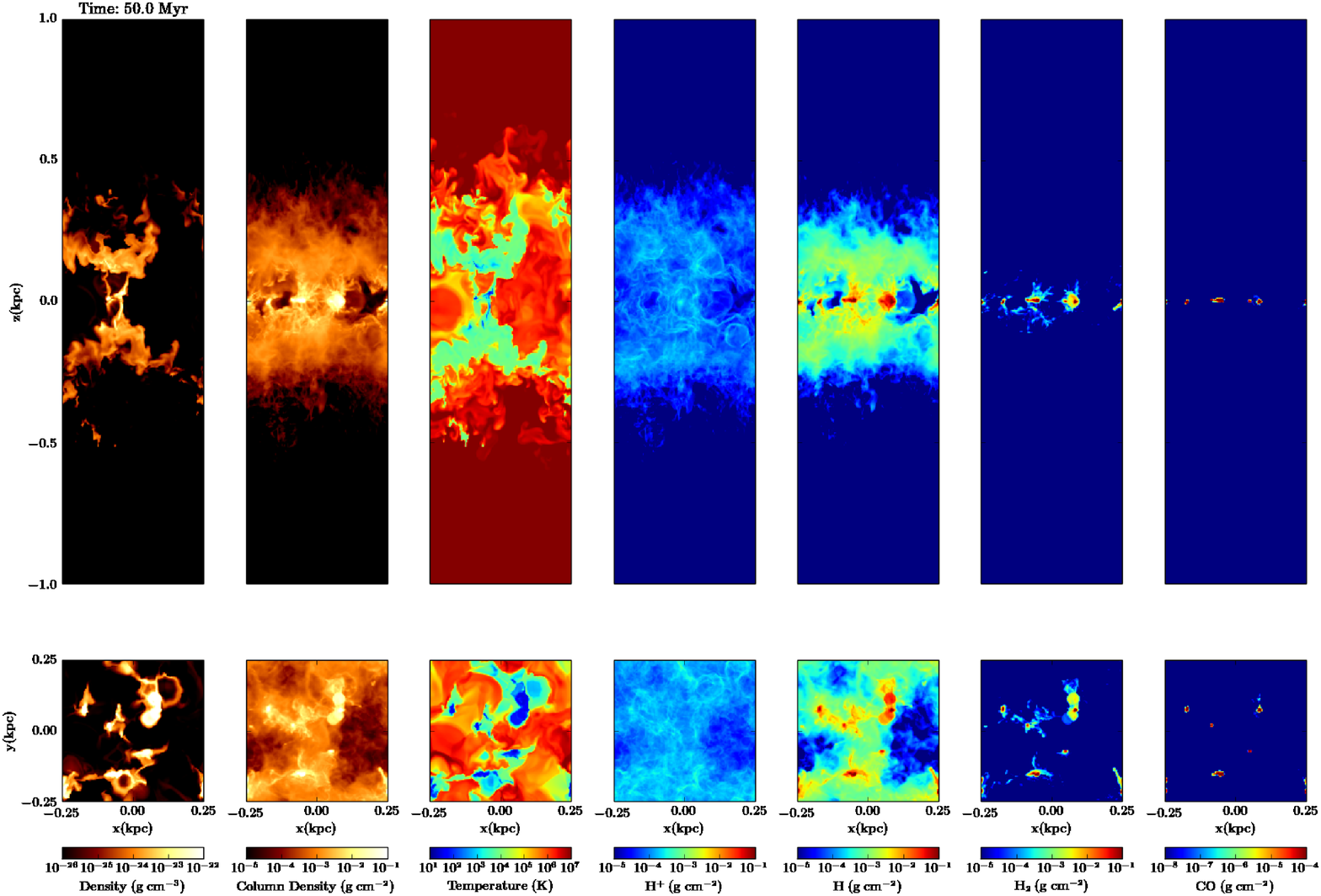} \\
  
  \includegraphics[width=150mm]{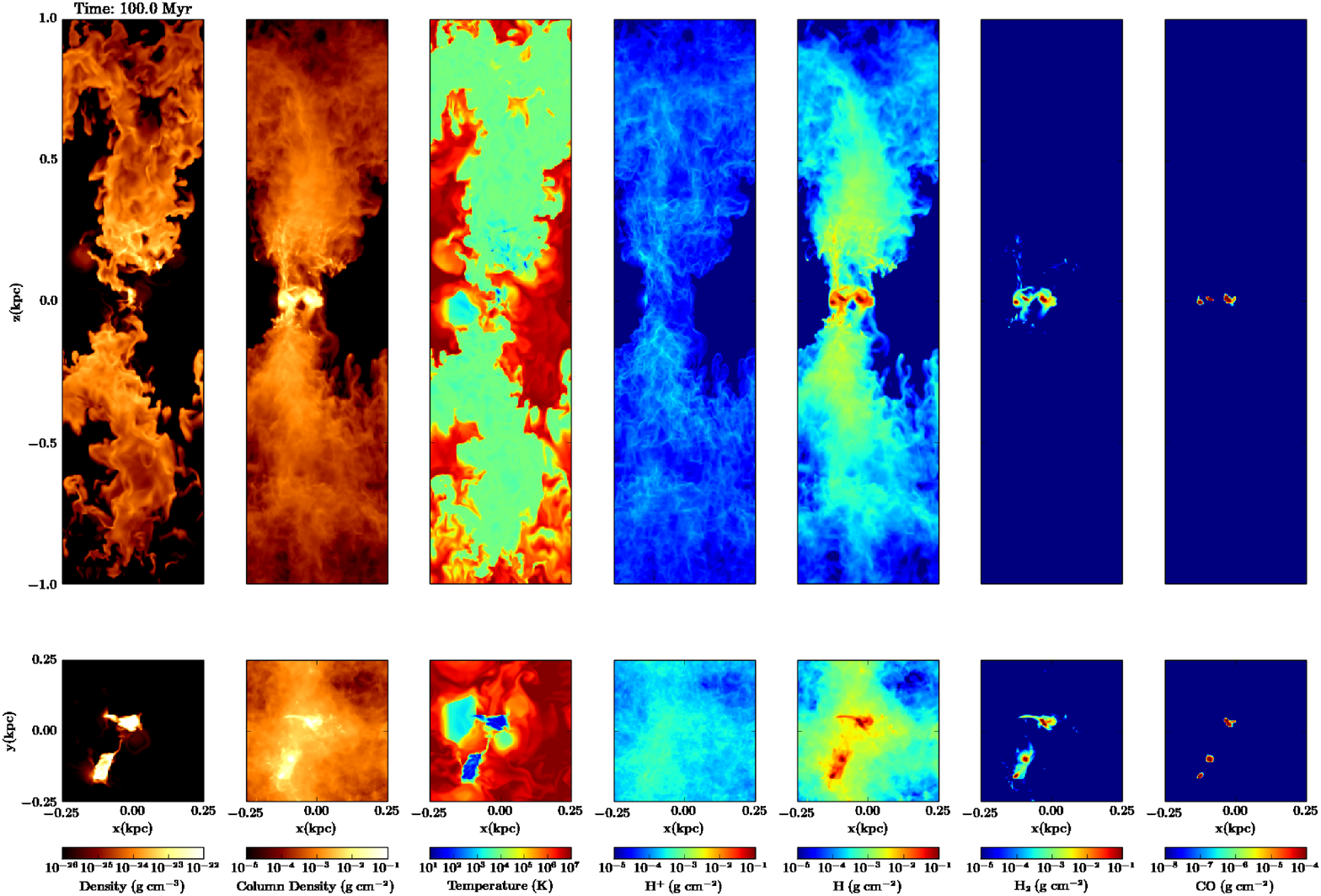} 
\end{center}
 \caption{Snapshot of run {\it S10-KS-rand} at $t=50$~Myr ({\it top}) and at $t=100$~Myr ({\it bottom}). Only a section of the simulation box, which measures 0.5~kpc $\times$ 0.5~kpc $\times$ 2~kpc, is shown. {\it From left to right:} Gas density (slice), column density (projection), gas temperature (slice), and column densities of H$^{+}$, H and H$_{2}$, and CO. All slices are taken at $y=0$ (elongated panels), or at $z=0$ (square panels), respectively. Accordingly, the column density is projected along the $y$- or $z$-axis.  } \label{FIG_S10-rand-KS}
\end{minipage}
\end{figure*}

In Fig. \ref{FIG_S10-rand-KS} we show run {\it S10-KS-rand} at $t=50$ Myr and $t=100$ Myr. From left to right we plot slices of the density (in the $x$-$z$-plane and in the disc mid plane), the column density, slices of the gas temperature, and column densities of ionised hydrogen, atomic hydrogen, molecular hydrogen, and CO. The gas density and temperature distributions span a large dynamic range. Most of the cold gas is located close to the disc mid plane, where multiple molecular clouds, which are visible in H$_2$ and CO, are forming and evolving. As time goes on, the clouds evolve continuously and do not come into an equilibrium configuration. They collide, merge, and are pushed by nearby SN explosions, thereby picking up bulk motion and angular momentum. CO is embedded within H$_2$ clumps at high column densities and thus, does not trace all of the H$_2$ within a particular clump \citep[see][for a detailed analysis on galactic scales]{Smith2014}. However, the total CO mass fraction and the total H$_2$ mass fraction evolve in a similar way in all simulations (see section \ref{SEC_MASS}). Furthermore, within the dense gas self-gravity is locally important (see section \ref{SEC_MORPH} for a further discussion). 

A large fraction of the ISM is filled with hot gas, which is injected in the SN explosions. The hot gas is nicely visible in the ${\rm H}^+$ component, which is dominated by thin filaments and bubbles, which are particularly apparent in the disc midplane. Outflowing gas is lifted off the disc, mostly in the form of atomic hydrogen. The bulk of this material moves with velocities of $\sim 10$ km/s (at maximum, with the sound speed of the respective gas component) and consequently reaches a distance of $\sim 1$ kpc after $\sim 100$ Myr. Since the gas velocity is smaller than the escape velocity of this galaxy it is expected to return and thus, form a galactic fountain. The galactic outflow rates and the multi-phase structure of the outflow are discussed in a companion paper \citep[][hereafter SILCC2]{Girichidis2014}. 

\begin{figure*}   

\begin{minipage}[b]{0.8\linewidth}
\begin{center}
  \includegraphics[width=150mm]{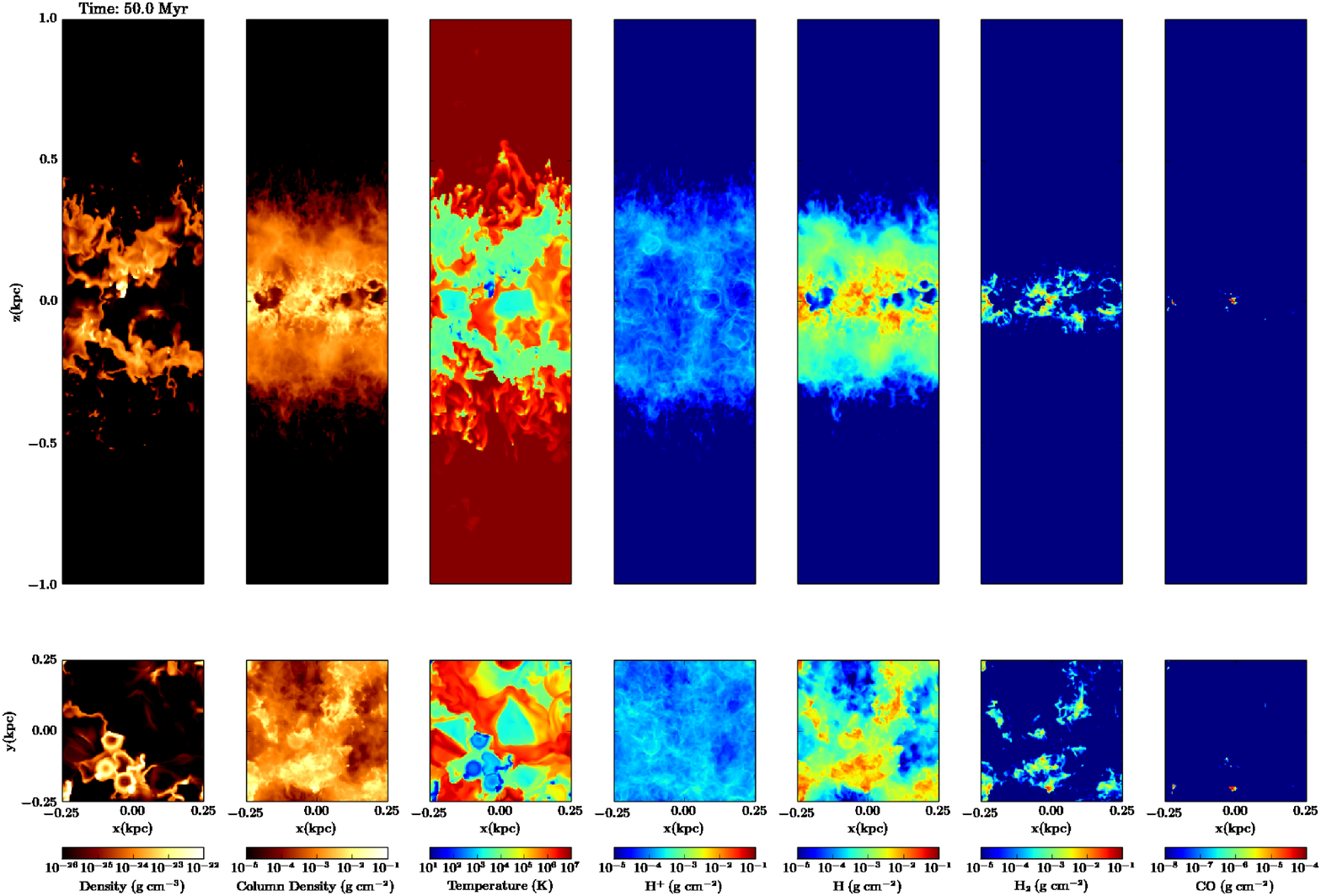} \\
  
  \includegraphics[width=150mm]{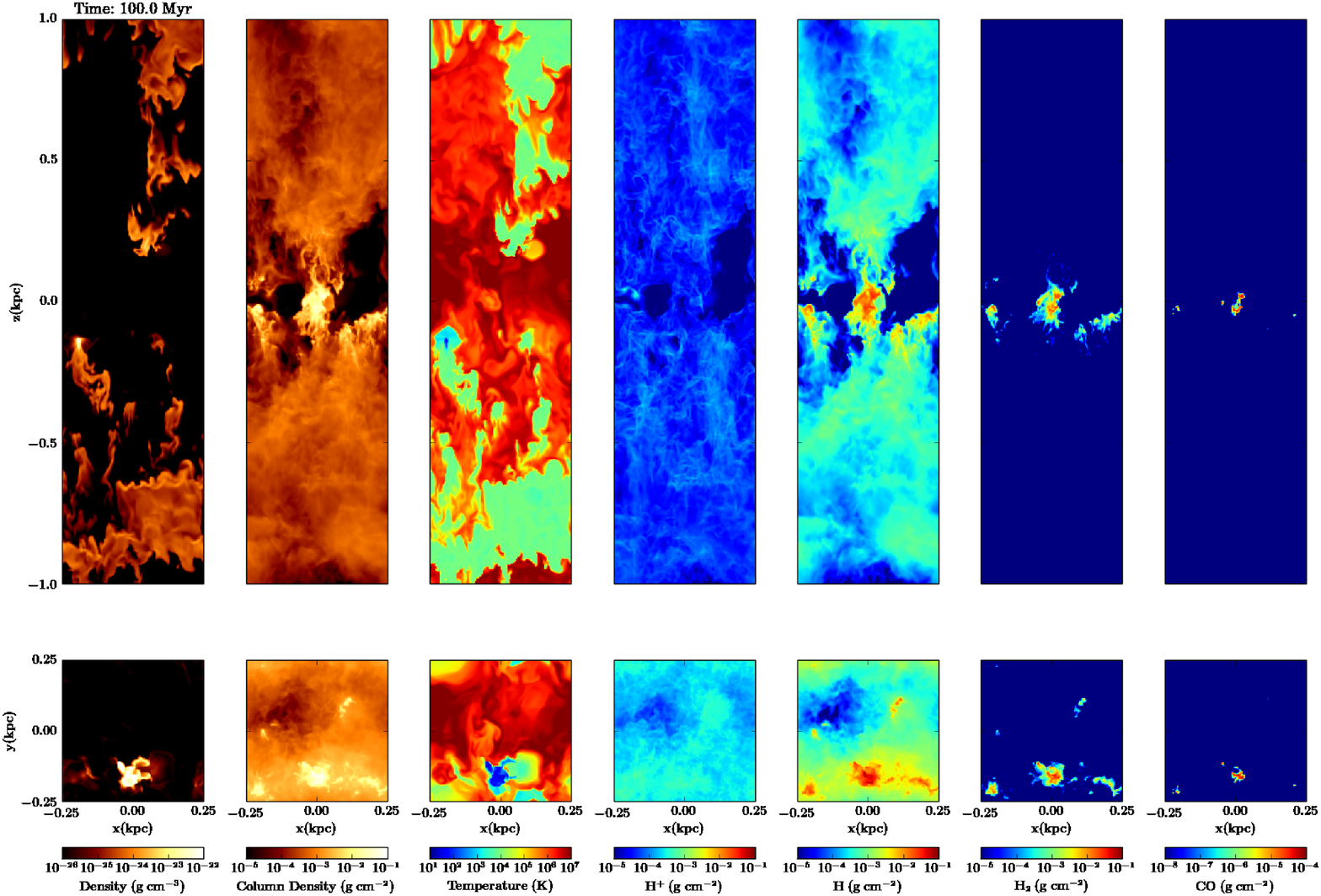} 
\end{center}
 \caption{Same as Fig. \ref{FIG_S10-rand-KS} but for run {\it S10-KS-rand-nsg} without self-gravity at $t=50$~Myr ({\it top}) and at $t=100$~Myr ({\it bottom}).  } \label{FIG_S10-rand-nsg-KS}
\end{minipage}
\end{figure*}

\begin{figure*}   

\begin{minipage}[b]{0.8\linewidth}
\begin{center}
  \includegraphics[width=150mm]{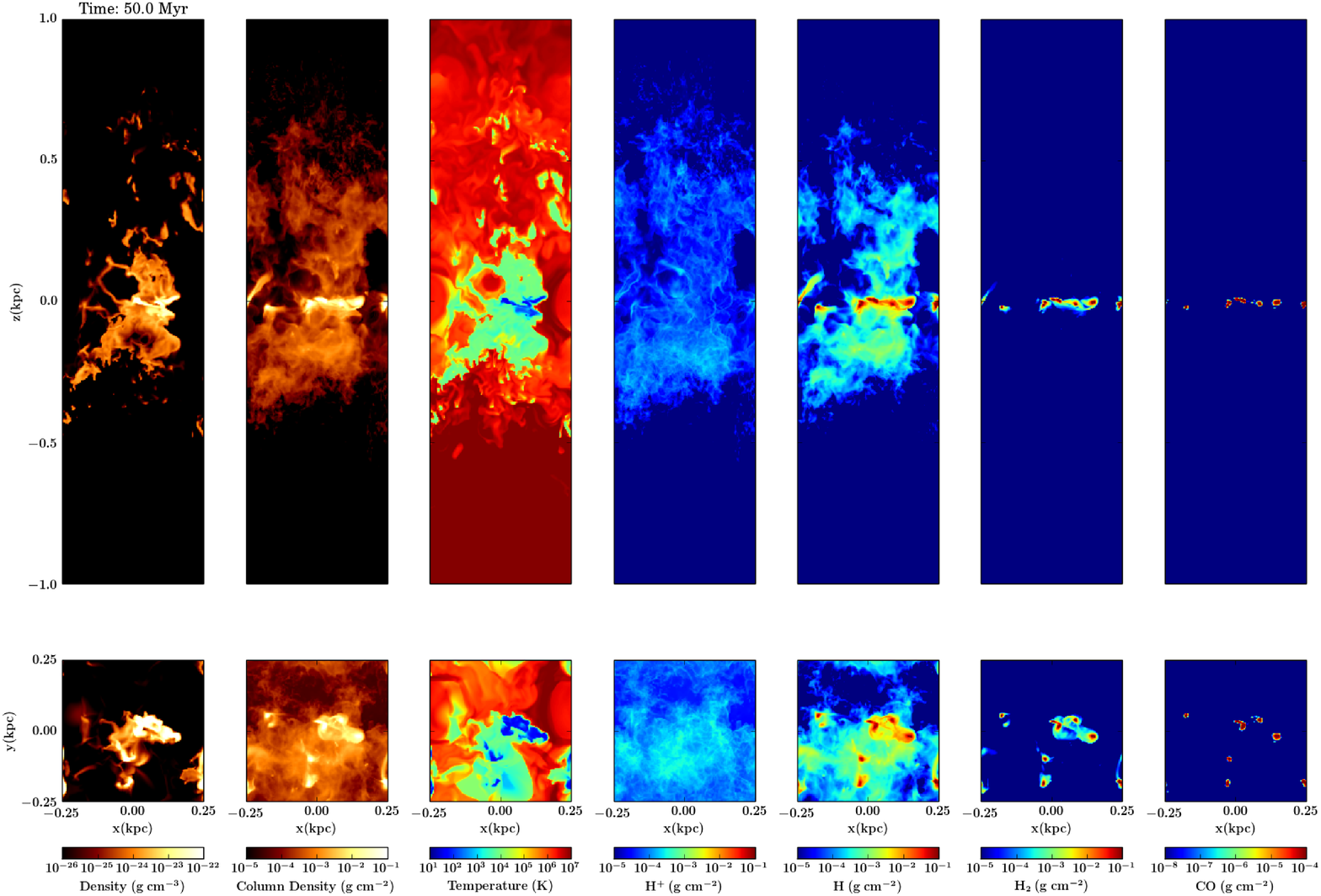} \\
  
  \includegraphics[width=150mm]{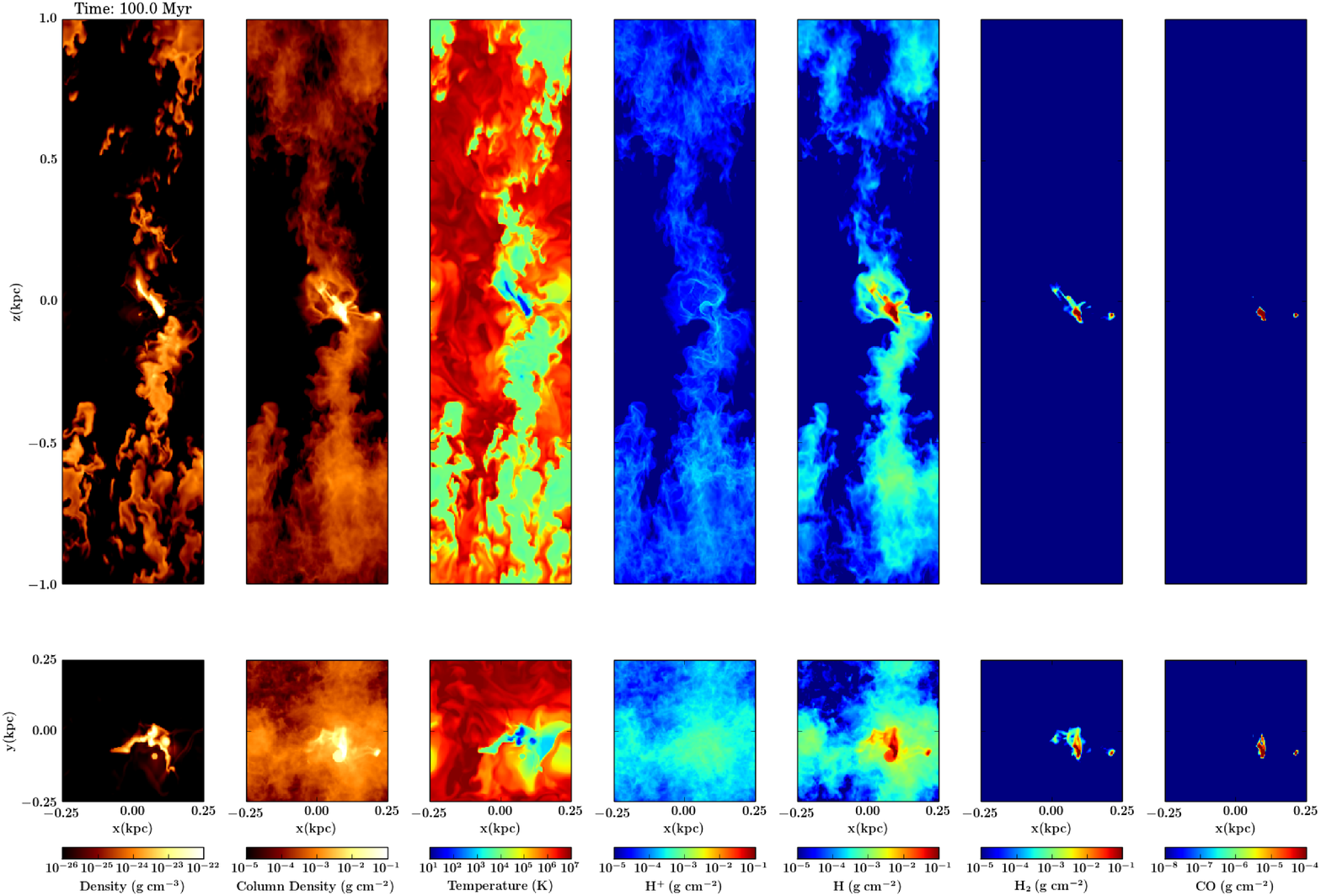} 
\end{center}
 \caption{Same as Fig. \ref{FIG_S10-rand-KS} but for run {\it S10-KS-clus} with clustered SN driving at $t=50$~Myr ({\it top}) and at $t=100$~Myr ({\it bottom}).  } \label{FIG_S10-clus-KS}
\end{minipage}
\end{figure*}
\begin{figure*}   

\begin{minipage}[b]{0.8\linewidth}
\begin{center}
  \includegraphics[width=150mm]{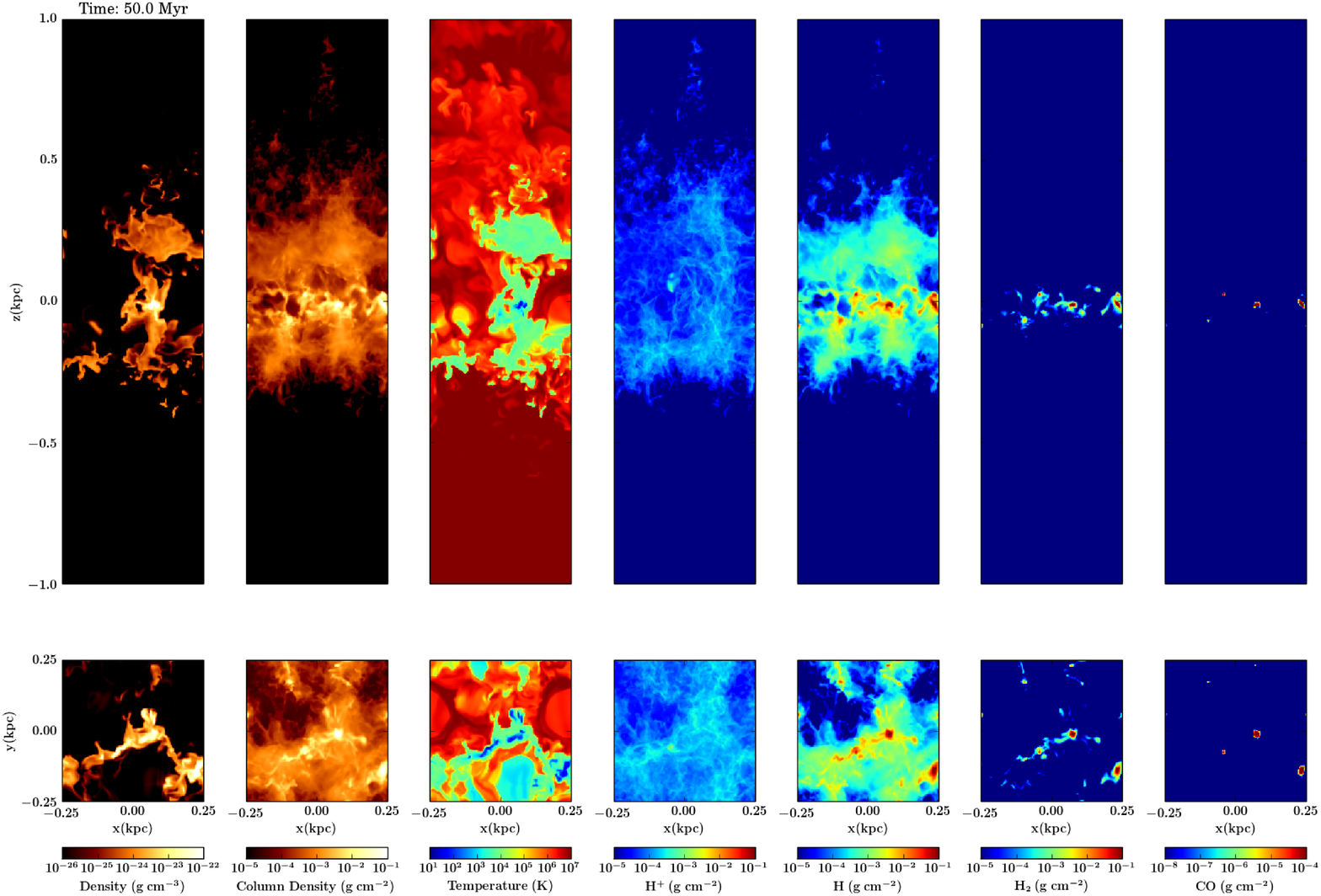} \\
  
  \includegraphics[width=150mm]{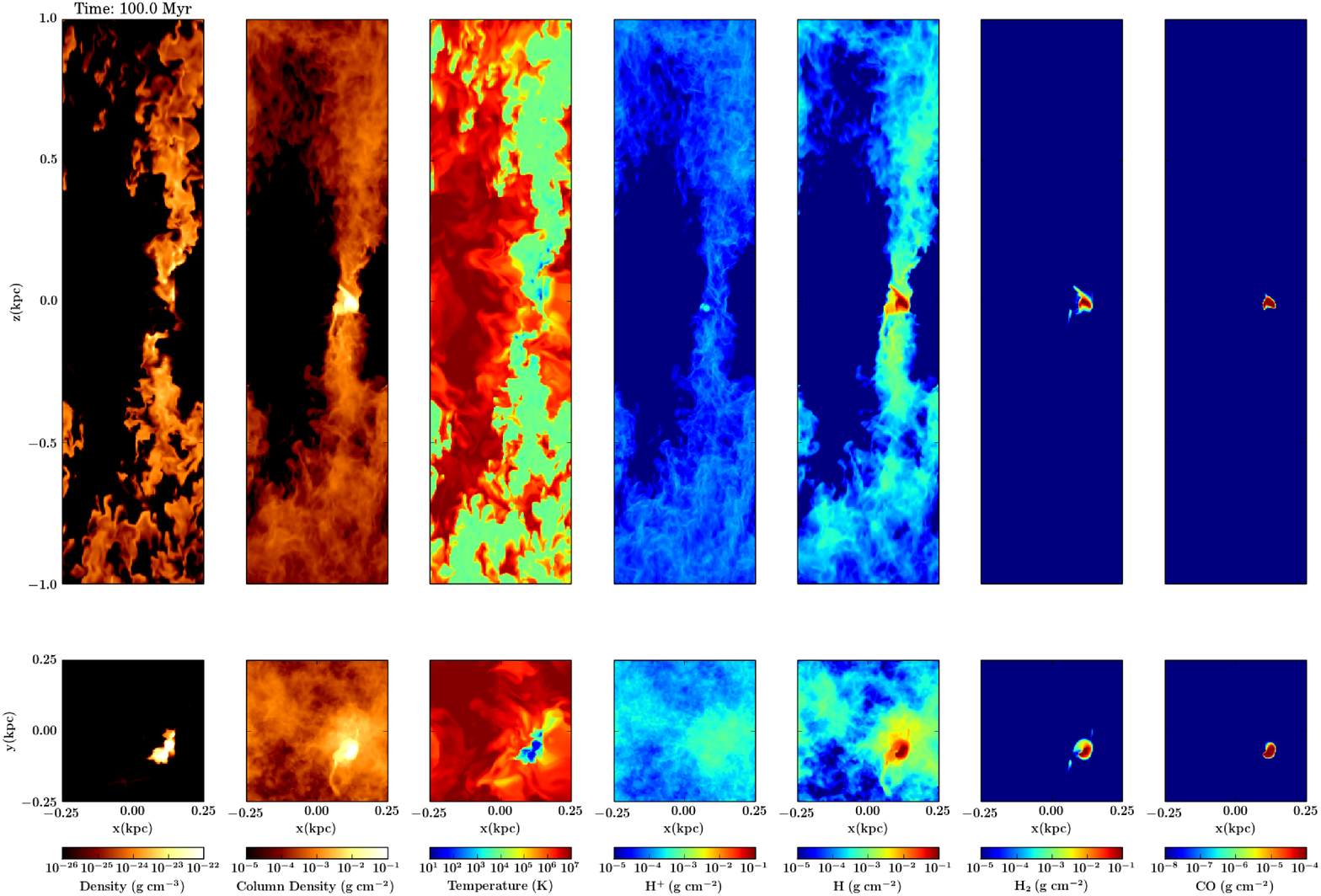} 
\end{center}
 \caption{Same as Fig. \ref{FIG_S10-rand-KS} but for run {\it S10-KS-clus-mag3} with clustered SN driving and magnetic fields at $t=50$~Myr ({\it top}) and at $t=100$~Myr ({\it bottom}).  } \label{FIG_S10-clus-mag-KS}
\end{minipage}
\end{figure*}

\begin{figure*}   
   \begin{minipage}[b]{0.8\linewidth}
   \begin{center}
  \includegraphics[ width=150mm]{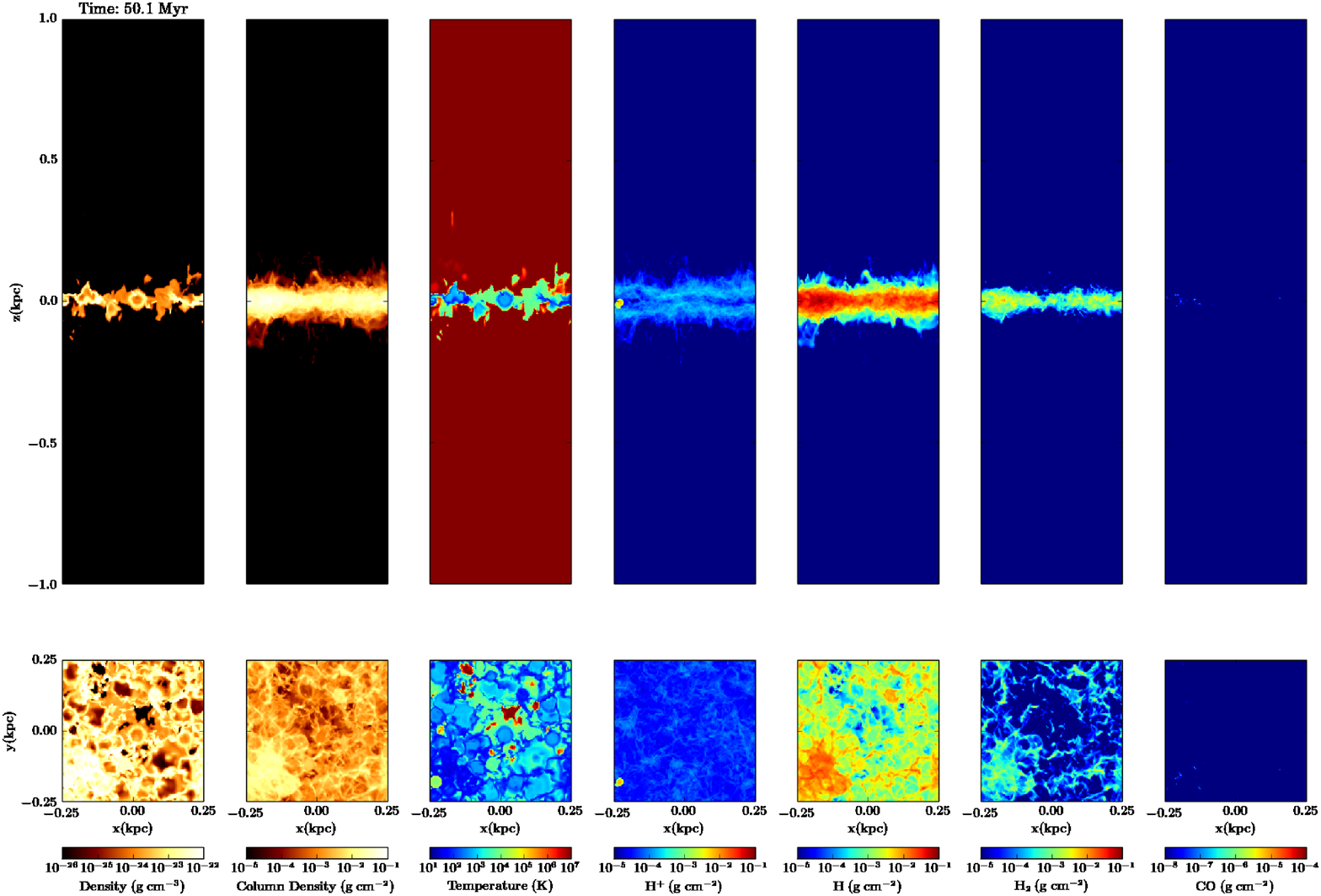} \\
  \vspace{0.5cm}
  \includegraphics[width=150mm]{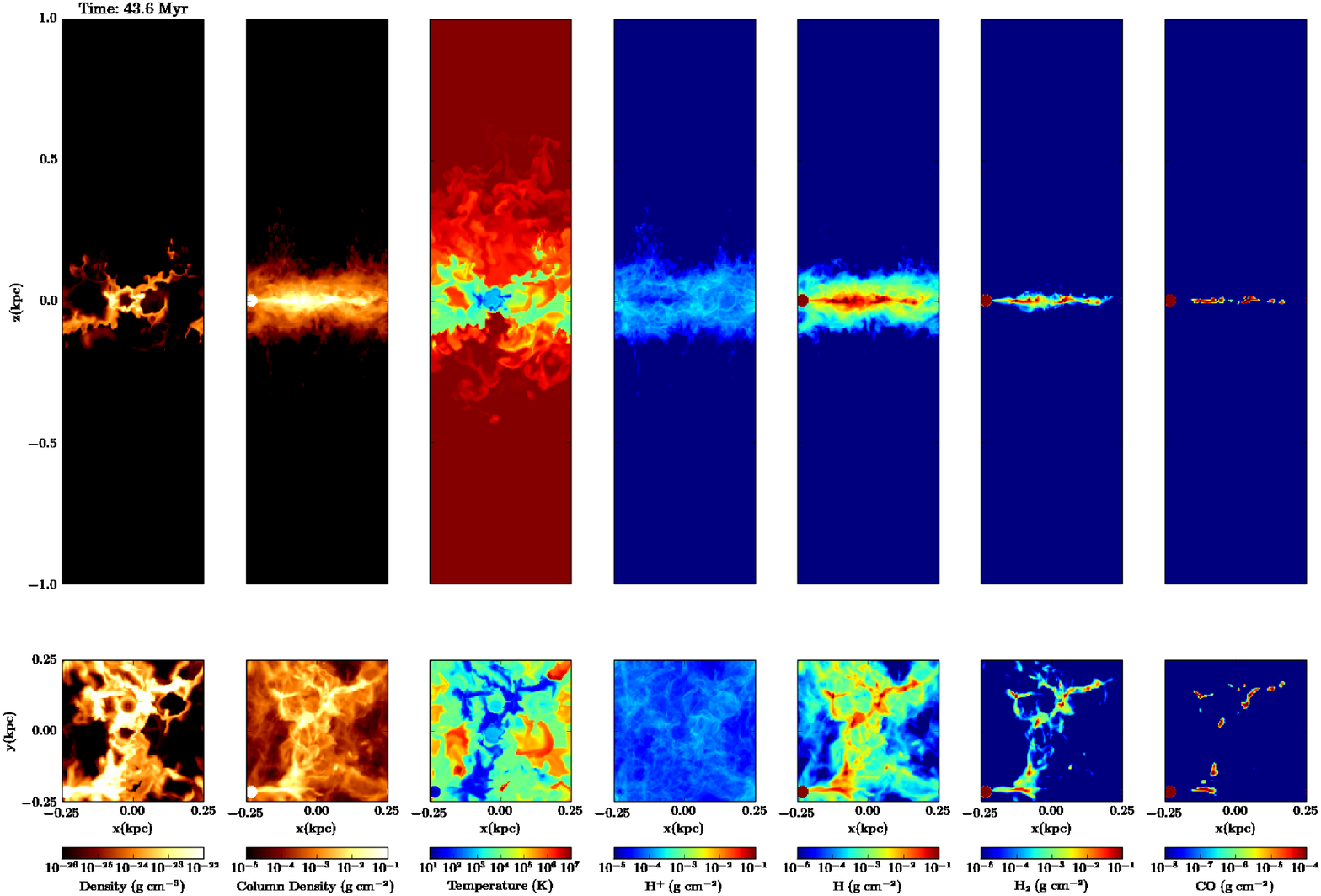} 
 \end{center}
  \caption{Same as Fig. the run with peak driving {\it S10-peak-KS} at t$=50$~Myr ({\it top}) and for mixed driving {\it S10-mix-KS} at the final snapshot of this run at $t=43.6$~Myr ({\it bottom}). \label{FIG_S10-peak-KS}} 
\end{minipage}
\end{figure*}

\subsection{Comparison: Qualitative evolution of all fiducial runs}
In Fig. \ref{FIG_S10-rand-KS}, we plot the disc structure for run {\it S10-KS-rand-nsg} at $t=50$ Myr and $t=100$ Myr in Fig. \ref{FIG_S10-rand-nsg-KS}; for run {\it S10-KS-clus} at $t=50$ Myr and $t=100$ Myr in Fig. \ref{FIG_S10-clus-KS}; for run {\it S10-KS-clus-mag3} at $t=50$ Myr and $t=100$ Myr in Fig. \ref{FIG_S10-clus-mag-KS}; and for runs {\it S10-KS-peak} and {\it S10-KS-mix} at $t\approx50$ Myr in Fig. \ref{FIG_S10-peak-KS}. The simulations with peaked and mixed driving did not continue until $t\gtrsim50$ Myr due to the accumulation of $>90$\% of the total mass in a single dense clump, which cannot be resolved. This will be discussed further below. 

Compared with {\it S10-KS-rand}, the run without self-gravity ({\it S10-KS-rand-nsg}) shows a disc structure which is less concentrated to the midplane. Consistently, also the larger-scale structure and the developing outflow appear more extended. There is molecular gas present, but there seems to be less of it, in particular less CO (see section \ref{SEC_MASS} for a more quantitative analysis). 

In run {\it S10-KS-clus} with clustered SN driving, we can see the opposite effect. Compared with {\it S10-KS-rand}, there is a bit more molecular gas and the disc is much more concentrated to the midplane with less mass in atomic hydrogen being entrained in the outflow. Although there are many molecular clumps at $t=50$ Myr, there is only one main molecular cloud left towards the end of the simulation at $t\approx 100$ Myr. Thus, with clustered instead of random driving, the clouds seem to merge more rapidly. With magnetic fields (run {\it S10-KS-clus-mag3}; Fig. \ref{FIG_S10-clus-mag-KS}), we see a very similar evolution of the molecular clumps as in {\it S10-KS-clus}. On larger scales, however, the disc is slightly more extended and similar to run {\it S10-KS-rand}.

With peak and mixed driving the discs are highly concentrated to the midplane. All components have a small scale height and there is no outflow developing, in particular in the case of peak driving. The top-down view in Fig. \ref{FIG_S10-peak-KS} shows a less clumpy and more filamentary distribution of the dense gas. For peak driving, there is little H$_2$ and basically no CO present at this point in time. In the mixed driving case, a network of filamentary, dense molecular clumps develops.


\begin{figure}  
   \begin{minipage}[b]{1.0\linewidth}
   \begin{center}
  \includegraphics[width=80mm]{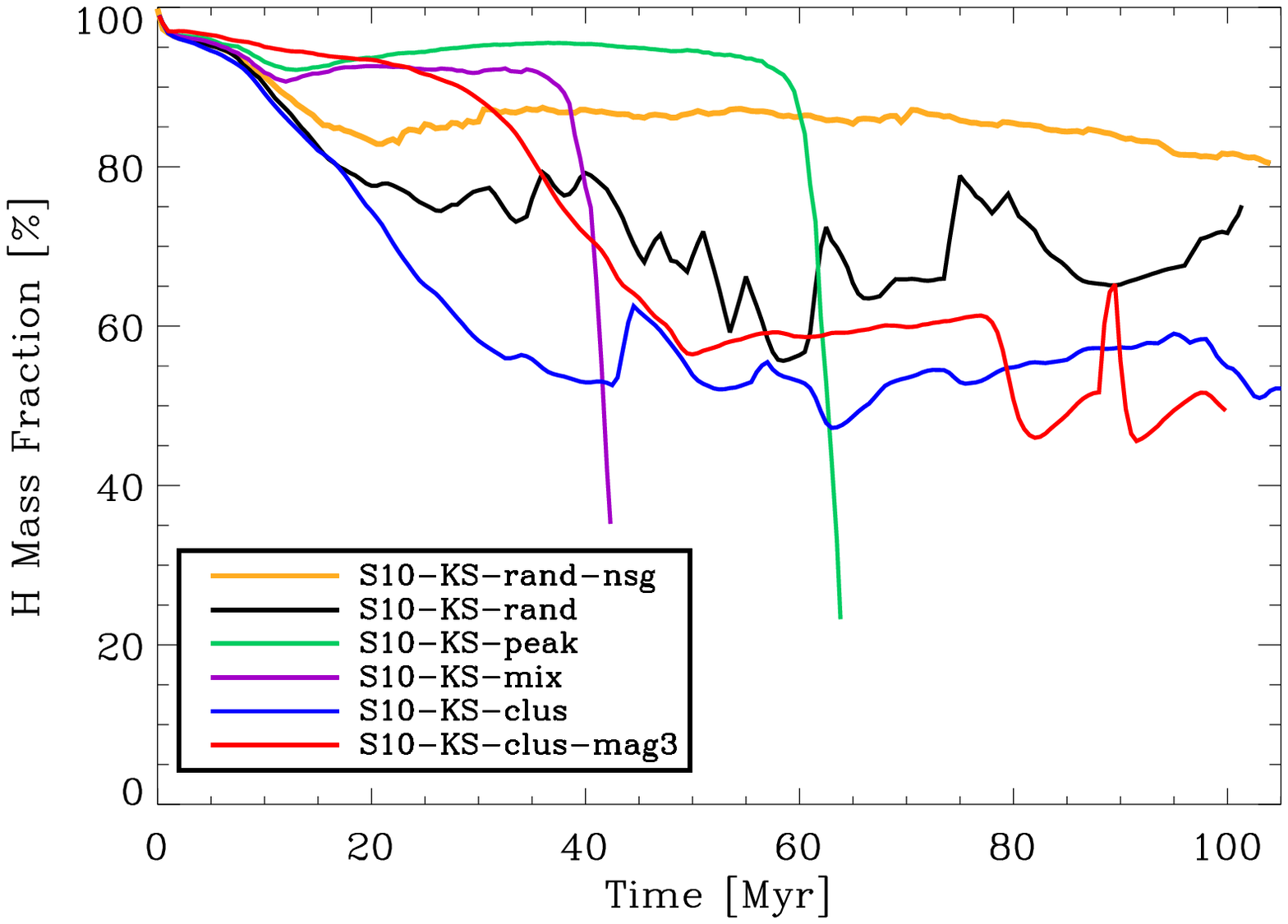}
  \includegraphics[width=80mm]{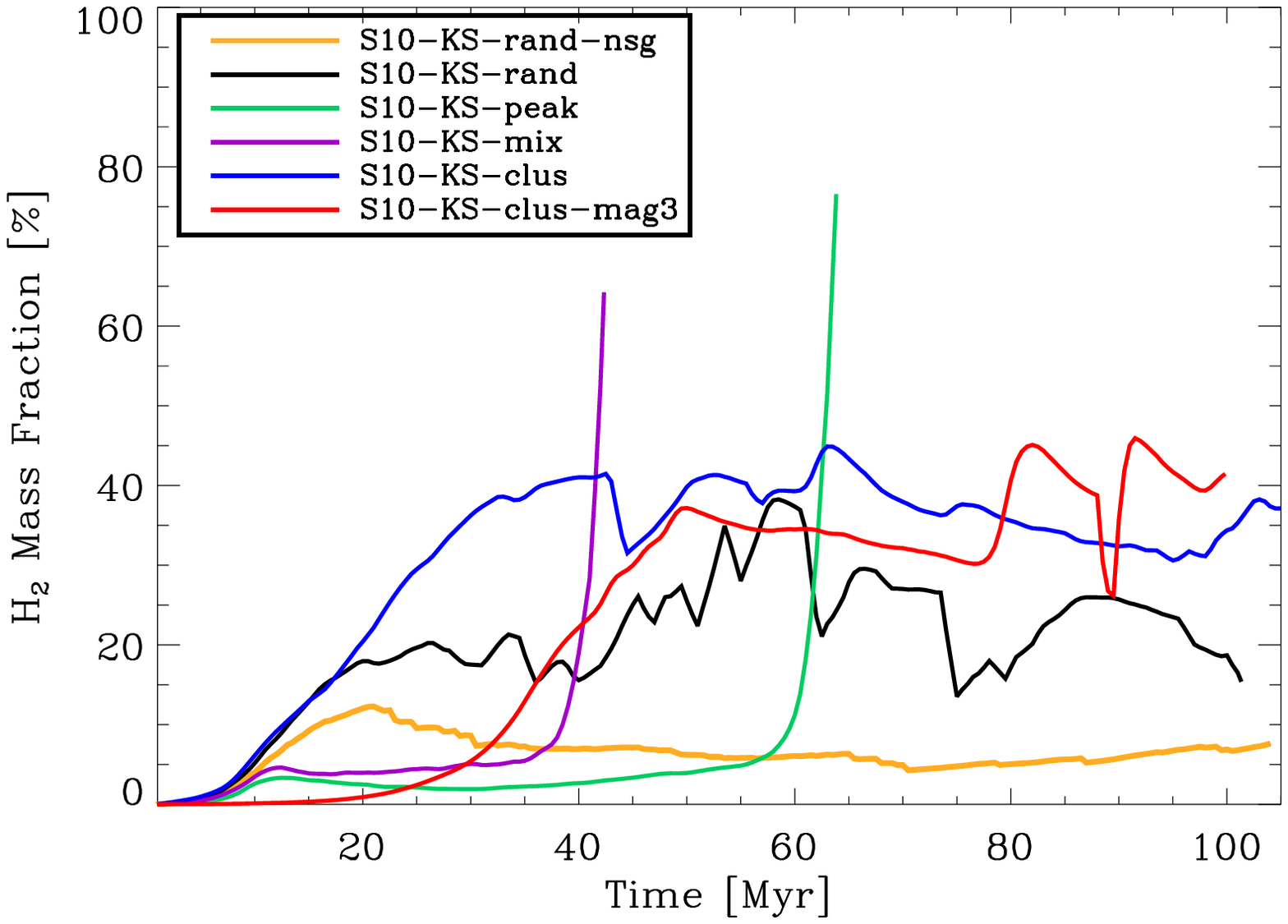}
    \includegraphics[width=80mm]{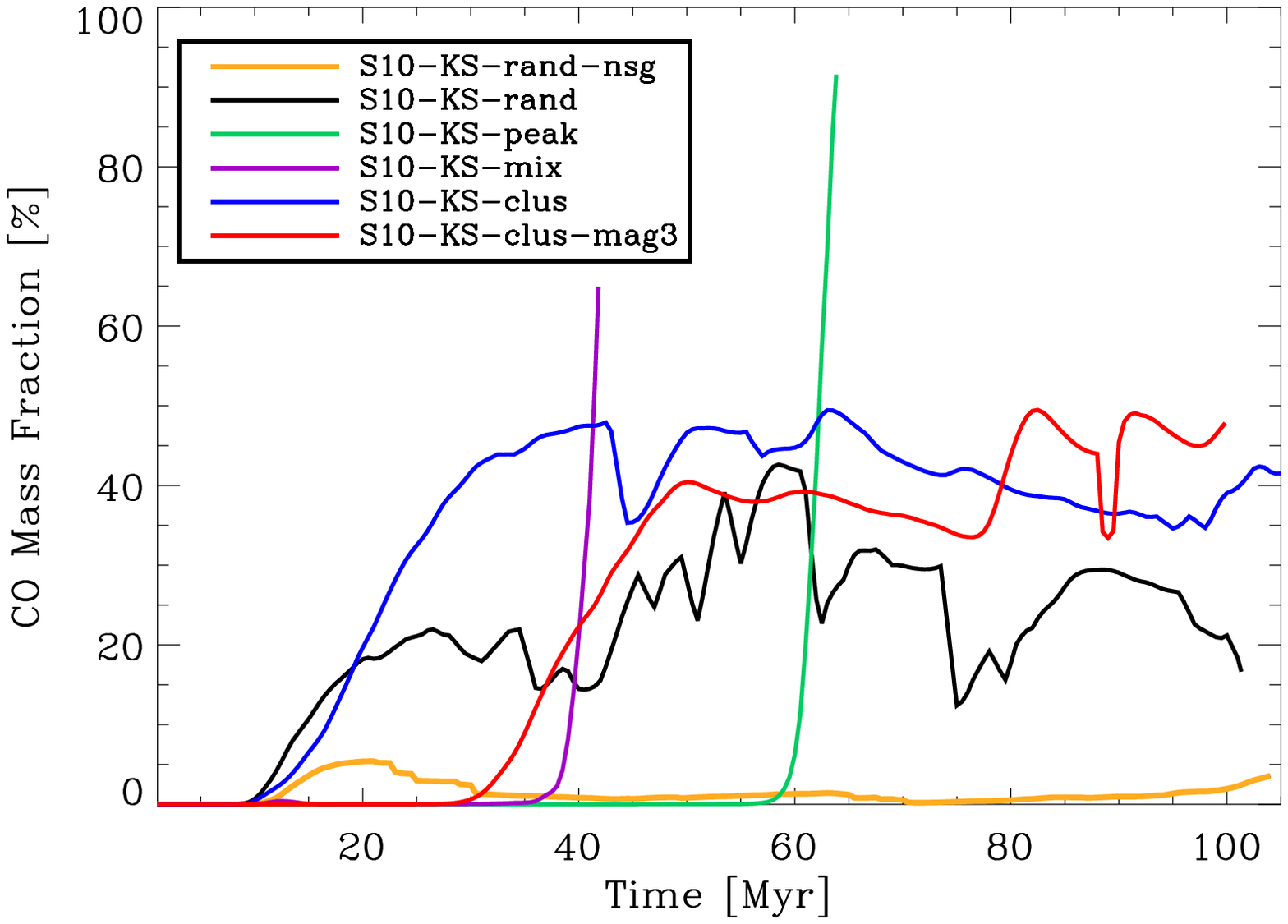}

 \end{center}
  \caption{H ({\it top}) and H$_2$ ({\it middle}) mass fractions (with respect to the total hydrogen mass) and CO mass fraction ({\it bottom}; with respect to the total mass in carbon) as a function of time for all main runs with $\Sigma_{_{\rm GAS}} = 10\;{\rm M}_\odot/{\rm pc}^2$ using the KS SN rate, i.e. 15/Myr. Random and clustered driving with/without magnetic fields evolve similarly. Runs with peak or mixed driving first form very little H$_2$, but later on diverge due to the formation of a single massive clump. Without self-gravity, the H$_2$ fraction stays below 10\%. \label{FIG_MASS1} }
\end{minipage}
\end{figure}

\subsection{Chemical evolution of the discs}\label{SEC_MASS}
In Fig. \ref{FIG_MASS1} we show the total mass fractions (with respect to the total hydrogen mass) in atomic hydrogen, H, molecular hydrogen, H$_2$, and carbon monoxide, CO, as a function of time. Overall, the H distribution is roughly the inverse of the H$_2$ plus CO mass fractions. A further analysis shows that the mass fraction of ionised gas (not shown here) remains roughly constant at $\sim$10\%. The total CO/H$_2$ fraction appears to be roughly constant in time for each of the simulations. 
 
For run {\it S10-KS-rand} (black line), $\sim$ 20\% of the total mass is converted to H$_2$ within the first 20 Myr. At $t\approx60$ Myr, the H$_2$ mass fraction reaches a local maximum of $\sim$ 40\%, and drops again to $\sim 20$\% at $t=100$ Myr. The mass fraction of atomic hydrogen decreases and increases inversely proportionally to H$_2$.
 
Comparing the different runs, we see that self-gravity significantly changes the structure of the dense interstellar medium and, therefore, should not be neglected in studies of stratified galactic discs. Without self-gravity (yellow line; see also \citealt{Gatto2014}), compared to all other simulations, the molecular gas mass fractions are reduced to a minimal amount of $\sim$5\%, whereas $\sim$80\% of the mass is in atomic gas. Consequently, $\sim$15\% of the total mass is in ionised hydrogen. After $t\approx 40-50$ Myr, the time evolution of the mass fractions is also very smooth compared to the simulations with self-gravity and random or clustered driving, which show fluctuations of approximately $\pm 10$\% around the respective mean values. This is a misleading feature, which falsely suggests that the simulations establish some sort of dynamical equilibrium after $\sim$50 Myr. However, this is not the case in any of the more realistic setups that include self-gravity.
 
Second, we find that if a significant number of SNe interact with dense gas (50\% in case of mixed driving -- violet lines; 100\% in case of peak driving -- green lines), the amount of H$_2$ which survives in the simulations is significantly reduced. If a SN explodes within a dense cloud, it locally dissociates all H$_2$. This is true as long as the environment is not too dense, i.e. the mean density is $\bar{n} \lesssim 10^3\;{\rm cm}^{-3}$. Once a massive, very dense, gravitationally collapsing cloud is forming, the SN remnants quickly become radiative \citep{Blondin1998, Gatto2014} and the leftover momentum input ($\sim 3\times 10^5\;{\rm M}_\odot\;{\rm km/s}$) is not strong enough to stop the collapse. At this point, the H$_2$ and CO mass fractions increase dramatically as a massive GMC goes into runaway gravitational collapse, forcing us to terminate the simulations. We believe that peak driving at a fixed SN rate is not a good way to model feedback in galactic discs as it fails to reproduce the observed structure of the multi-phase ISM. Also, a significant delay of a few Myr between the formation of a massive star and its SN explosion would decouple the explosions from the densest peaks, leading to a different positioning of the SNe with respect to the dense gas. Therefore, a regulation of the star formation rate by SN feedback alone seems artificial.

Third, runs with clustered SN driving and a small fraction of Type Ia SNe (with (red lines) and without magnetic fields (blue lines)), result in higher mean mass fractions of H$_2$ and CO than with random driving. Although the simulations are similar at $t\approx 60$ Myr, the H$_2$ mass fraction in run {\it S10-KS-rand} decreases to $\sim$ 20\% at $t=100$ Myr. With an initial magnetic field, the formation of H$_2$ is delayed by $\sim20$ Myr, but increases with a similar slope to the case of clustered SNe without magnetic fields once it is initiated. 

Overall, the strong variations in the mass fractions suggest that the discs do not establish a robust (chemo-)dynamical equilibrium, even at late times ($t>$ 100 Myr). The reason for this is the on-going and irreversible merging of the H$_2$-clouds. 

\begin{figure*}   

\begin{minipage}[b]{0.8\linewidth}
{\hspace{1cm} \bf S10-KS-rand-nsg}\\
  \includegraphics[trim = 0mm 13mm 0mm 4mm, clip,width=140mm]{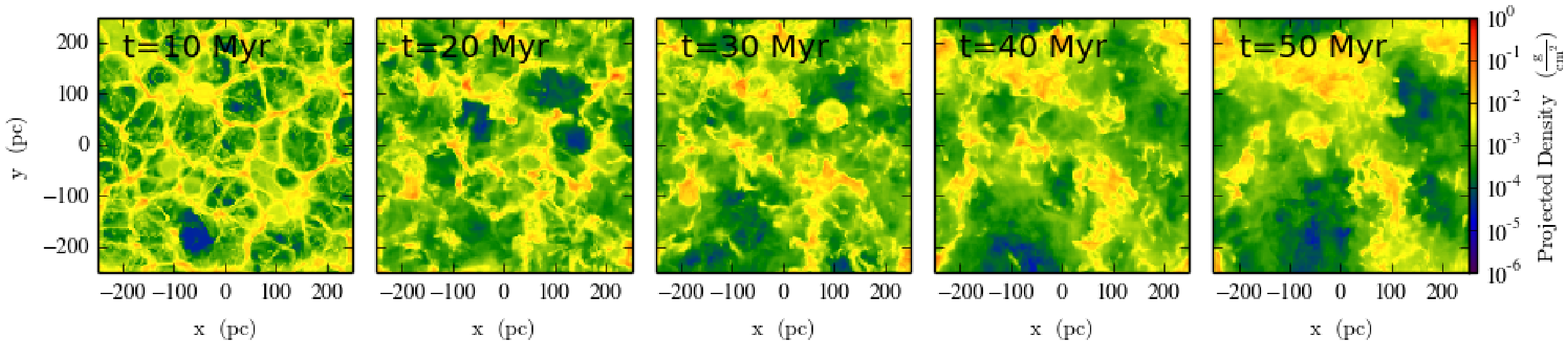} \\
  \includegraphics[trim = 0mm 3mm 0mm 4mm, clip,width=140mm]{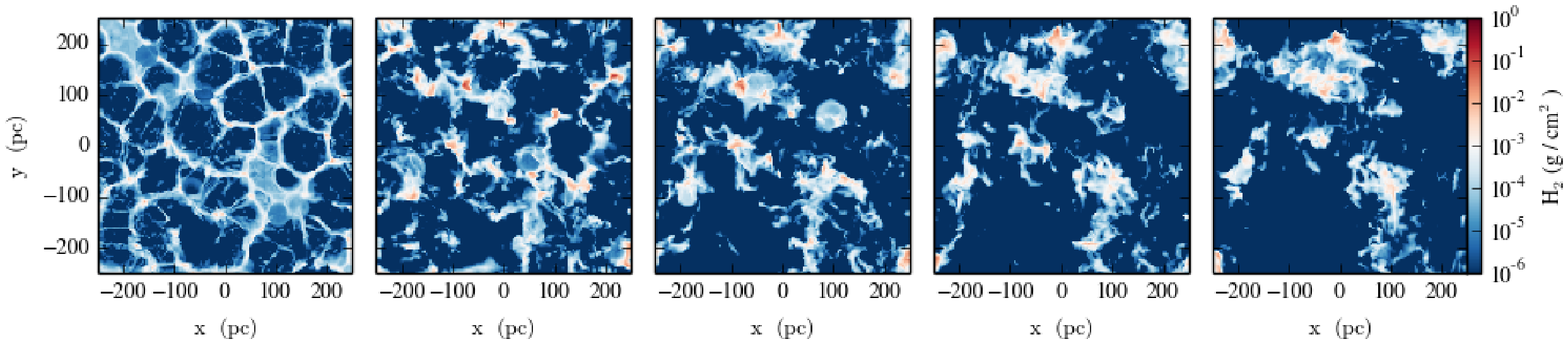} 
  
  {\hspace{1cm} \bf S10-KS-rand}\\
  \includegraphics[trim = 0mm 13mm 0mm 4mm, clip,width=140mm]{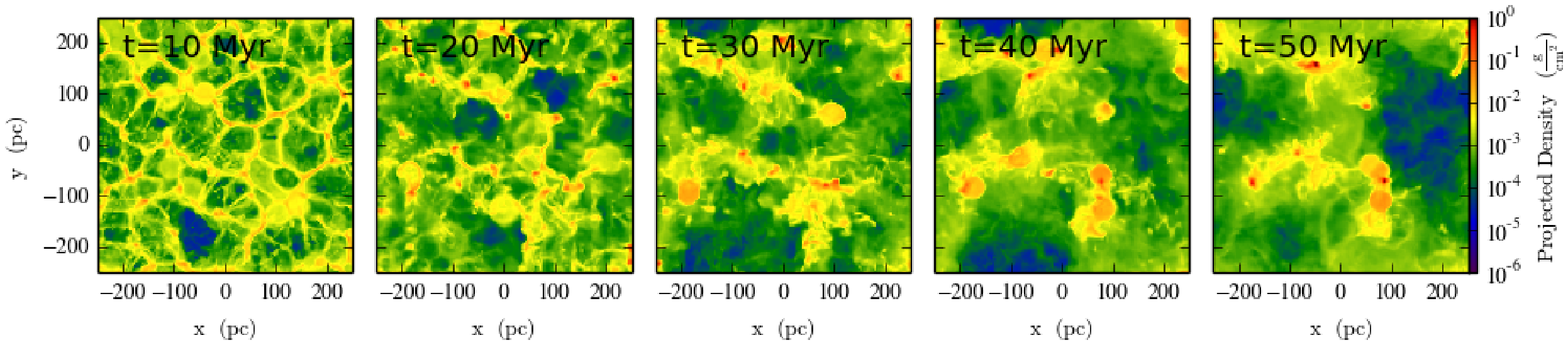} \\
  \includegraphics[trim = 0mm 3mm 0mm 4mm, clip,width=140mm]{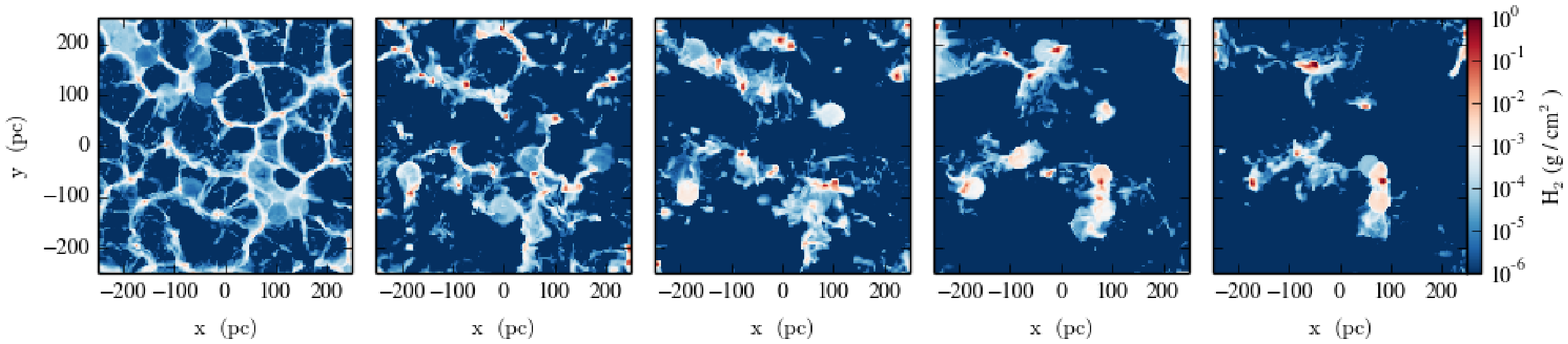} 
  
  {\hspace{1cm} \bf S10-KS-clus}\\
  \includegraphics[trim = 0mm 13mm 0mm 4mm, clip,width=140mm]{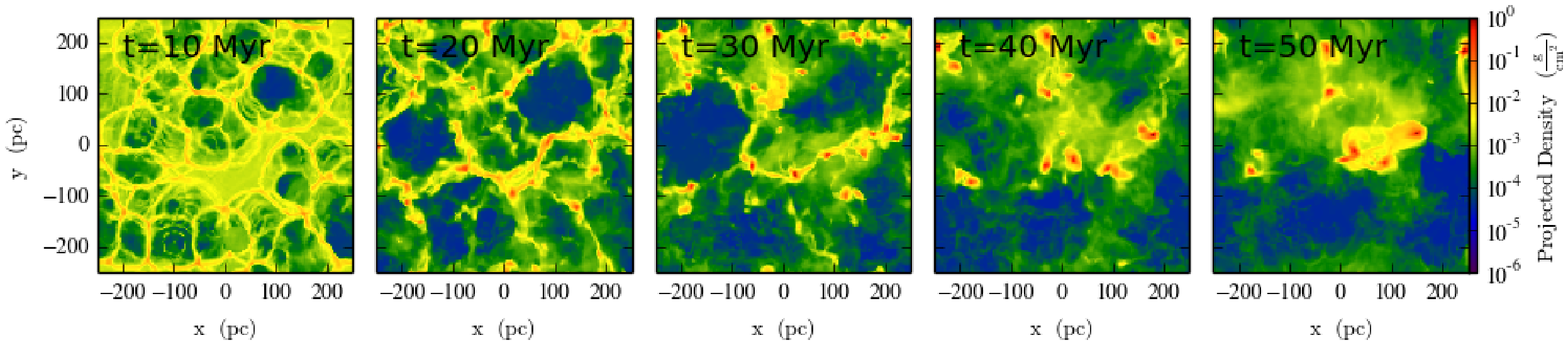} \\
  \includegraphics[trim = 0mm 3mm 0mm 4mm, clip,width=140mm]{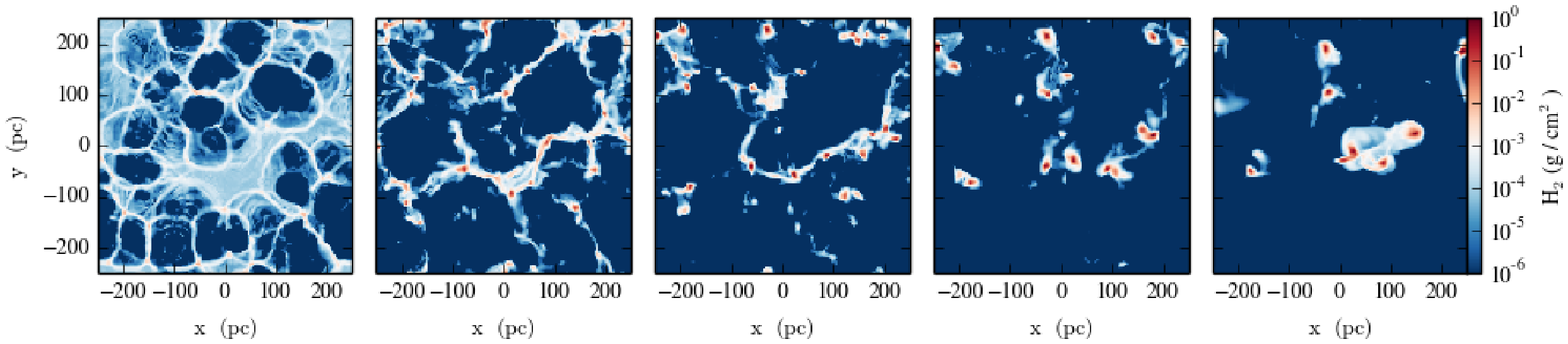} 
  
 {\hspace{1cm} \bf S10-KS-clus-mag3}\\
  \includegraphics[trim = 0mm 13mm 0mm 4mm, clip,width=140mm]{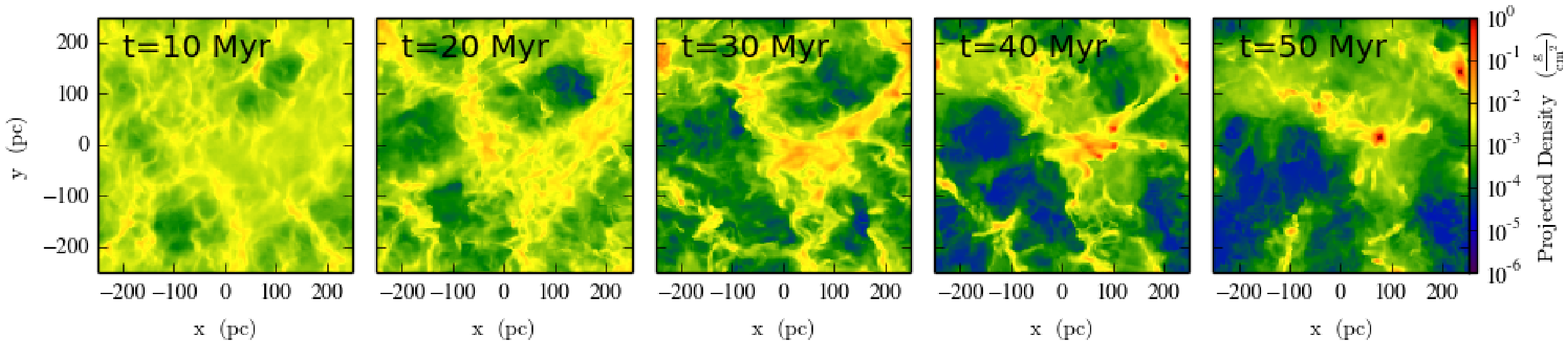} \\
  \includegraphics[trim = 0mm 3mm 0mm 4mm, clip,width=140mm]{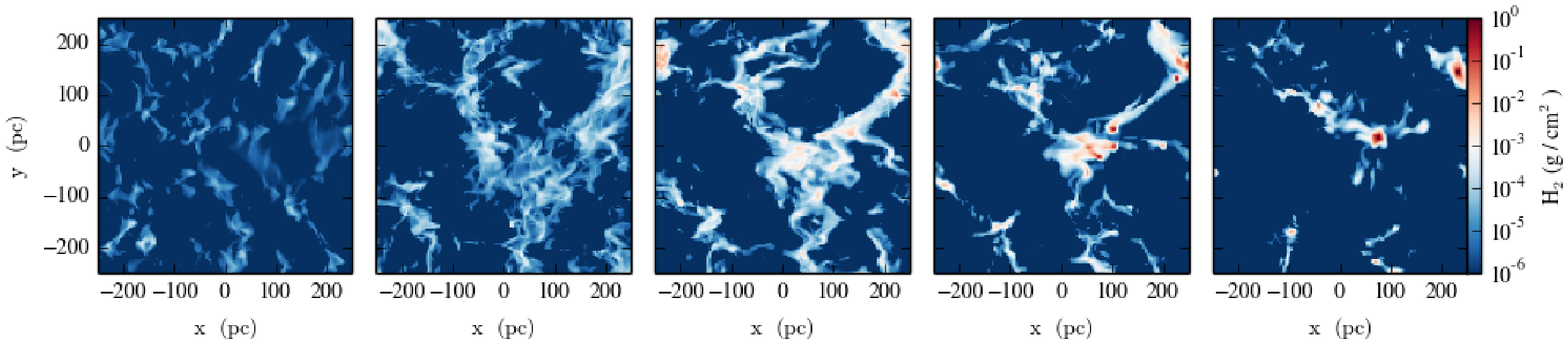}

 \caption{Time evolution of total and H$_2$ column densities (as seen face-on) for runs with different levels of physical complexity, i.e. {\it S10-KS-rand-nsg} (top), {\it S10-KS-rand} ($2^{\rm nd}$ row), {\it S10-KS-clus} (3$^{\rm rd}$ row), and {\it S10-KS-clus-mag3} (bottom). Without self-gravity the molecular clouds appear to be more diffuse. The difference between clustered and completely randomly distributed SNe is small. However, in {\it S10-KS-clus} the clouds appear to be slightly more concentrated. With magnetic fields ({\it S10-KS-clus-mag3}) there are fewer clouds and some more diffuse H$_2$ is present.} \label{FIG_MORPH2}
\end{minipage}
\end{figure*}

\subsection{The distribution of molecular gas}\label{SEC_MORPH}

\begin{figure}  
   \begin{center}
  \includegraphics[width=80mm]{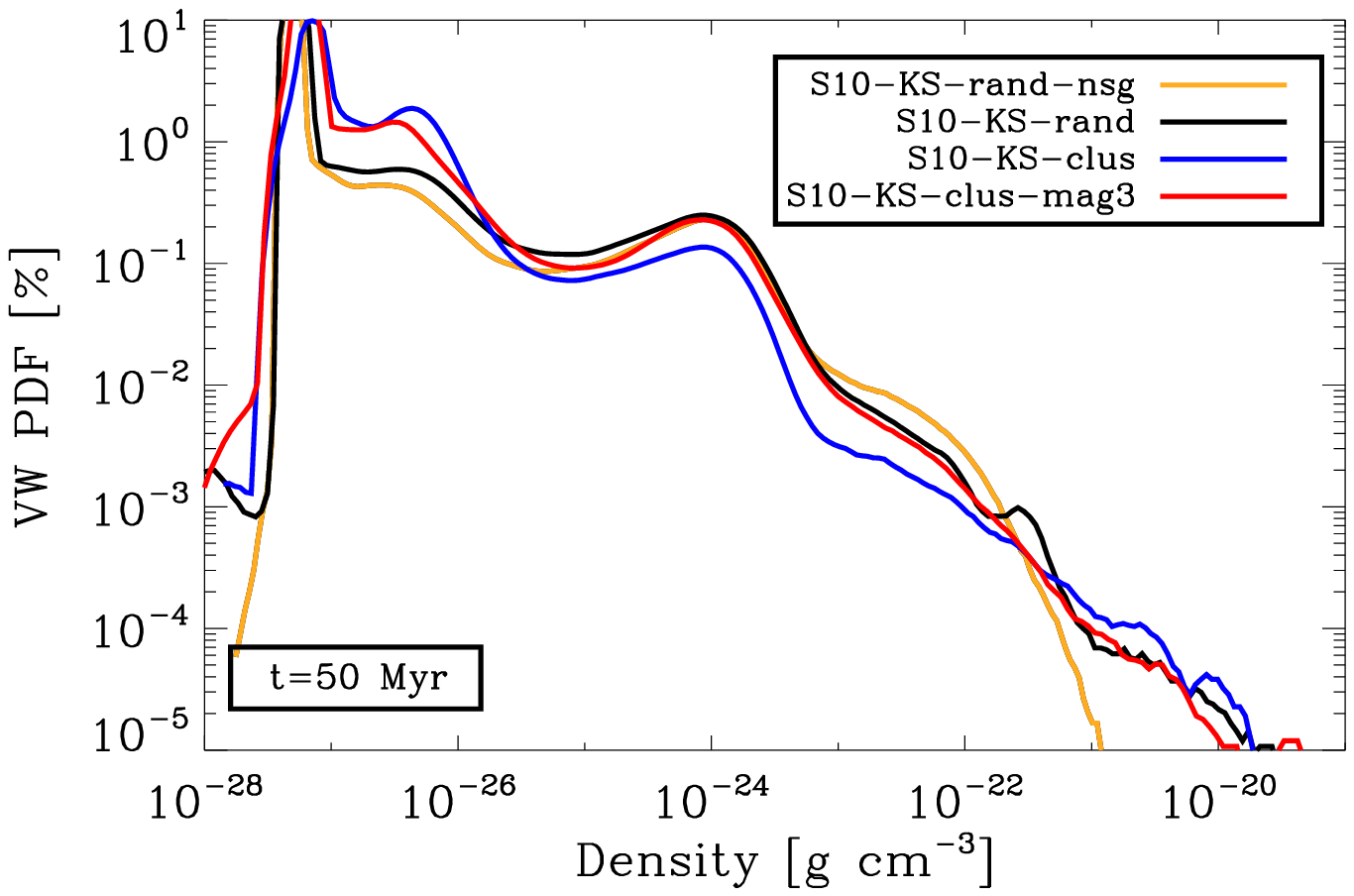}
  \includegraphics[width=80mm]{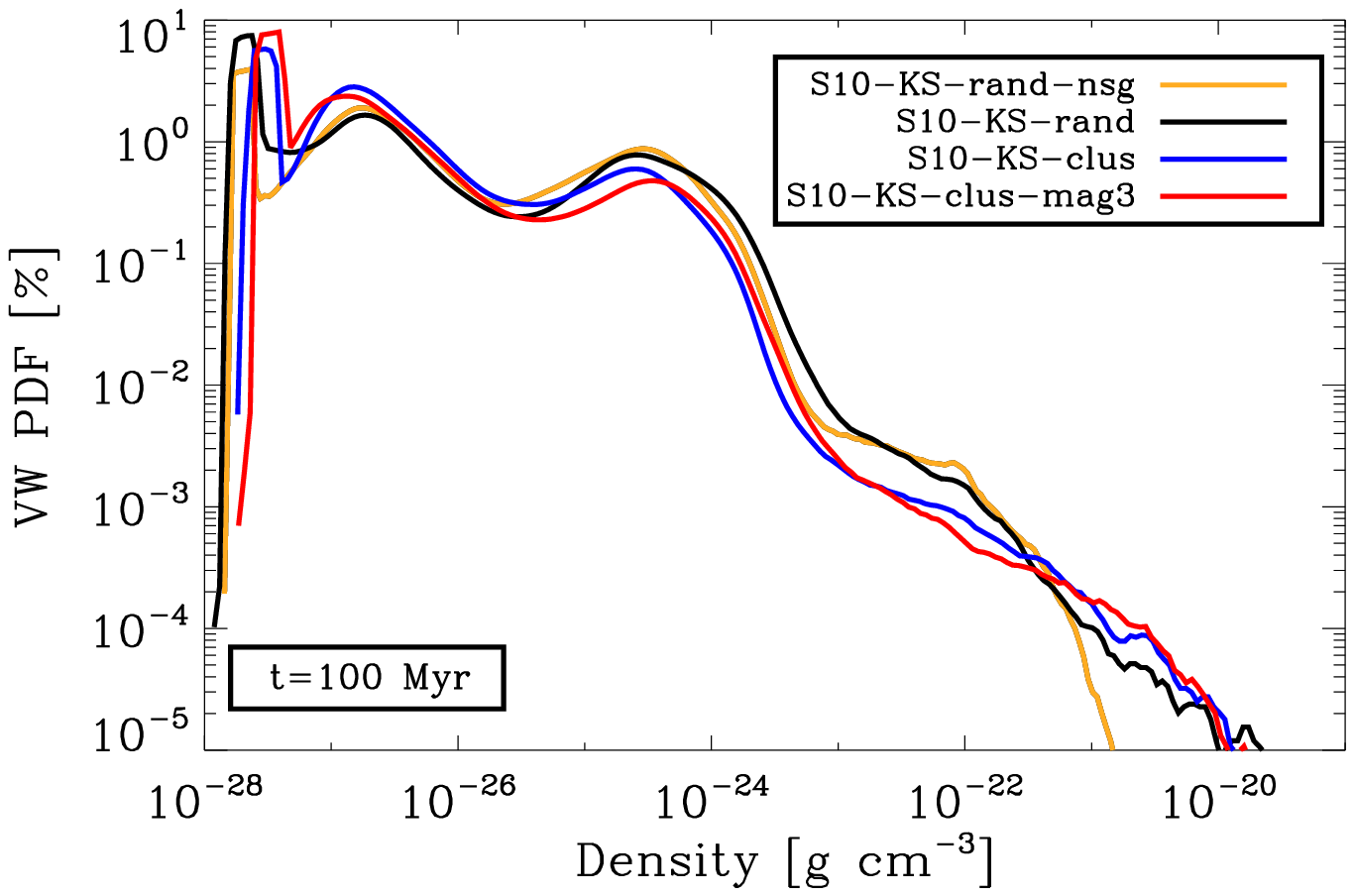}

 \end{center}
  \caption{ Volume-weighted density PDFs of runs {\it S10-KS-rand-nsg}, {\it S10-KS-rand}, {\it S10-KS-clus}, and {\it S10-KS-clus-mag3} at $t=50$ Myr ({\it top panel}) and $t=100$ Myr ({\it bottom panel}). The PDFs are very similar, in particular for the more evolved snapshot. The major difference is the lack of dense gas without self-gravity. \label{FIG_rhoPDF_VW} }
\end{figure}

We discuss the morphology and evolution of the forming molecular gas as a function of time. Due to our limited resolution, we do not consider the CO abundance to be fully converged. Therefore, we focus on the morphology of gas in the form of H$_2$. In Fig. \ref{FIG_MORPH2}, we show the time evolution of the total and the H$_2$ surface density for runs {\it S10-KS-rand-nsg}, {\it S10-KS-rand}, {\it S10-KS-clus}, and {\it S10-KS-clus-mag3}.

In all simulations, the molecular gas forms in shock-compressed filaments which evolve into clumps that merge into larger clouds during the course of the simulation. Without self-gravity, the small amount of H$_2$ can only form in shock-compressed structures. These are sometimes pushed together by the SN feedback, which results in a clumpy medium. However, due to the absence of self-gravity, the clumps do not feel each others' gravitational attraction and the H$_2$ mass fraction saturates as the distribution comes to a dynamical equilibrium. 

In Fig. \ref{FIG_rhoPDF_VW} we show the corresponding volume-weighted density PDFs for these four runs at $t=50$ Myr and $t=100$ Myr. With the exception that the non-gravitating run lacks high density gas, the density PDFs are very similar and do not reflect the structural changes in the gas surface density distribution.

\begin{figure*}   

\begin{minipage}[b]{0.8\linewidth}
{\hspace{1cm} \bf S10-KS-rand}\\
  \includegraphics[trim = 0mm 13mm 0mm 4mm, clip,width=140mm]{multiplot_dens_rand.eps} \\
  \includegraphics[trim = 0mm 3mm 0mm 4mm, clip,width=140mm]{multiplot_rand.eps} 
  
  {\hspace{1cm} \bf S10-KS-peak}\\
    \includegraphics[trim = 0mm 13mm 0mm 4mm, clip,width=140mm]{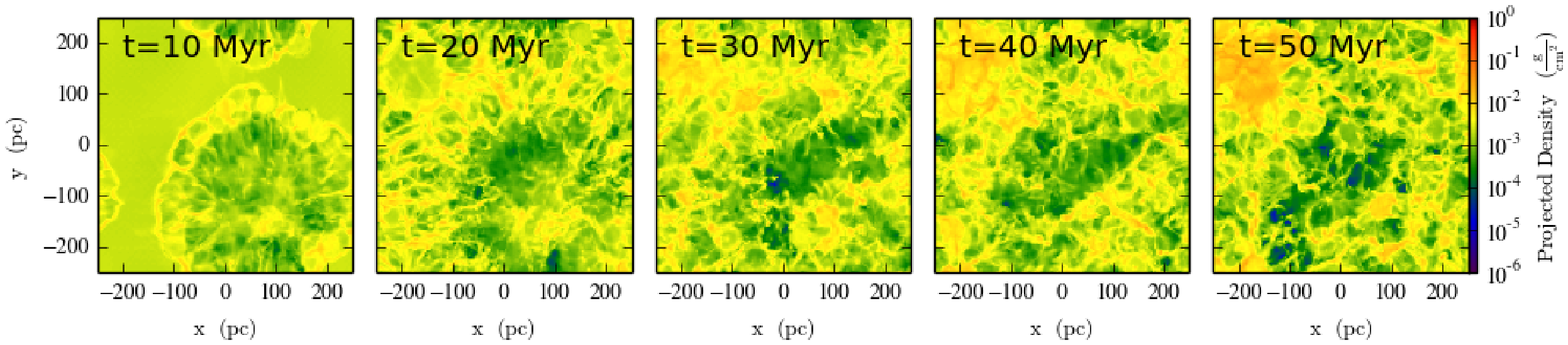} \\
  \includegraphics[trim = 0mm 3mm 0mm 4mm, clip,width=140mm]{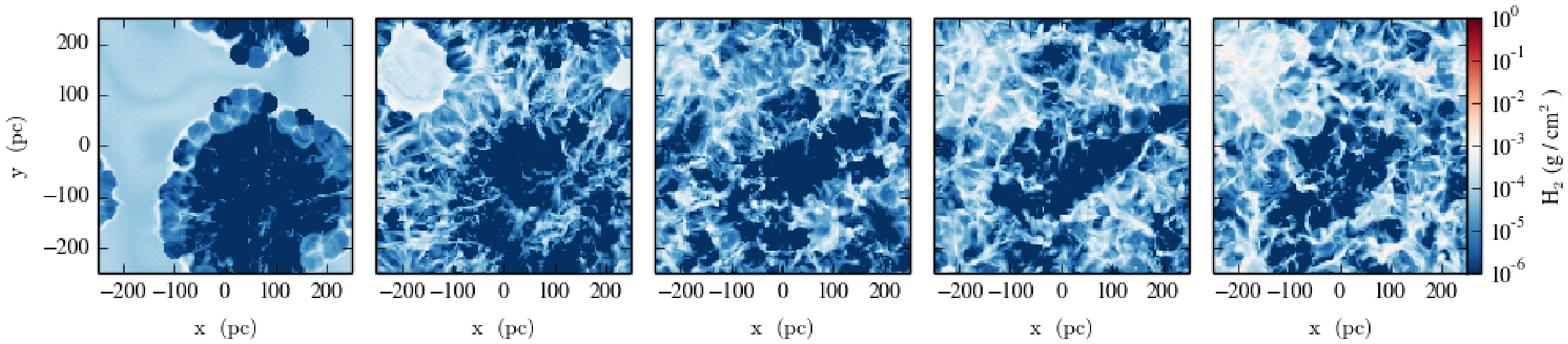}
   
  {\hspace{1cm} \bf S10-KS-mix}\\
  \hspace{1.5cm}  \includegraphics[trim = 0mm 13mm 0mm 4mm, clip,width=113mm]{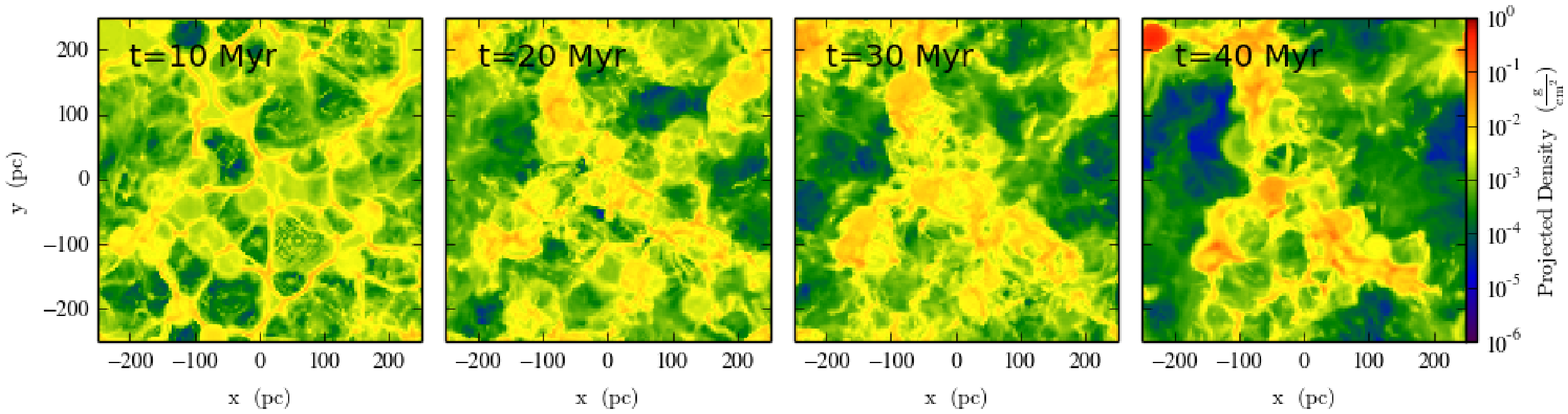} \\
  \hspace{1.5cm}  \includegraphics[trim = 0mm 3mm 0mm 4mm, clip,width=113mm]{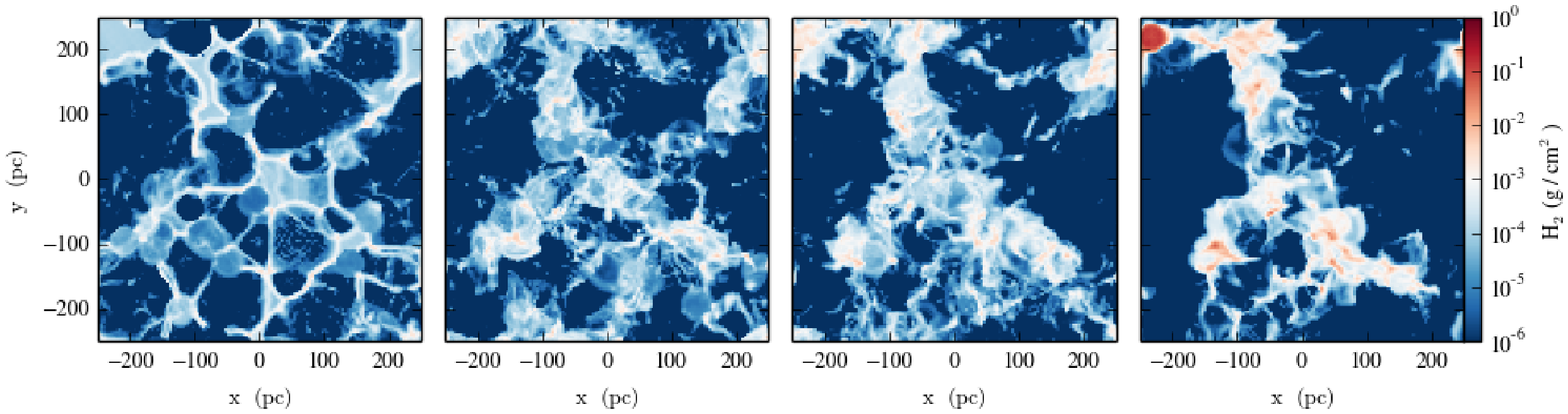} 
 \caption{Time evolution of the total and H$_2$ column densities for runs with different SN placement, i.e. {\it S10-KS-rand} (top), {\it S10-KS-peak} (center), and {\it S10-KS-mix} (bottom). Run {\it S10-KS-mix} can only be shown up to $t=40$ Myr, since this simulation fails due to the gravitational collapse of a single, massive clump forming in the upper left corner at $t\approx 44$ Myr. The placement of the SNe strongly affects the gas distribution and thus also the H$_2$ distribution and total mass fraction (see Fig. \ref{FIG_MASS1}). } \label{FIG_MORPH1}
\end{minipage}
\end{figure*}

In Fig. \ref{FIG_MORPH1}, we show the impact of the SN positioning on the H$_2$ distribution. At first, peak and mixed driving (middle and bottom rows) form more coherent, large-scale bubbles than random driving (top row). As the bubble(s) expand the H$_2$ distributions become more filamentary. However, as the bubbles converge, the simulations fail since they rapidly form a molecular clump whose collapse cannot be stopped by the SN feedback, which is too inefficient in dense environments due to rapid radiative cooling of the remnants. Including star formation, this clump would turn into a dense star cluster, which would eat up most of the gas mass in the volume. Stellar feedback in the form of SNe could not stop this collapse since the SN feedback is acting on too long timescales (the first SNe explode $\gtrsim 5$ Myr after stars start to form). Therefore, it seems that peak driving is not a good method to treat the SN feedback in galactic discs on the simulated scales if self-gravity is involved. It would lead to a highly variable star formation rate and overall fails to reproduce the mean observed properties of the ISM in the Milky Way (at first the molecular gas fraction is too low, then it suddenly diverges). The situation is equally bad for mixed driving, where the collapse of the isolated clump proceeds even faster since only 50\% of the SNe are used to stop the collapse of the dense clump, whereas the other 50\% keep on compressing the gas from within the cavities. Overall we see that large-scale bubbles can sweep up a lot of gas, leading to a quick conversion of atomic to molecular hydrogen once self-gravity starts to dominate locally.

\section{Impact of the star formation rate}\label{SEC_SFR}

We decrease/increase the SN rate by a factor of 3 to mimic higher and lower star formation rates in the discs (runs {\it S10-rand-lowSN} and {\it S10-rand-highSN}). The runs therefore explore the disc structure below/above the KS relation. 

\begin{figure}  
   \begin{center}
  \includegraphics[width=80mm]{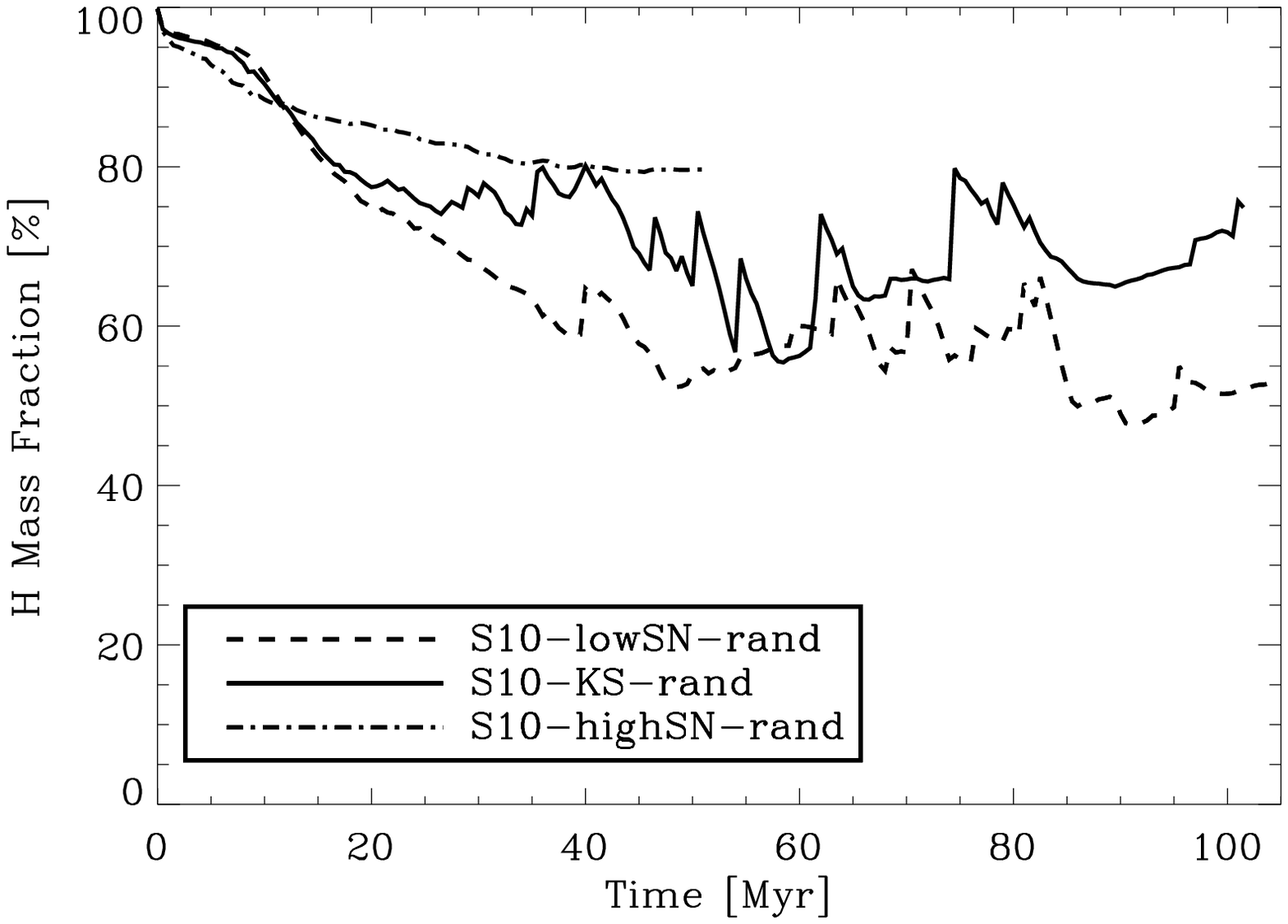}
  \includegraphics[width=80mm]{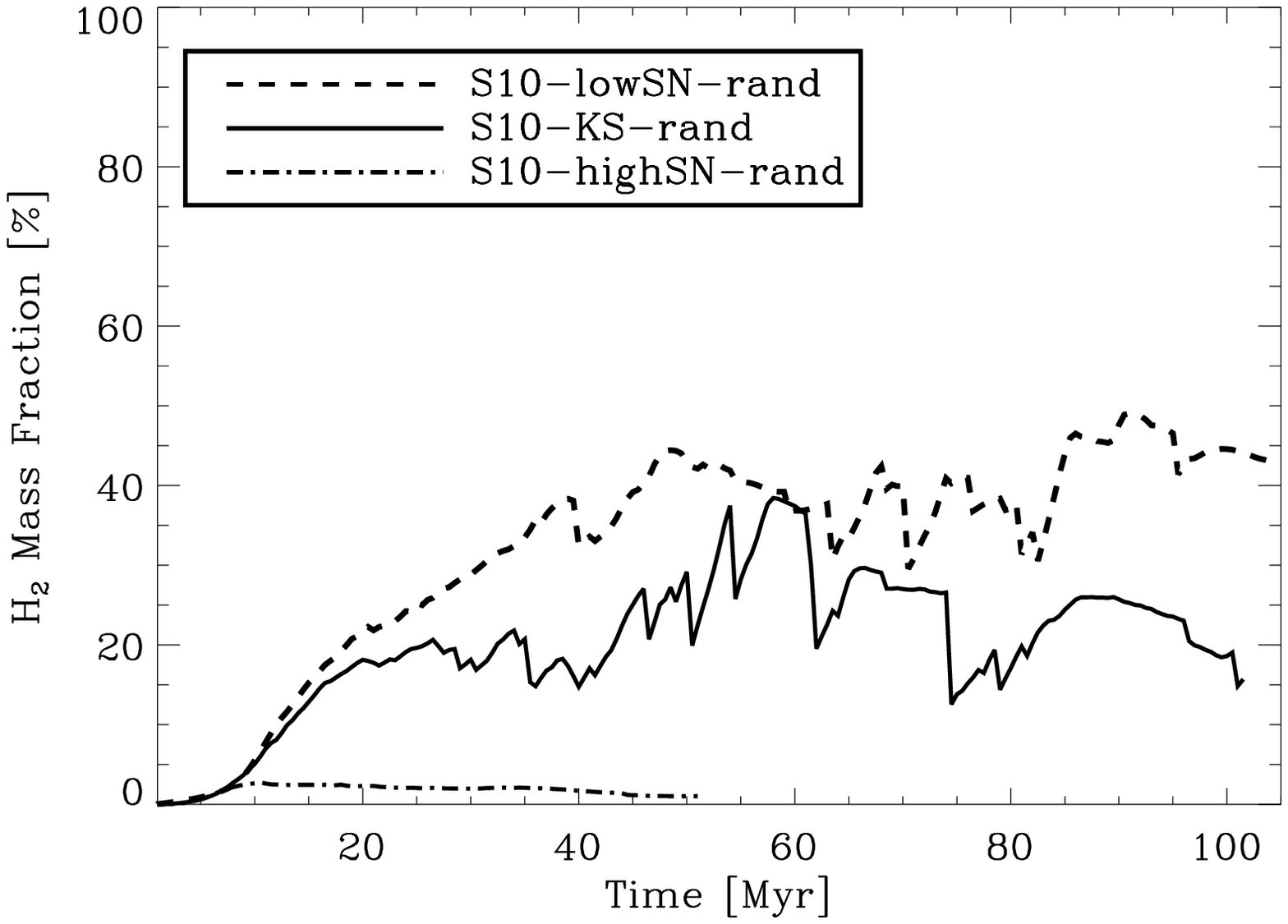}
    \includegraphics[width=80mm]{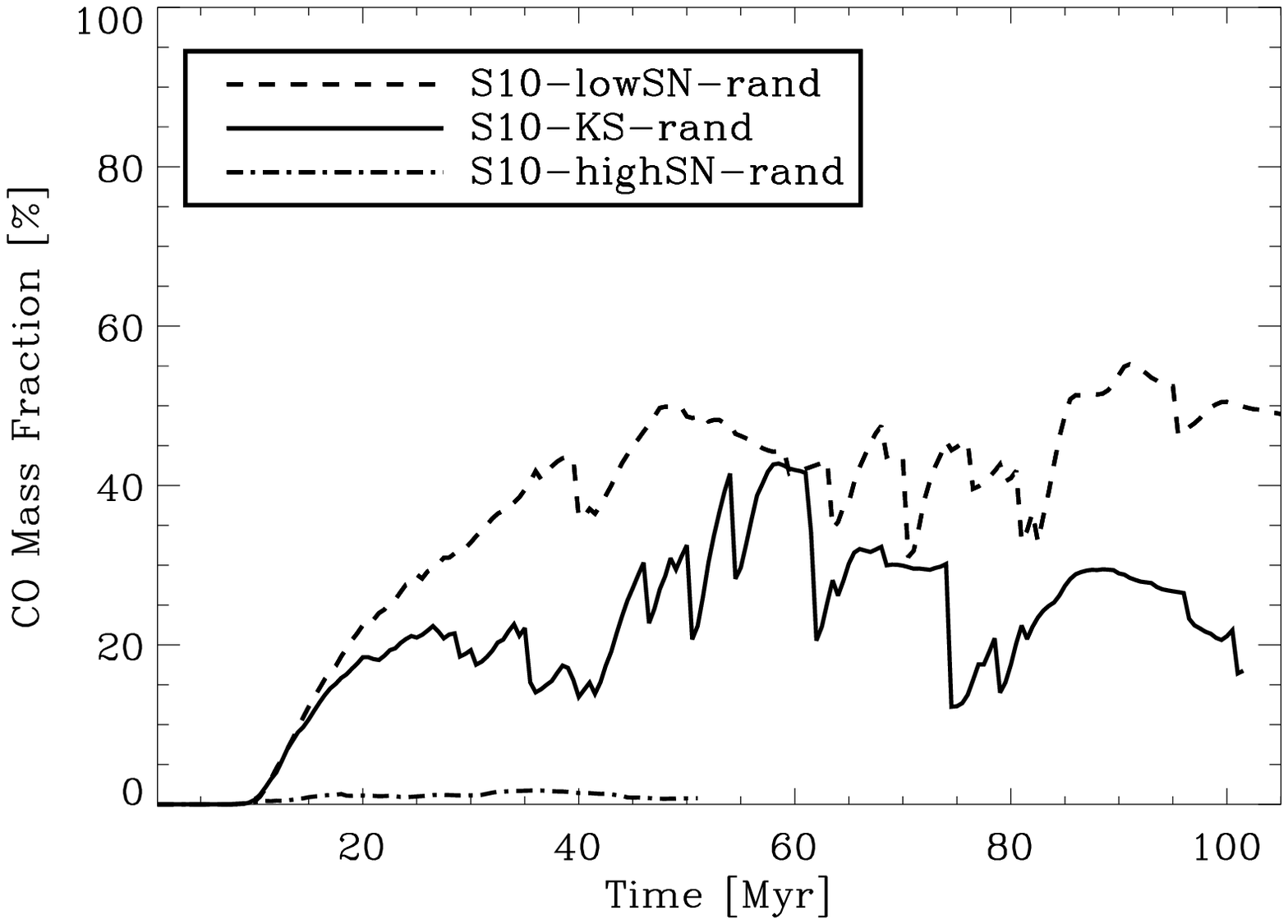}

 \end{center}
  \caption{H ({\it top}) and H$_2$ ({\it middle}) mass fractions (with respect to the total hydrogen mass) and CO mass fraction ({\it bottom}; with respect to the total mass in carbon) as a function of time for runs with random driving using three different SN rates to estimate the influence of the star formation feedback efficiency on the chemical composition of the gas. We show run {\it S10-lowSN-rand} where the SN rate is a factor of 3 times lower than the KS rate, i.e. 5 SNe per Myr (dashed line), run {\it S10-KS-rand} which uses the KS SN rate at 15/Myr (solid line), and {\it S10-highSN-rand} with a SN rate of 45/Myr (dash-dotted line). We find significantly less cold and molecular gas with increasing SN rates. \label{FIG_MASS_HILO} }
\end{figure}

We find that the molecular gas mass fractions strongly depend on the SN rate. In Fig. \ref{FIG_MASS_HILO} we show that the H$_2$ mass fraction is reduced from on average $\sim$40\% in case of a low SN rate (5/Myr; dashed curve) to $\sim$20\% for the KS rate (15/Myr; solid curve) to almost 0\% for the high SN rate (45/Myr; dash-dotted curve). Also, for {\it S10-highSN-rand}, the mass in H$^+$ is about 20\%, whereas it is around 10\% in all other simulations. The CO mass fractions closely follow the trend of the H$_2$ mass fractions. 

\begin{figure*}   

\begin{minipage}[b]{0.8\linewidth}
  {\hspace{1cm} \bf S10-lowSN-rand}\\
    \includegraphics[trim = 0mm 12mm 0mm 4mm, clip,width=140mm]{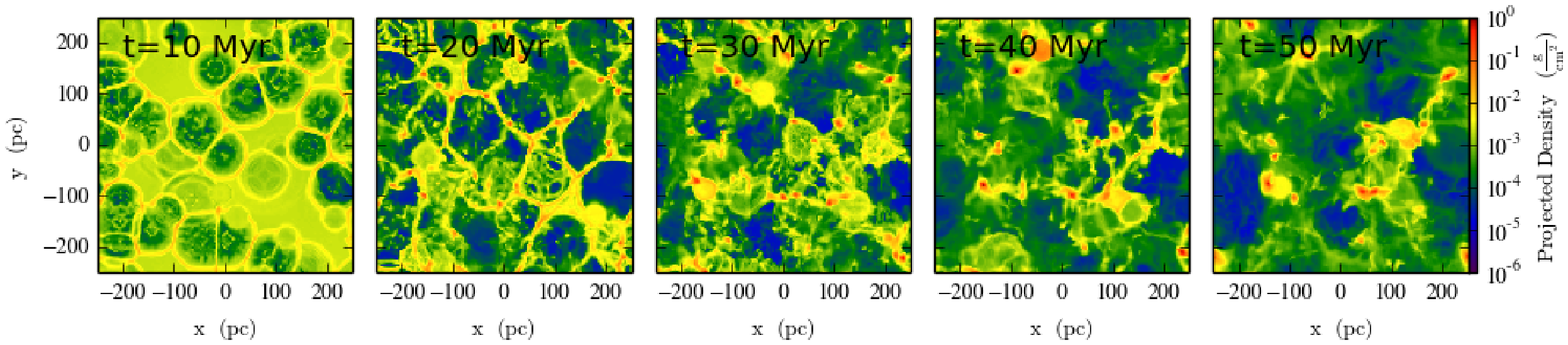} \\
  \includegraphics[trim = 0mm 3mm 0mm 4mm, clip,width=140mm]{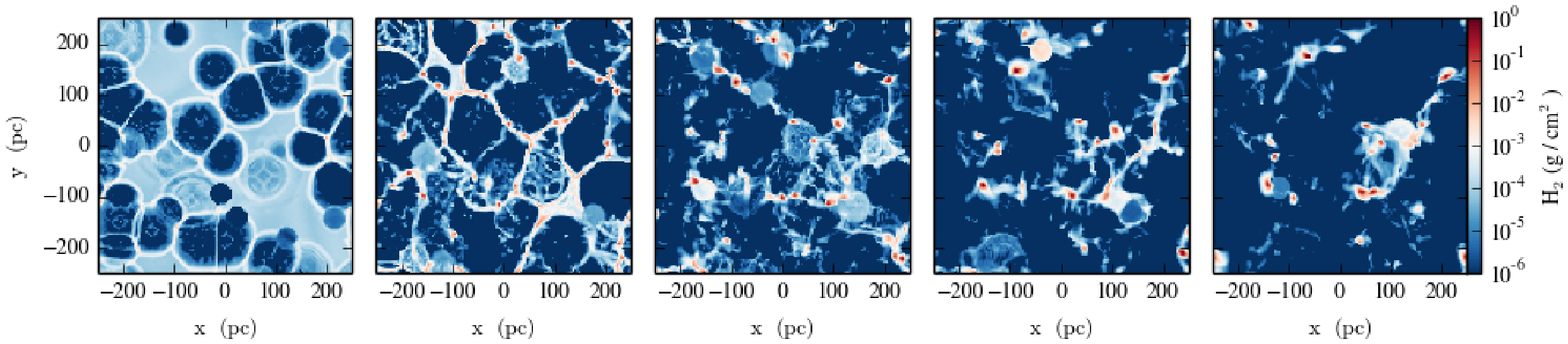} 

{\hspace{1cm} \bf S10-KS-rand}\\
  \includegraphics[trim = 0mm 12mm 0mm 4mm, clip,width=140mm]{multiplot_dens_rand.eps} \\
  \includegraphics[trim = 0mm 3mm 0mm 4mm, clip,width=140mm]{multiplot_rand.eps} 
    
  {\hspace{1cm} \bf S10-highSN-rand}\\
    \includegraphics[trim = 0mm 12mm 0mm 4mm, clip,width=140mm]{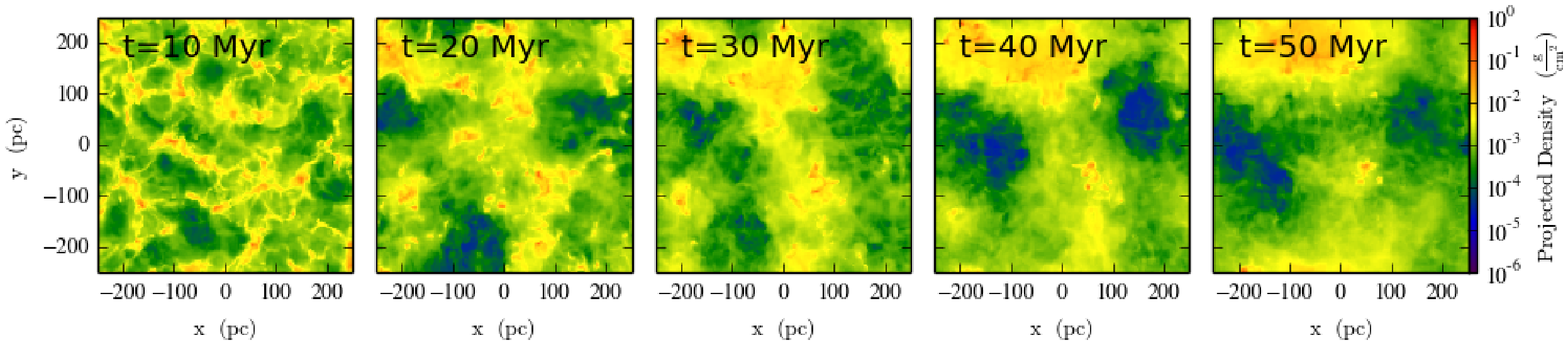} \\
  \includegraphics[trim = 0mm 3mm 0mm 4mm, clip,width=140mm]{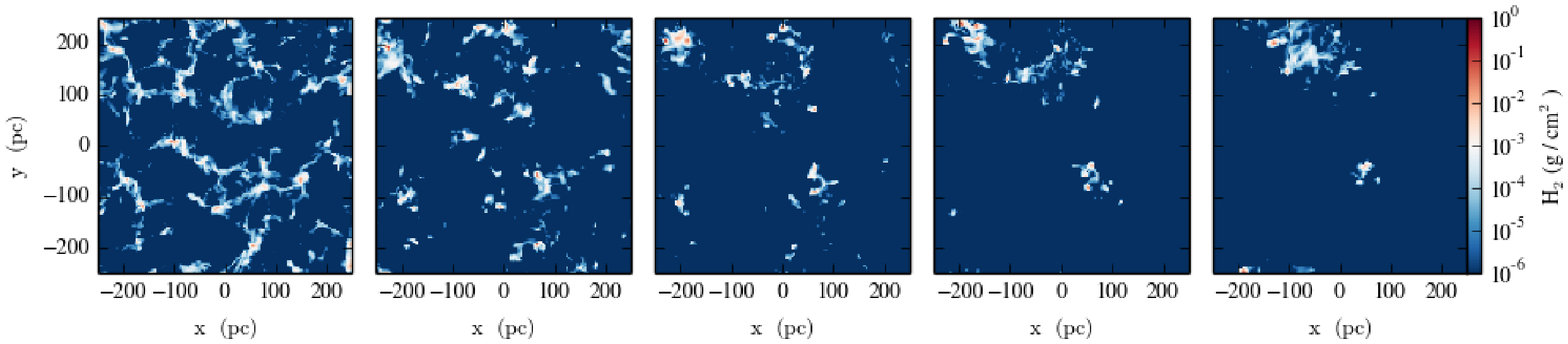} 
  
 \caption{Time evolution of total and H$_2$ column densities (as in Fig. \ref{FIG_MORPH1}) for runs with different SN rates, i.e. {\it S10-lowSN-rand} (top), {\it S10-KS-rand} (middle), and {\it S10-highSN-rand} (bottom). The SN rate strongly influences the morphology of the gas as well as the total amount of formed molecular gas (see Fig. \ref{FIG_MASS_HILO}). With increasing SN rate the structure of the ISM becomes more diffuse.} \label{FIG_MORPH3}
\end{minipage}
\end{figure*}

\begin{figure}  
   \begin{minipage}[b]{1.0\linewidth}
   \begin{center}
  \includegraphics[width=80mm]{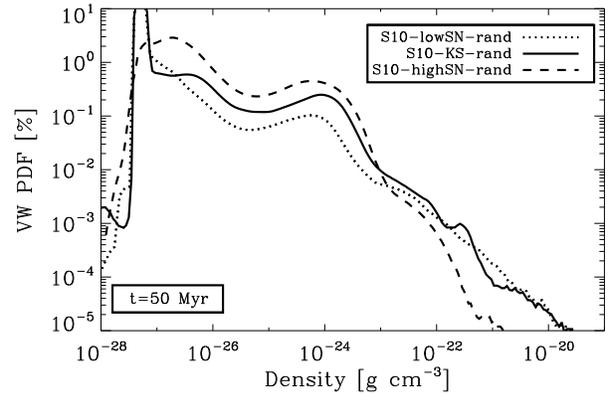}

 \end{center}
  \caption{ Volume-weighted density PDFs of runs with different SN rates and random driving, i.e. {\it S10-lowSN-rand}, {\it S10-KS-rand}, and {\it S10-highSN-rand}. With increasing SN rate the PDF becomes more narrow since the distribution is truncated at low densities, which essentially leaves two phases (warm at around $\rho \approx 10^{-24}\;{\rm g\; cm}^{-3}$ and hot at $\rho \approx 10^{-27.5}\;{\rm g\; cm}^{-3}$).  \label{FIG_rhoPDF_VW_HILO} }
\end{minipage}
\end{figure}

We compare the morphology of the ISM for different SN rates and random driving in Fig. \ref{FIG_MORPH3}. We find that, with increasing SN rate (from top to bottom), the ISM becomes more diffuse and less molecular. From this analysis we may conclude that SN rates that are as high as three times the KS value could not be maintained in a self-consistently star-forming ISM because all molecular gas is dispersed due to the high input of SN energy and momentum. Interestingly, the different SN rates have a measurable effect on the volume-weighted density PDFs, which are shown in Fig. \ref{FIG_rhoPDF_VW_HILO}. For high SN rates the amount of dense gas is significantly reduced and the volume is filled with two distinct peaks of intermediate ($n \approx 1\;{\rm cm}^{-3}$) and low density ($n \approx 3\times10^{-3}\;{\rm cm}^{-3}$) gas. This PDF appears more narrow than the ones derived from runs with a low or KS SN rate, which seems to be in contradiction with a simple broadening of the density and column density distributions with increasing levels of turbulence. This behaviour will be investigated more thoroughly in a subsequent paper.
 
\section{Vertical distributions}\label{SEC_VERTICAL}

\begin{figure*}   
   \begin{tabular}{@{}c@{\hspace{0.5cm}}c@{}c@{}c@{}}
   & Density profile at $t=50$ Myr & VFF at $t=50$ Myr & Density profile at $t=100$ Myr\\
\begin{sideways}{\bf $\;\;\;\;\;\;\;\;\;\;\;\;\;\;\;$ S10-KS-rand-nsg}  \end{sideways}&
    \includegraphics[trim = 7mm 4mm 3mm 1mm, clip,width=58mm]{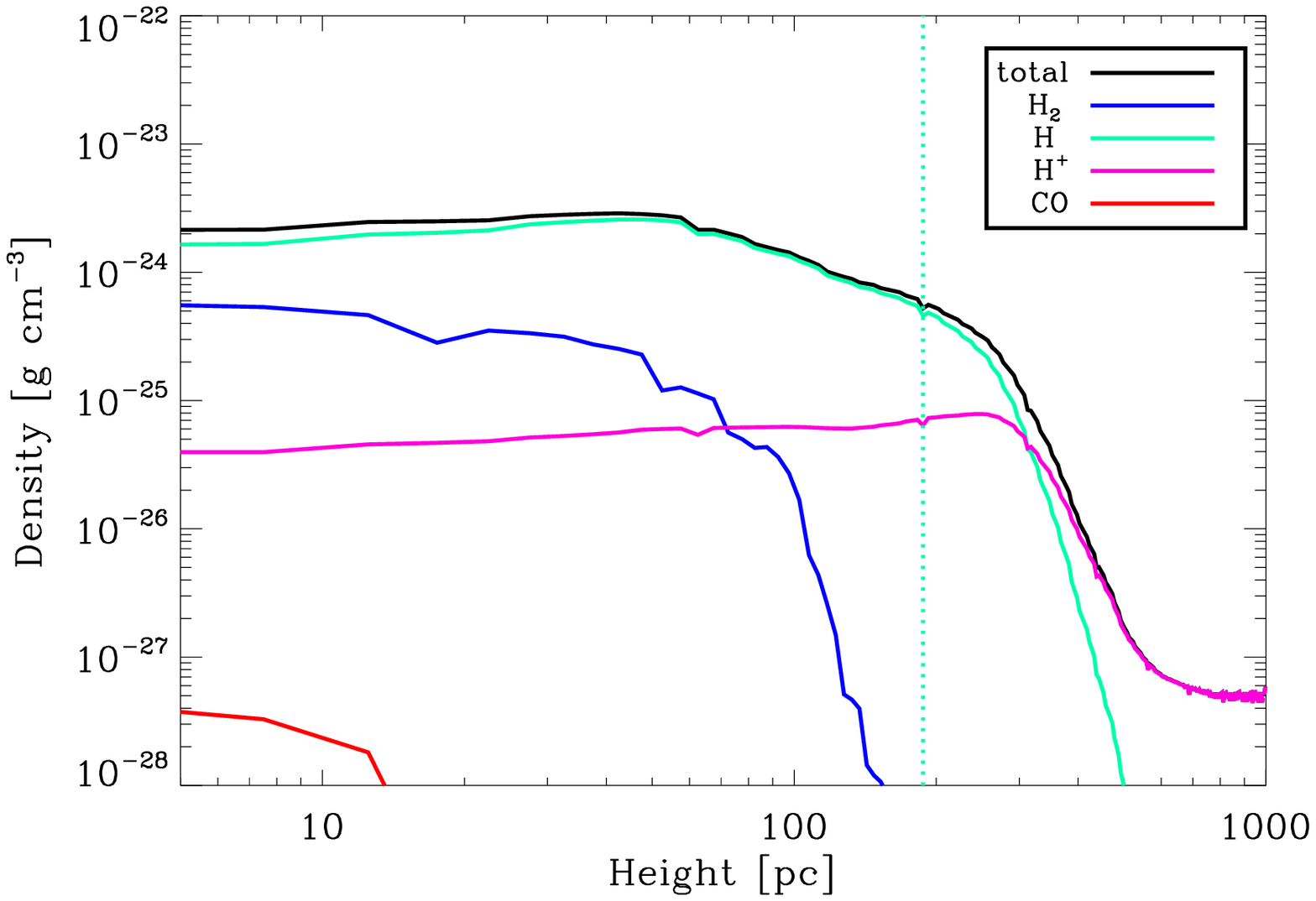} &
      \includegraphics[trim = 7mm 4mm 3mm 1mm, clip,width=58mm]{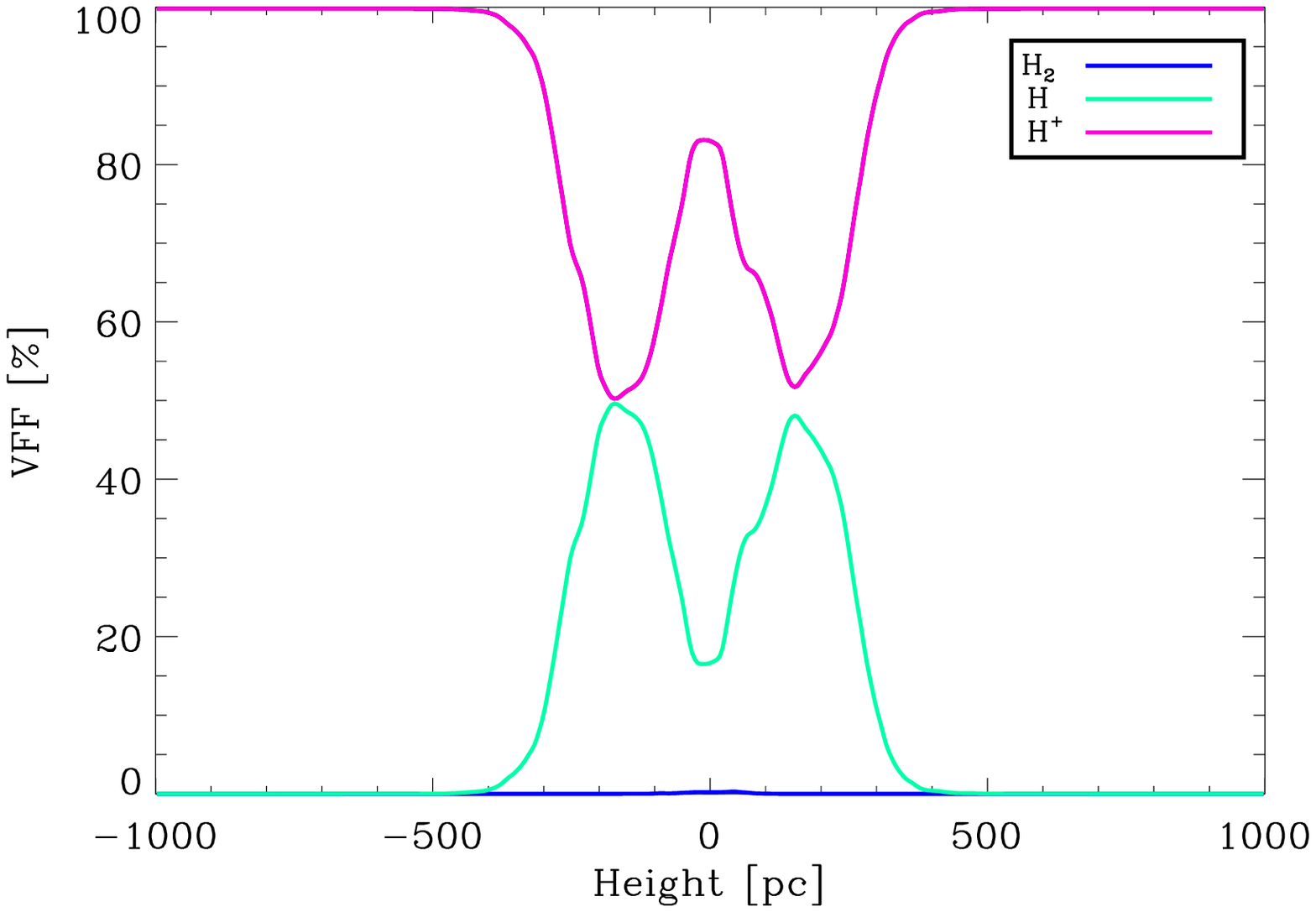}&
 \includegraphics[trim = 7mm 4mm 3mm 1mm, clip,width=58mm]{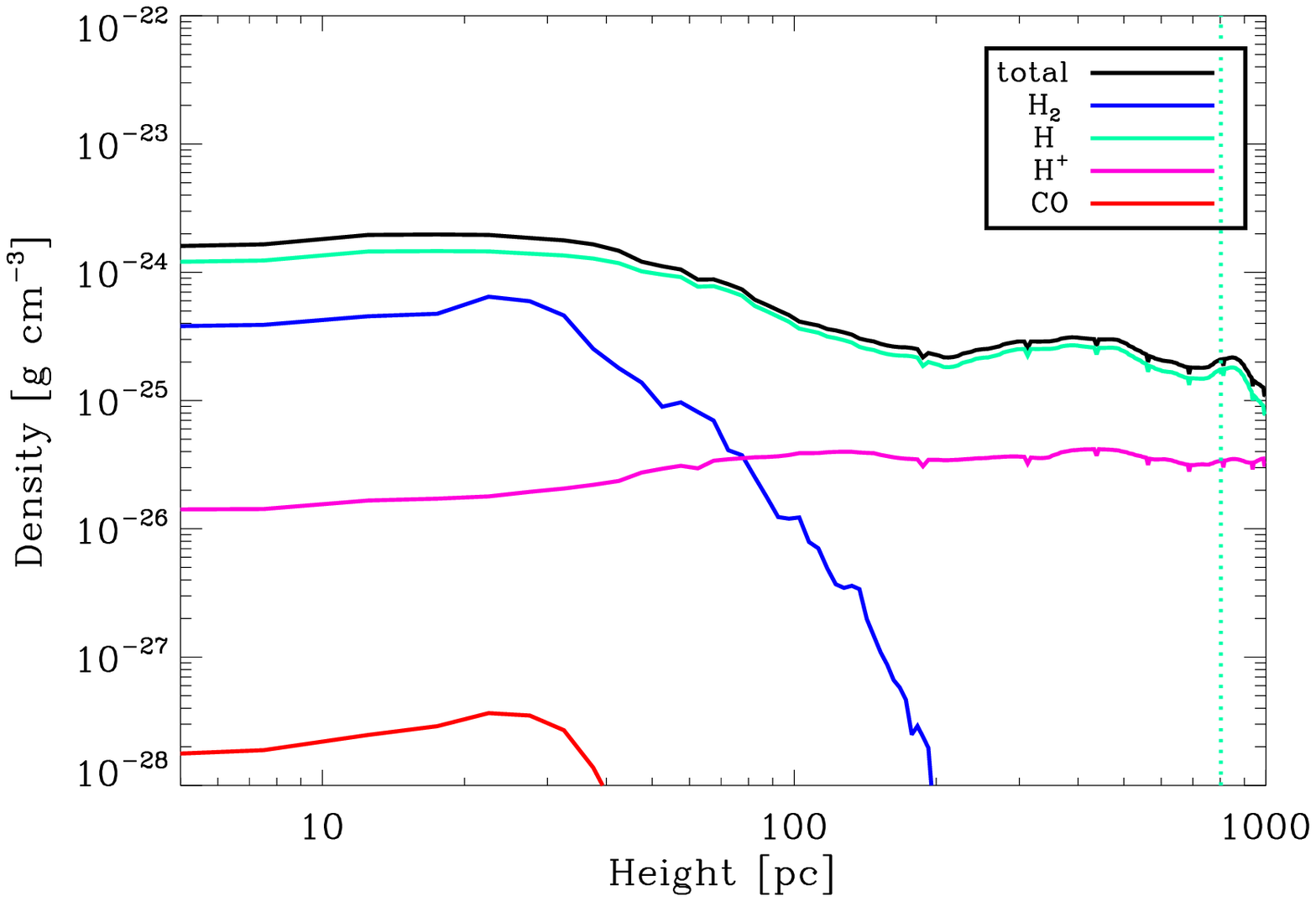}\\
      
\begin{sideways}{\bf $\;\;\;\;\;\;\;\;\;\;\;\;\;\;\;\;\;\;\;$S10-KS-rand} \end{sideways}&
    \includegraphics[trim = 7mm 4mm 3mm 1mm, clip,width=58mm]{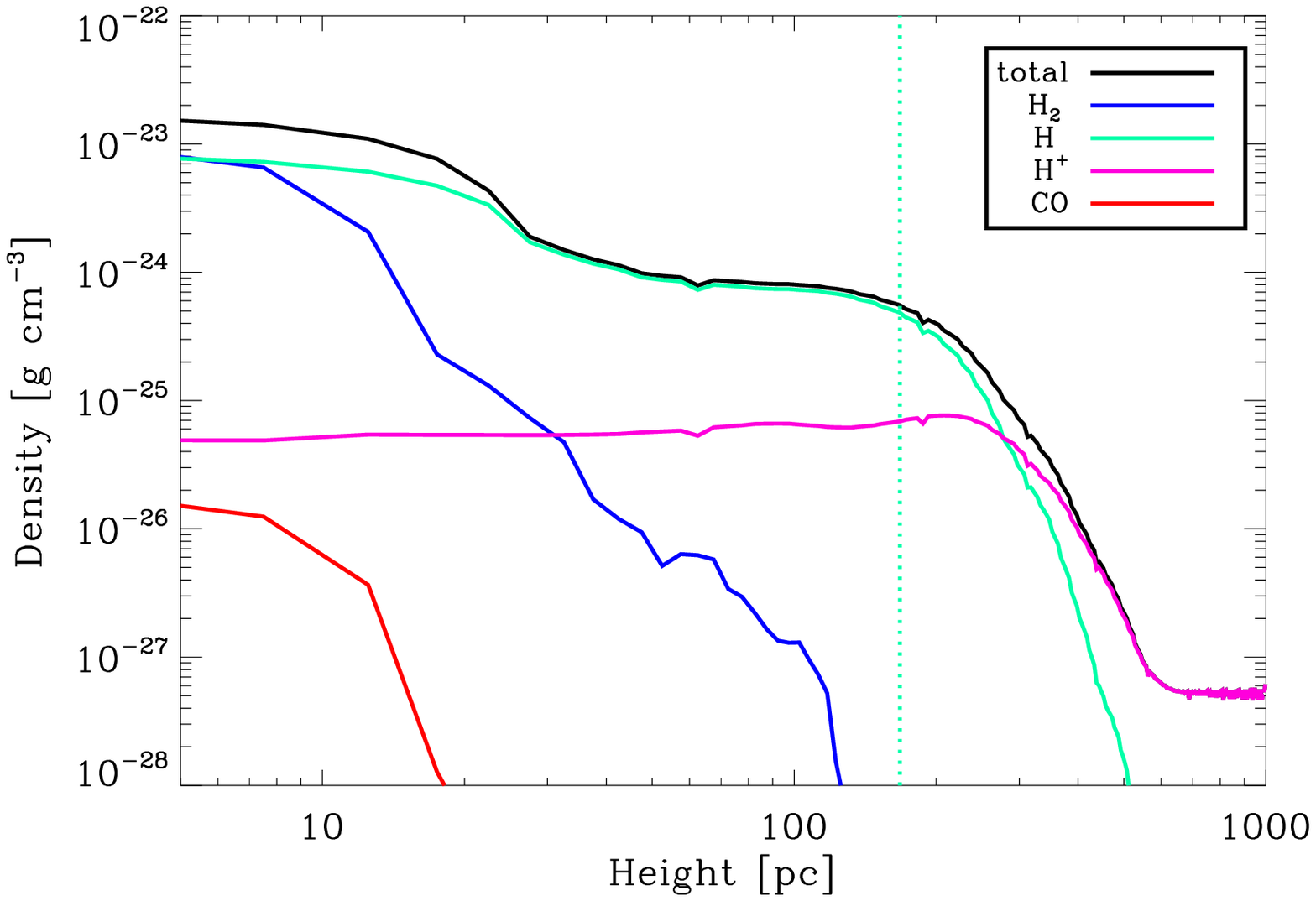}&
        \includegraphics[trim = 7mm 4mm 3mm 1mm, clip,width=58mm]{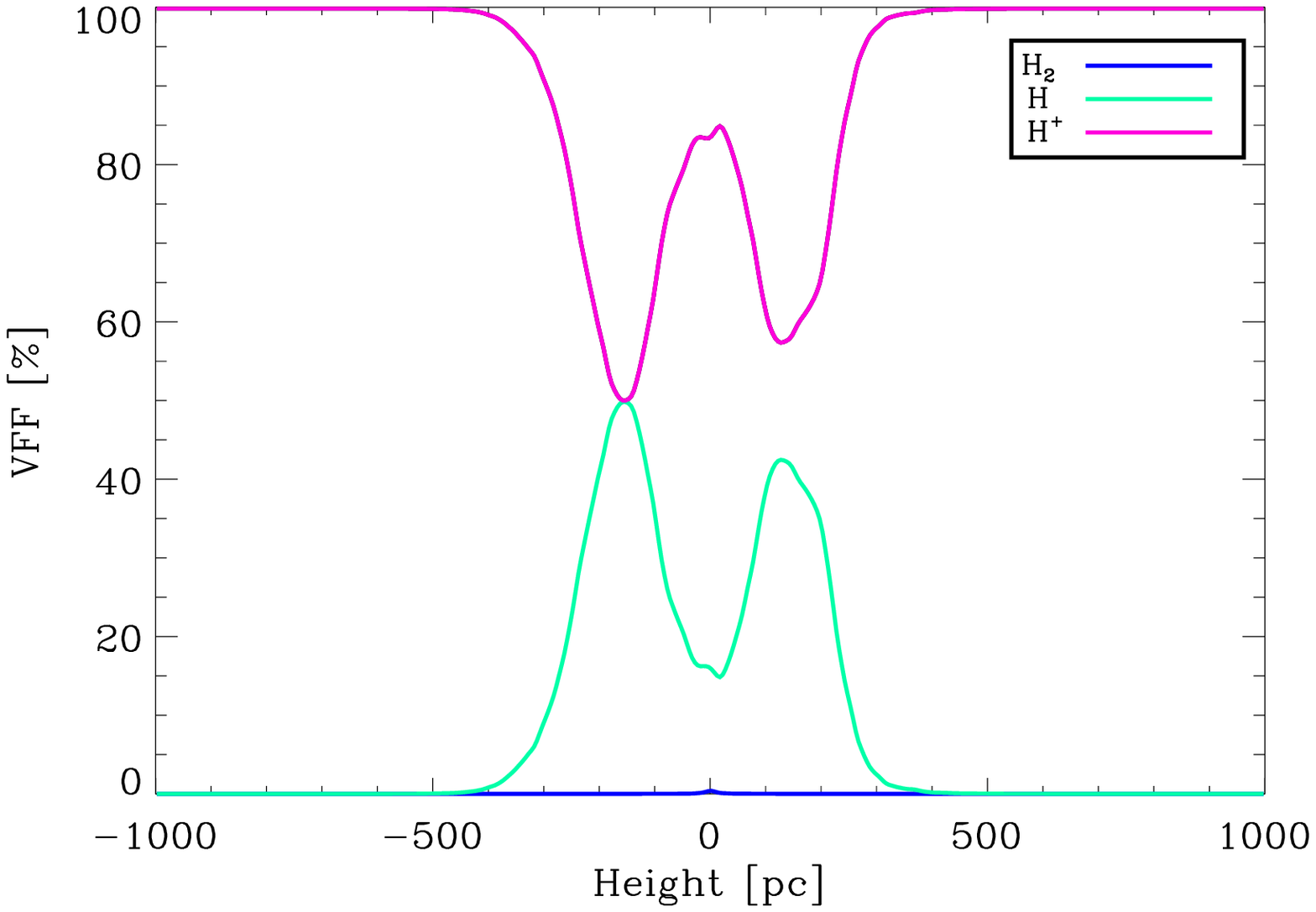} &
                  \includegraphics[trim = 7mm 4mm 3mm 1mm, clip,width=58mm]{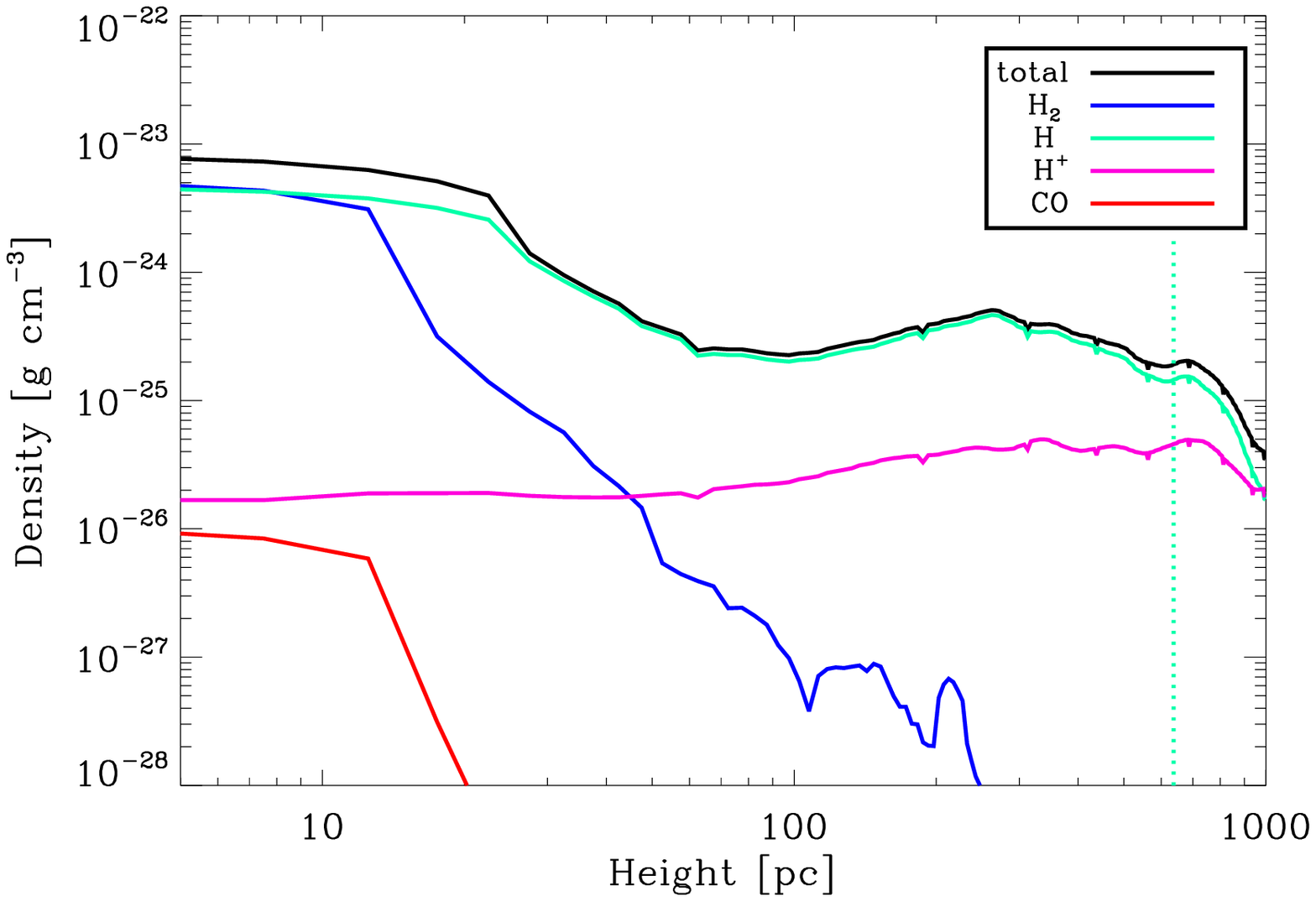}\\

  \begin{sideways}{\bf $\;\;\;\;\;\;\;\;\;\;\;\;\;\;\;\;\;\;\;$S10-KS-peak} \end{sideways}&
    \includegraphics[trim = 7mm 4mm 3mm 1mm, clip,width=58mm]{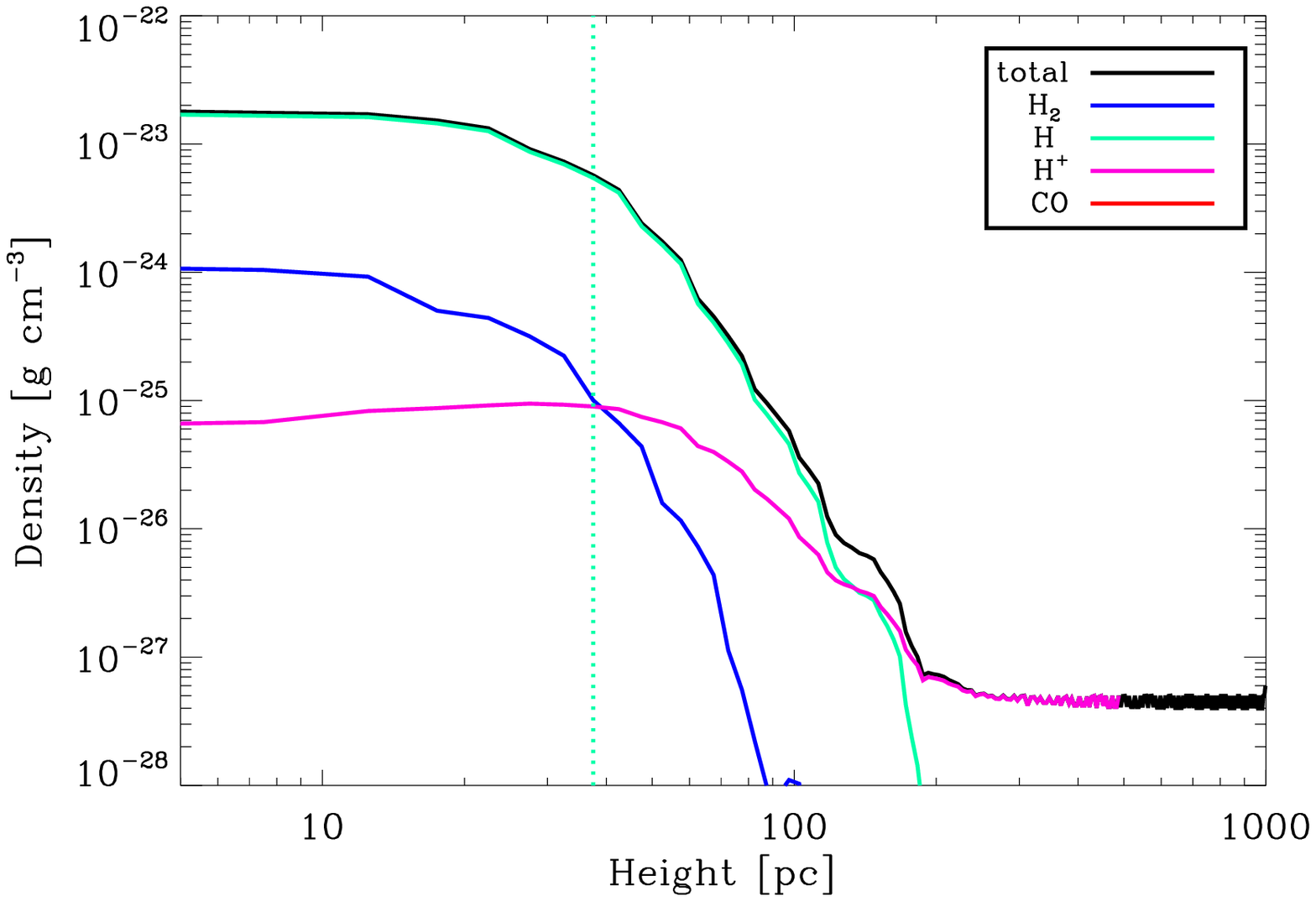}&
       \includegraphics[trim = 7mm 4mm 3mm 1mm, clip,width=58mm]{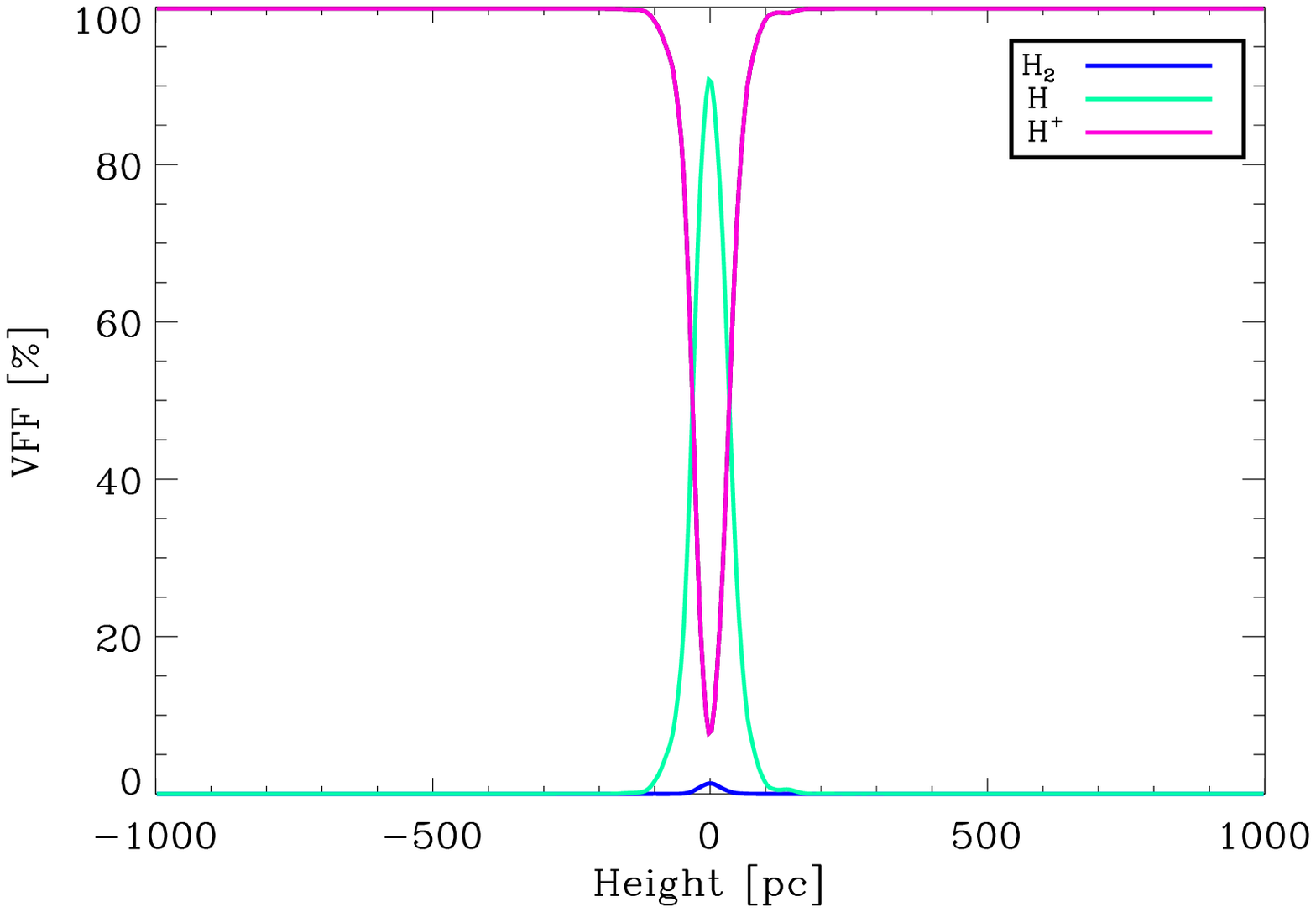}&\\

 \begin{sideways}{\bf $\;\;\;\;\;\;\;\;\;\;\;\;\;\;\;\;\;\;\;$S10-KS-clus}\end{sideways}&
    \includegraphics[trim = 7mm 4mm 3mm 1mm, clip,width=58mm]{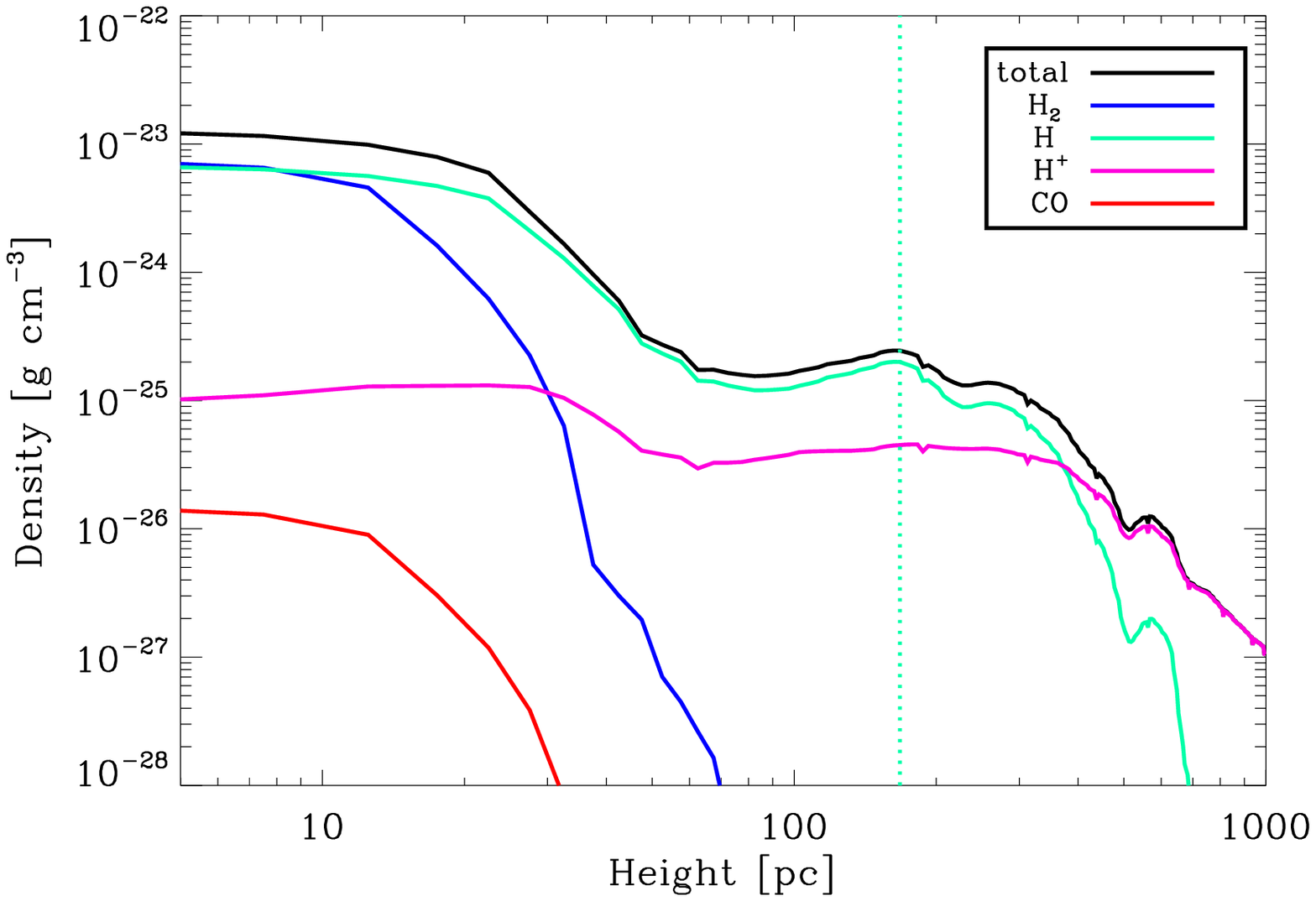}&
     \includegraphics[trim = 7mm 4mm 3mm 1mm, clip,width=58mm]{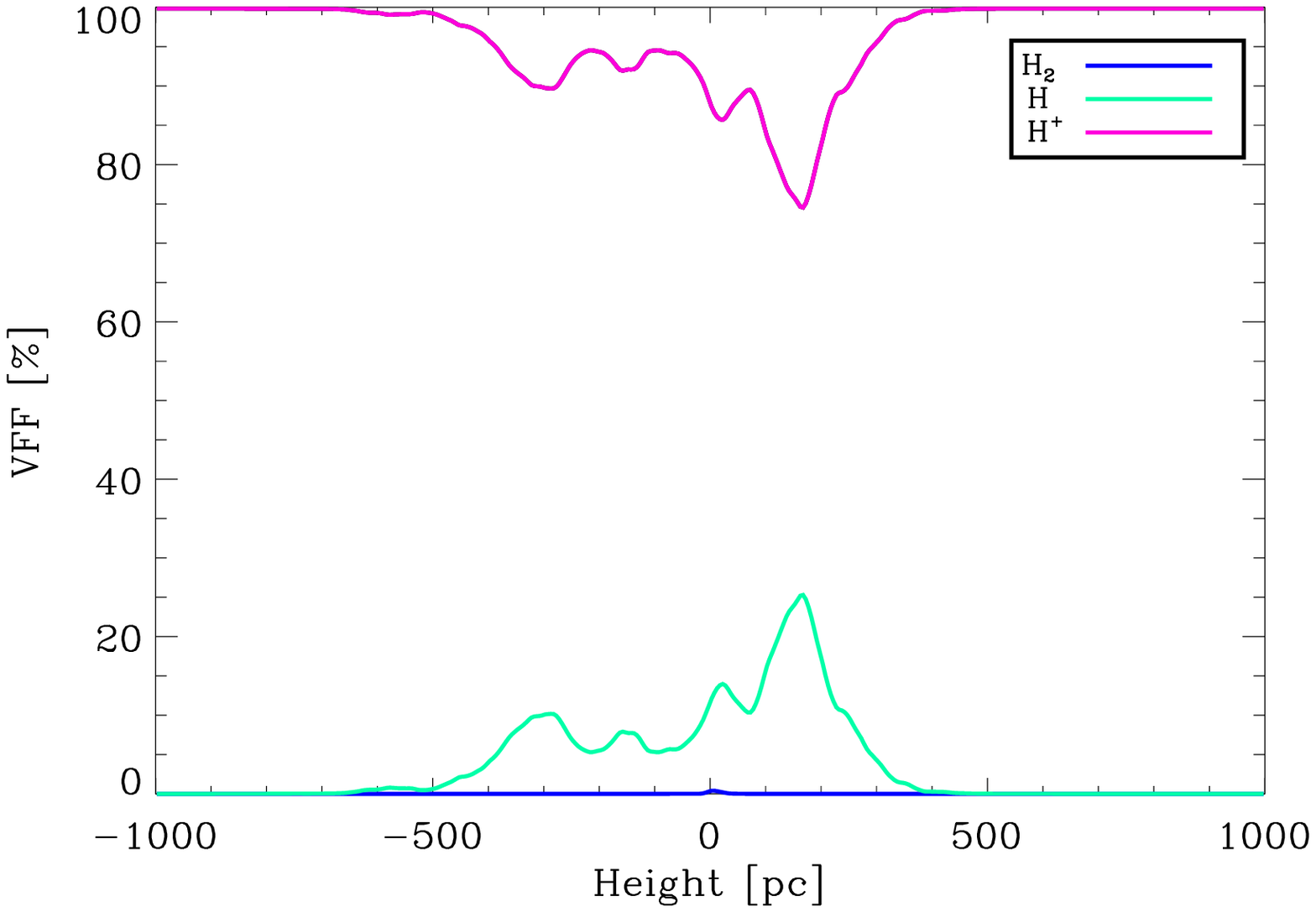} &
     \includegraphics[trim = 7mm 4mm 3mm 1mm, clip,width=58mm]{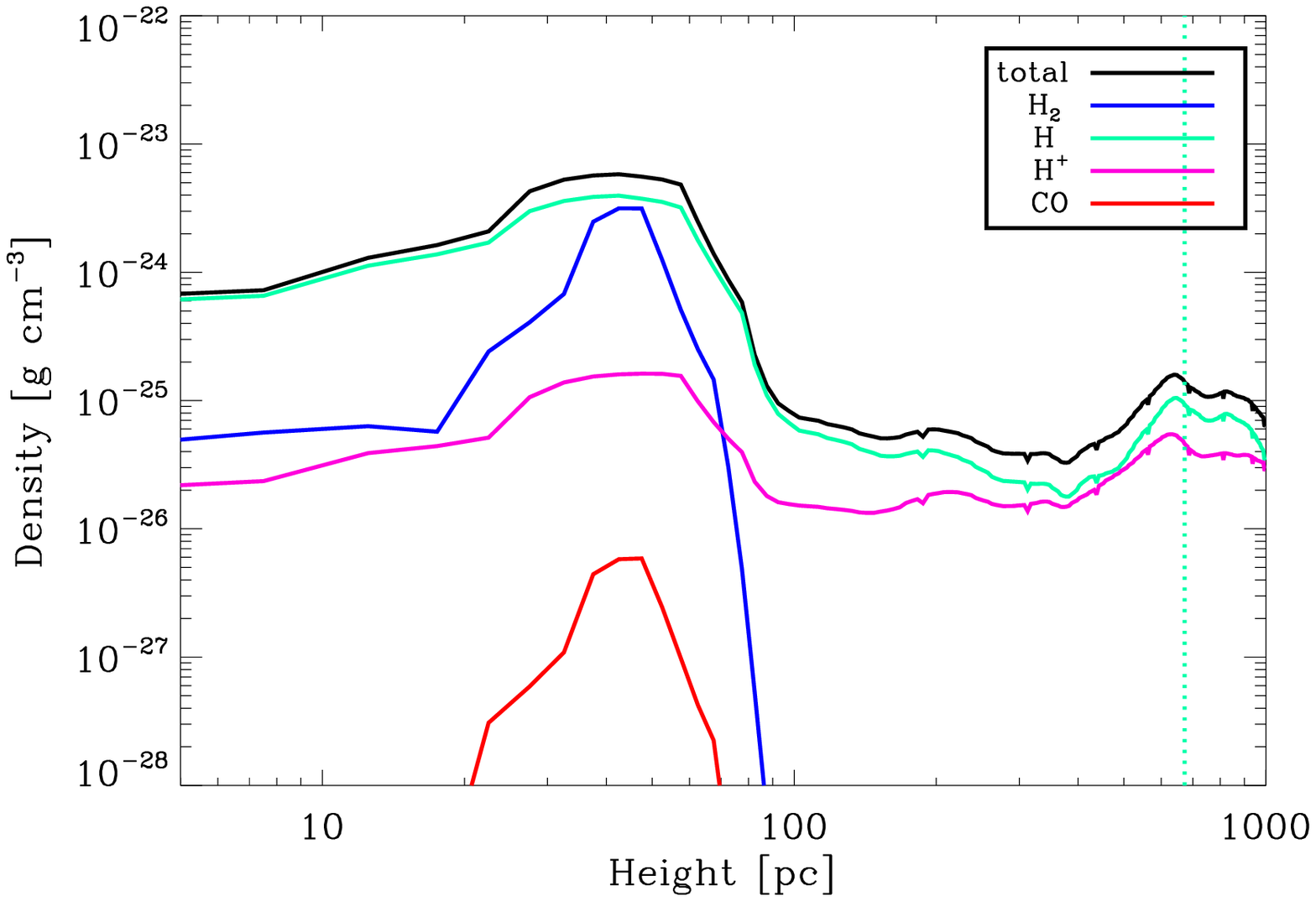}\\
  \begin{sideways} {\bf $\;\;\;\;\;\;\;\;\;\;\;\;\;\;\;\;$S10-KS-clus-mag3} \end{sideways}&
    \includegraphics[trim = 7mm 4mm 3mm 1mm, clip,width=58mm]{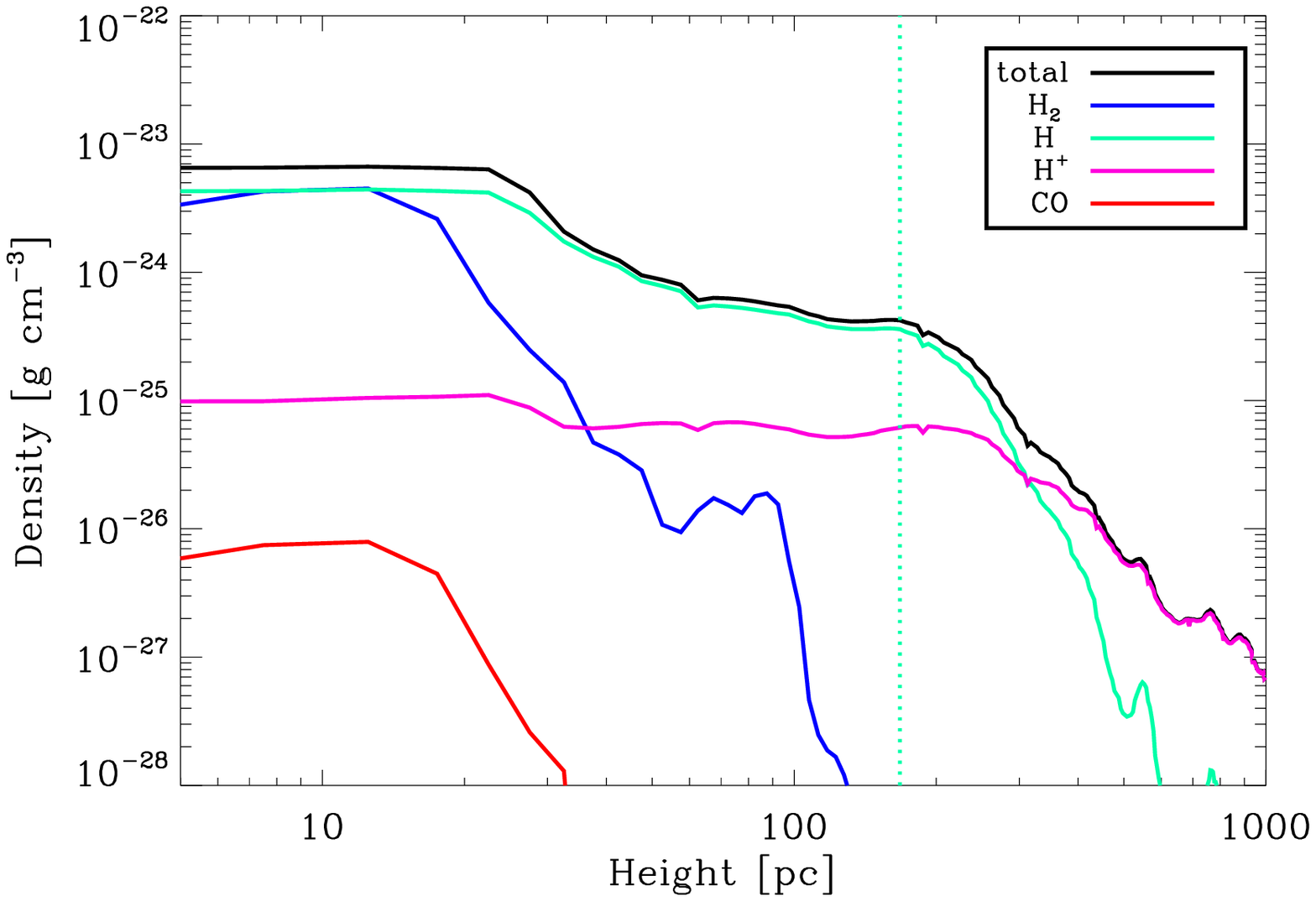}&
     \includegraphics[trim = 7mm 4mm 3mm 1mm, clip,width=58mm]{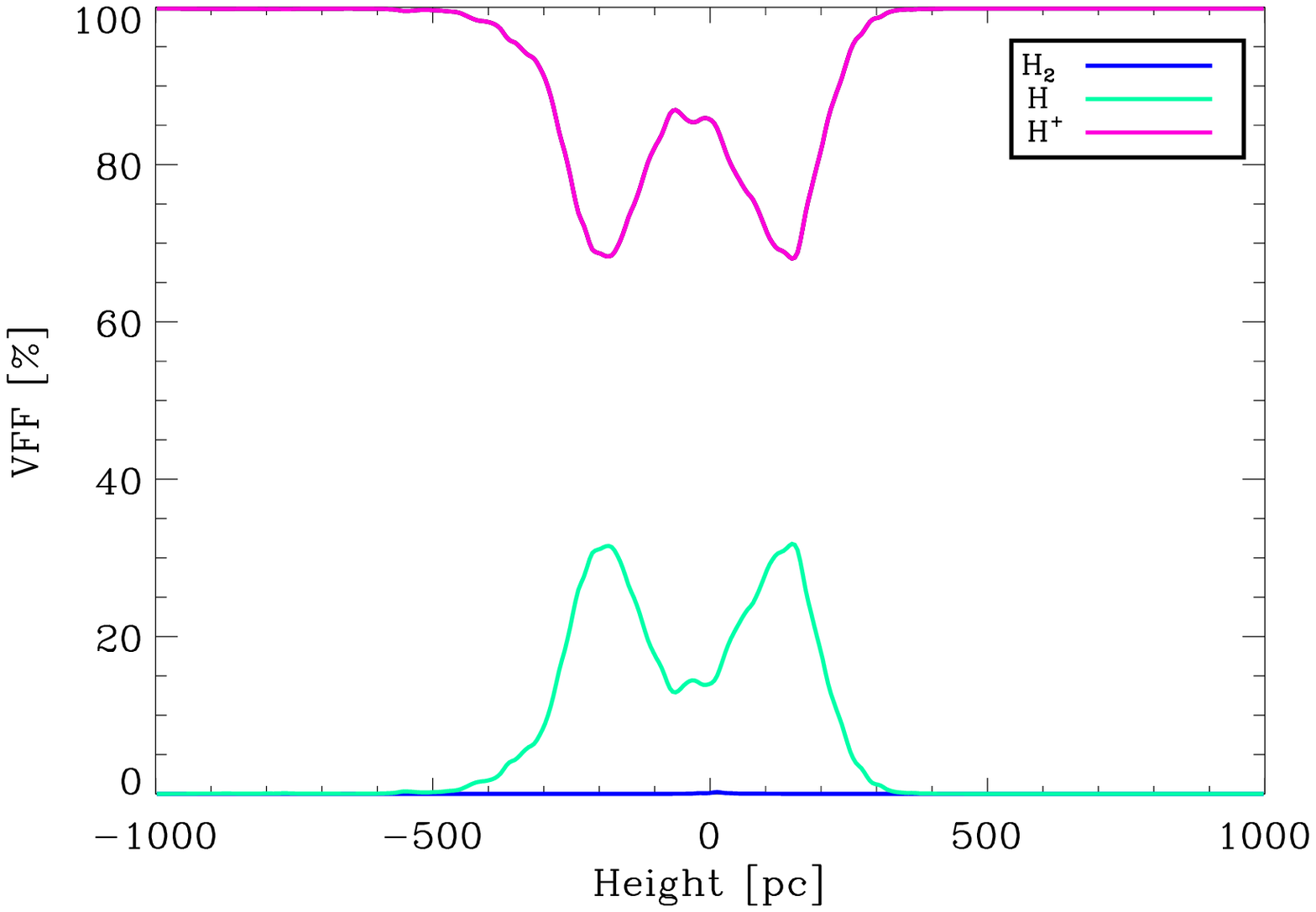} &
     \includegraphics[trim = 7mm 4mm 3mm 1mm, clip,width=58mm]{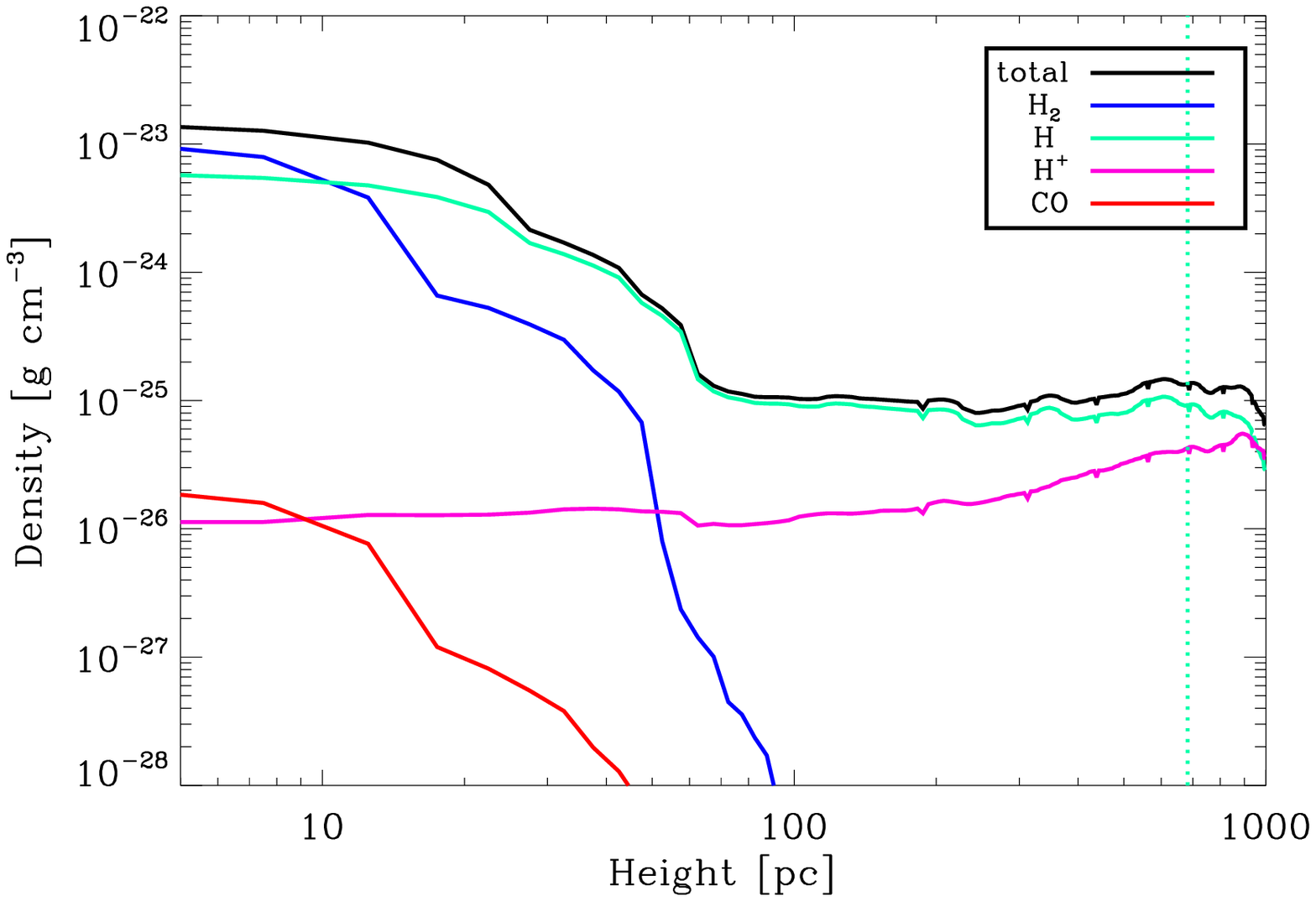}\\

\end{tabular}
 \caption{{\it Left column:} Mean vertical density profiles of the total density, H$^+$, H, H$_2$, and CO taken at $t=50\pm2.0$ Myr. We also average over positive and negative z-direction. The dotted green line indicates the vertical height, which encloses 90\% of the atomic hydrogen mass (see also Table \ref{TAB_HI_THICK}). {\it Middle column:} Vertical profiles of the corresponding VFFs of at $t=50\pm2.0$ Myr. Usually, the VFF of the molecular gas is negligible and H and H$^+$ provide the main contributions.  {\it Right column:} Density profiles at $t=100 \pm 2$ Myr. The disc appears to be significantly more extended than at $t=50$ Myr as a significant amount of mass (primarily in HI) is entrained in an outflow. {\it From top to bottom} we compare runs {\it S10-KS-rand-nsg}, {\it S10-KS-rand}, {\it S10-KS-peak}, {\it S10-KS-clus}, and {\it S10-KS-clus-mag3}.\label{FIG_VERTICAL} }
\end{figure*}

As inferred from Fig. \ref{FIG_S10-rand-KS}--\ref{FIG_S10-peak-KS}, the vertical distributions of the different chemical species are strongly influenced by the SN rate and position, where a higher peak fraction leads to a vertical density distribution which is more concentrated to the disc midplane.
In Fig. \ref{FIG_VERTICAL}, we show the vertical density profiles of H$^+$, H, H$_2$, and CO at $t=50$ Myr (left column) and $t=100$ Myr (right column). In the central column we plot the volume filling fractions (VFFs) as a function of the vertical height at $t=50$ Myr. From top to bottom we show run {\it S10-KS-rand-nsg} without self-gravity (top row), run {\it S10-KS-rand} with random driving (2$^{\rm nd}$ row), run {\it S10-KS-peak} with peak driving (3$^{\rm rd}$ row), and runs {\it S10-KS-clus} (4$^{\rm th}$ row) and {\it S10-KS-clus-mag3} (bottom row) with clustered driving and without/with magnetic fields.

The vertical profiles confirm that the concentration of the disc towards the midplane is strongly influenced by the type of SN driving. The discs are thicker for random than for peak driving, which we quantify by computing the vertical height that encloses 90\% of the total atomic hydrogen mass in the disc (green dashed vertical lines). This measure turns out to be robust, whereas we find that fitting the vertical profiles with a given function (e.g. a Gaussian) is not very educative since any choice of a Gaussian, log-normal, power law, or exponential function is possible but, at the same time, may only provide a good fit to a small part of the distribution. We list the derived values of the disc 'scale height', which encloses 90\% of the atomic hydrogen mass at $t=50$ Myr in Table \ref{TAB_HI_THICK}.
\begin{table}
\begin{tabular}{c | c c c}
  SN implementation     & lowSN & KS & highSN \\
  \hline
 rand & 33 pc & 168 pc  &  -- \\
 clust & -- & 168 pc & -- \\
 ng-rand & -- & 188 pc & -- \\
 peak &  13 pc & 38 pc & 173 pc \\ 
 \hline
\end{tabular}
\caption{Vertical height which encloses 90\% of the atomic hydrogen mass at $t\sim 50$ Myr. From top to bottom we list runs with random driving, the run with clustered SNe, with random driving but without self-gravity, and peak driving.  \label{TAB_HI_THICK}}
\end{table}

Concerning the vertical VFF profiles of the chemical species, runs {\it S10-KS-rand} and {\it S10-KS-rand-nsg} seem to give the most realistic values for the hot, ionised and the warm, atomic ISM towards the disc midplane (about 50\% each). The peak driving runs generally produce a very high atomic hydrogen VFF ($\sim$90\% for {\it S10-KS-peak}), which seems inconsistent with observations of the Milky Way. 

Following \citet{McKee1977} and \citeauthor{Tielens2005} (2005), we can estimate the volume filling fraction of the hot phase that we expect for a given SN rate and mean gas density.
The porosity parameter
\begin{equation}
Q \approx 0.12 N_\mathrm{SN} \left( \frac{E}{10^{51}\; {\rm erg}}\right)^{-44/45} n_0^{-44/45}
\end{equation}
is an estimate of the probability that a randomly distributed SN explodes within the bubble blown by a previous remnant. Here, $N_\mathrm{SN}$ is the SN rate within the whole galaxy per 100 years and $n_0$ is the mean density of the ISM. For $n_0 = 1\;{\rm cm}^{-3}$ and $N_\mathrm{SN}\simeq 0.19$, which is roughly the KS SN rate of 15/Myr scaled up to a full disc with radius 10 kpc, we derive $Q\simeq 0.228$. Note that the scaling of $Q$ with mean density is almost linear. Therefore, $n_0=0.2 \;{\rm cm}^{-3}$ gives $Q\simeq 1.0$, which results in $f_\mathrm{hot}=Q/(1+Q) = 0.5 $. This corresponds to the hot gas volume filling fraction close to the disc mid plane for model {\it S10-KS-rand}. The two runs with clustered driving have higher hot gas VFFs, which is consistent with a higher spatial correlation of the SN events. However, $Q$ can only be computed approximately since the H$^+$ density distribution is broad (see section \ref{SEC_CHEM}). 
We discuss the VFFs of different temperature phases for all runs in section \ref{SEC_SUMMARY}.

How variable is the vertical disc structure as a function of time? The answer is, highly.
From the right column of Fig. \ref{FIG_VERTICAL} one can see that the disc density profiles are much more extended at $t=100$ Myr. For instance, for run {\it S10-KS-rand}, the 90\% atomic hydrogen scale height is increased from $\sim 168$ pc at $t=50$ Myr to $\sim 650$ pc, which is an increase of a factor of $\sim$4. Typically, the VFFs change significantly because the atomic hydrogen component is driven to large heights above the midplane as it traces the developing galactic outflow. The outflows have a complex multi-phase structure and entrain significant amounts of ionised gas and warm atomic gas.

\begin{figure}  
  \includegraphics[width=88mm]{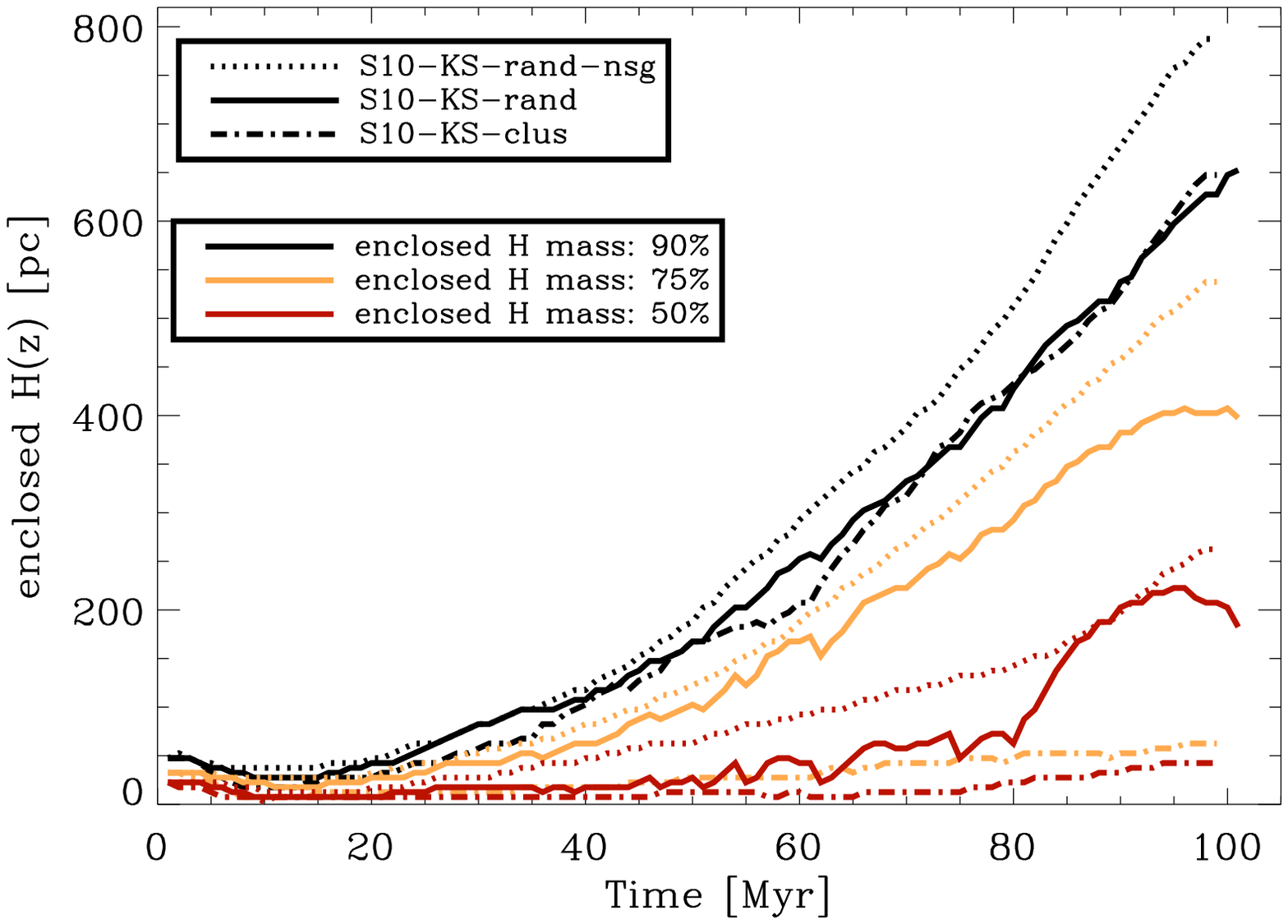}
  \caption{Time evolution of the disc height, which encloses 90\% (black lines), 75\% (yellow), or 50\% (red) of the mass in HI gas. We compare the effect of {\bf clustered SNe} (run {\it S10-clus-rand}; dotted lines), and the differences when running the exact same setup as run {\it S10-rand-KS} (solid lines) {\bf without} self-gravity (run {\it ng-S10-rand-KS}; dash-dotted lines). \label{FIG_HEIGHT} }
\end{figure}

In Fig. \ref{FIG_HEIGHT}, we show the time evolution of the atomic hydrogen `scale heights', which enclose 90\% (black curves), 75\% (yellow), or 50\% (red) of the total atomic hydrogen mass for three runs ({\it S10-KS-rand-nsg} -- dash-dotted lines; {\it S10-KS-rand} -- solid lines; and {\it S10-KS-clus} -- dotted lines). The atomic hydrogen distributions become more extended with time as the three 'scale heights' diverge slowly but continuously. We do not see a turnover in the 90\% distributions (black lines), which are still increasing at 100 Myr. The run without self-gravity shows the largest `scale heights', and all three of them are still increasing at $t=100$ Myr. For runs with self-gravity, the gas in the inner disc (75\% and 50\% distributions) seems to be on a cycle with a smaller turnover length since the 75\% distributions become flat after $\sim 90$ Myr, where the 50\% distributions already decrease. Random and clustered driving have similar `scale heights' for the 90\% 
 distribution, but the clustered driving run is more concentrated towards the disc midplane and shows small 'scale heights' for the smaller mass thresholds (75\% and 50\% are both at $\sim $50 pc at $t=100$ Myr).

However, we point out that the H distribution is not completely different with and without self-gravity, i.e. it is not off by an order of magnitude. Therefore, gas self-gravity cannot be the main agent for setting the scale height of atomic hydrogen. This analysis suggests that self-gravity is rather a secondary effect for gaseous discs with a relatively low surface density, while the global gravitational potential of these discs is dominated by the stellar population (here represented by the external gravitational potential), which mainly determines the vertical disc structure. It provides the balance for the energy and momentum input by SNe (see paper SILCC2 for a discussion on the global pressure balance).

\section{The phases of the ISM}\label{SEC_PHASES}

In the following, we discuss the probability density functions (PDFs) of the different chemical species or, respectively, gas temperature phases, in more detail. Previous studies of stratified discs \citep[e.g.][]{Joung2006, Gent2013a, Kim2014}, have not employed a chemical network to distinguish between molecular and atomic or ionised gas, and have therefore used temperature cuts to estimate the amount of molecular hydrogen that forms in the simulations. Here we show that the chemical composition is complicated and cannot be easily separated into distinct temperature/density phases. We note that, traditionally, the temperature phases of the ISM are defined as follows \citep[e.g.][]{Mihalas1981}:
\begin{itemize}
\item Molecular ($T \le 30$ K): very cold and dense gas, which is most likely in molecular form.
\item Cold ($T < 300$ K): thermally stable cold gas.
\item Warm ($300\; {\rm K} \le T < 10^4$ K): warm atomic and ionised gas.
\item Warm-hot ($10^4 \;{\rm K} \le T < 3 \times 10^5$ K): highly ionised gas in the thermally unstable regime.
\item Hot ($T \ge 3 \times 10^5$ K): very hot gas, mostly in hot SN remnants.
\end{itemize}


\begin{figure*}   
   \begin{minipage}[b]{1.0\linewidth}
   \begin{center}
  \includegraphics[width=80mm]{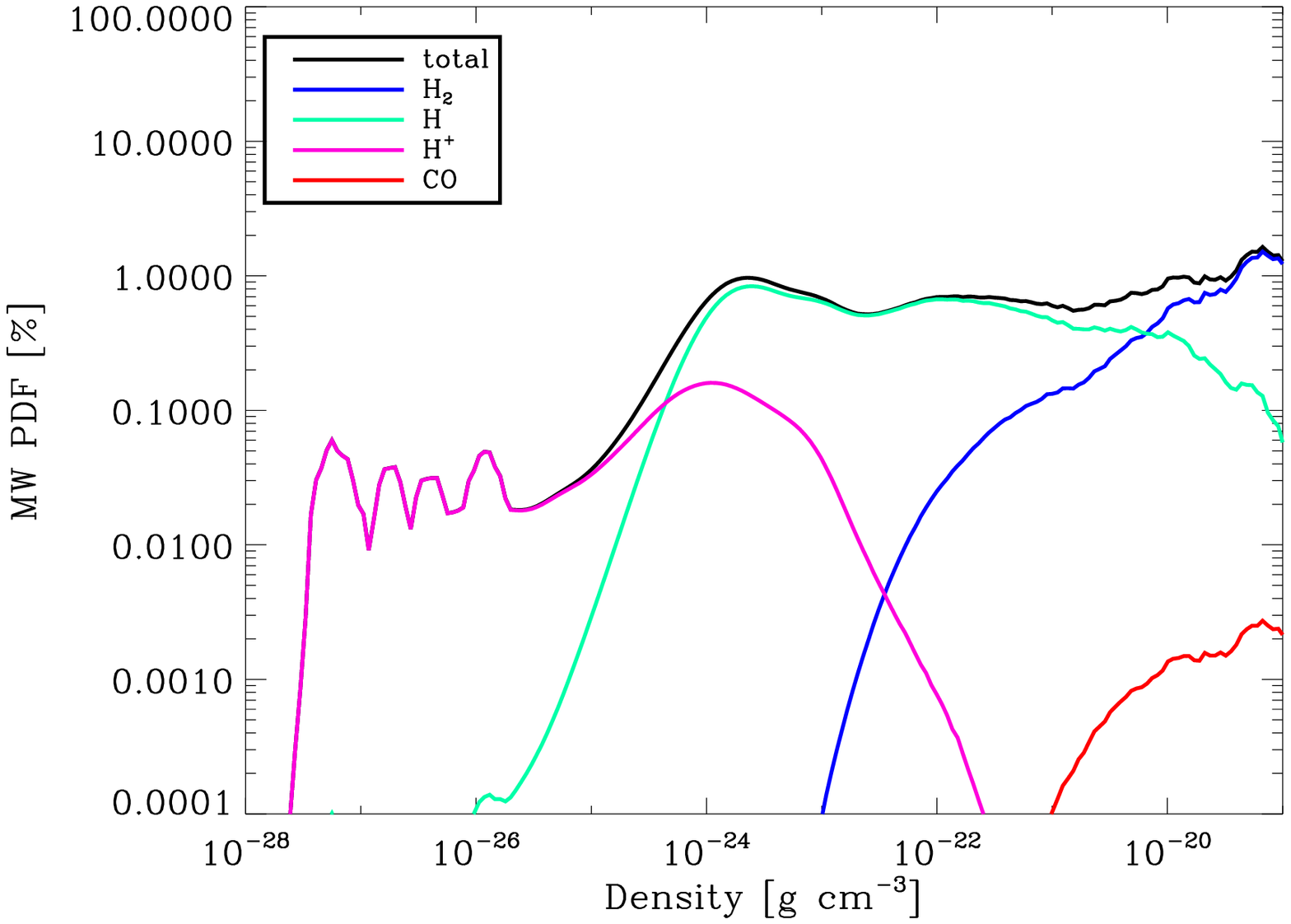} 
 \includegraphics[width=80mm]{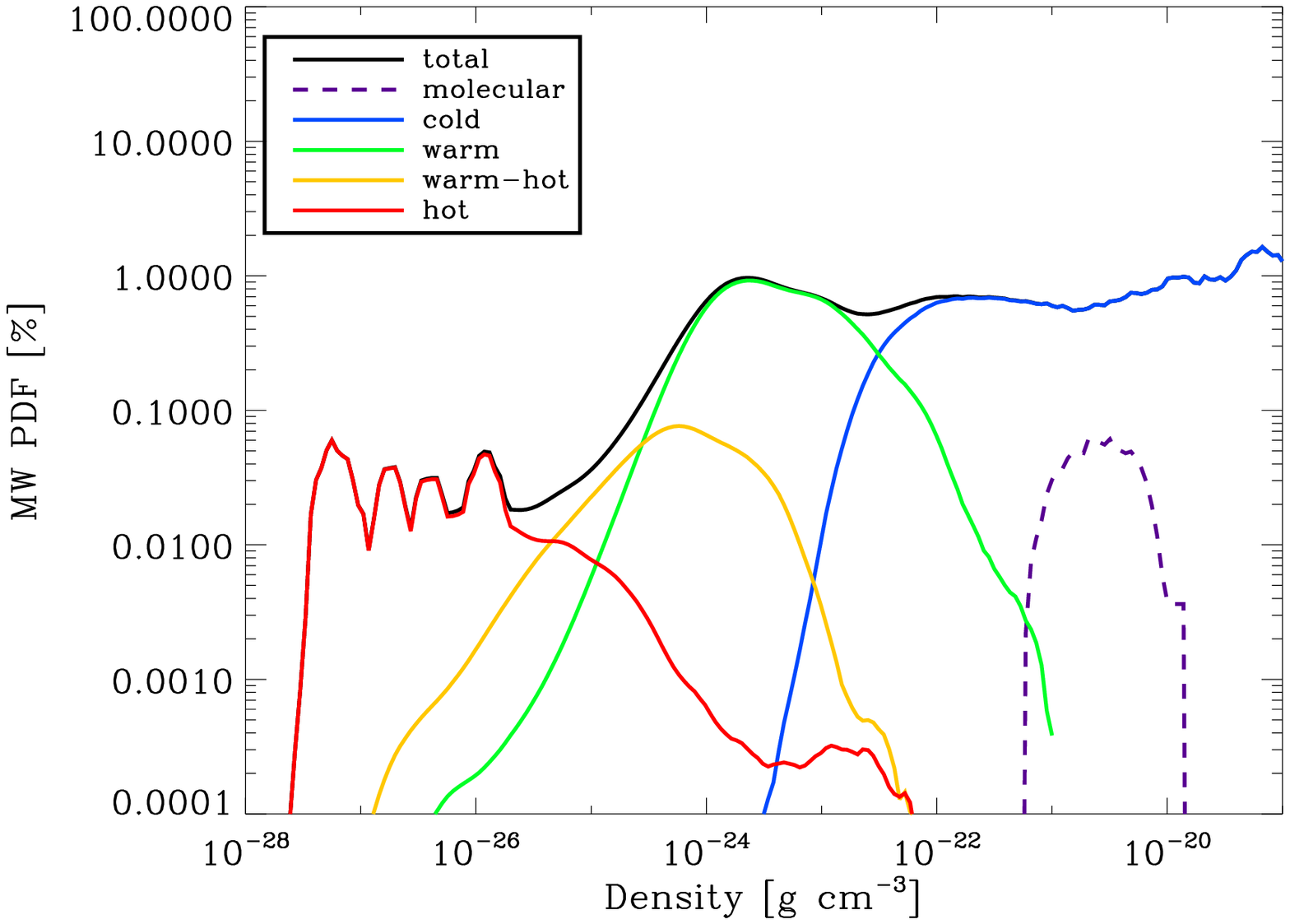} \\
  \includegraphics[width=80mm]{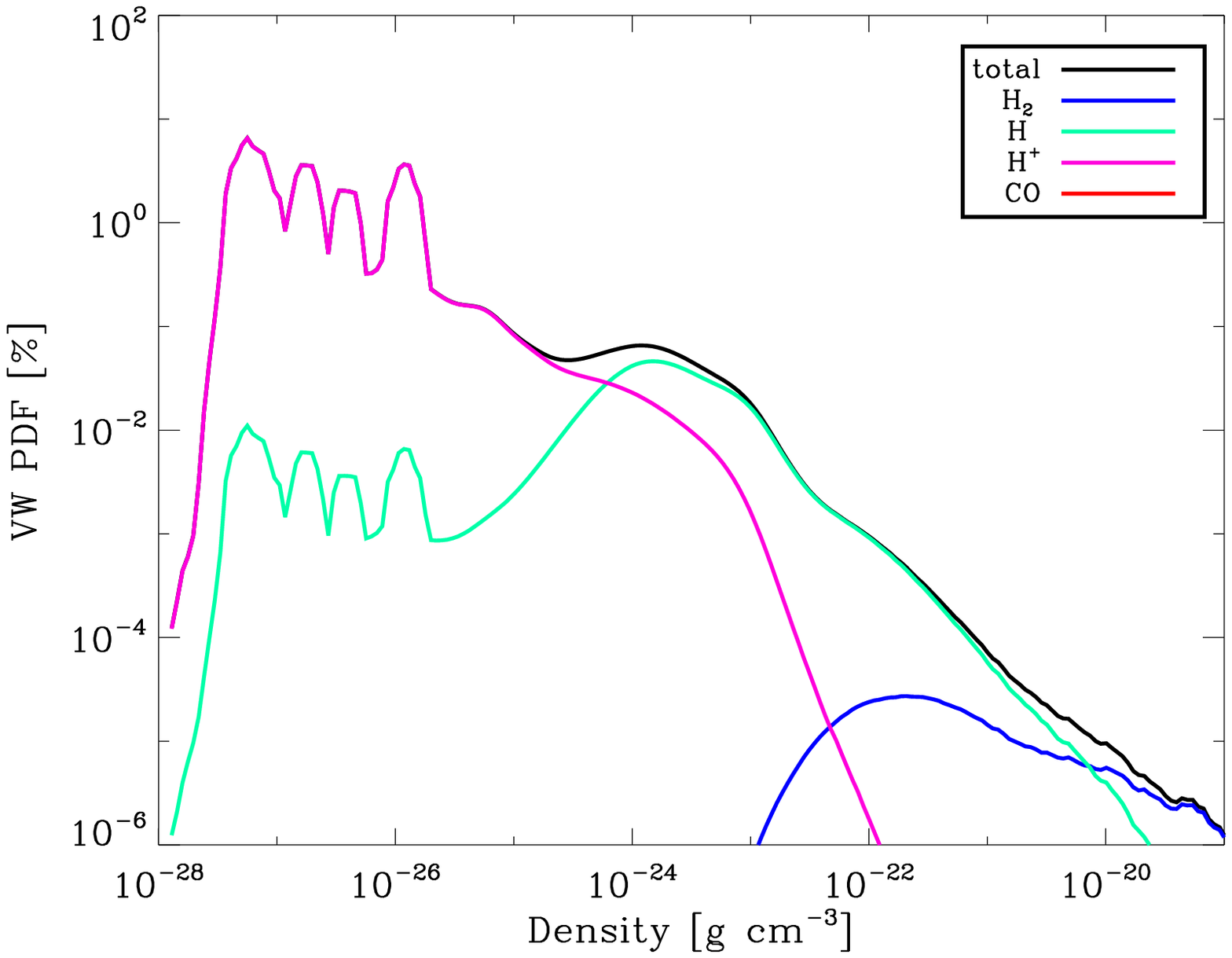} 
  \includegraphics[width=80mm]{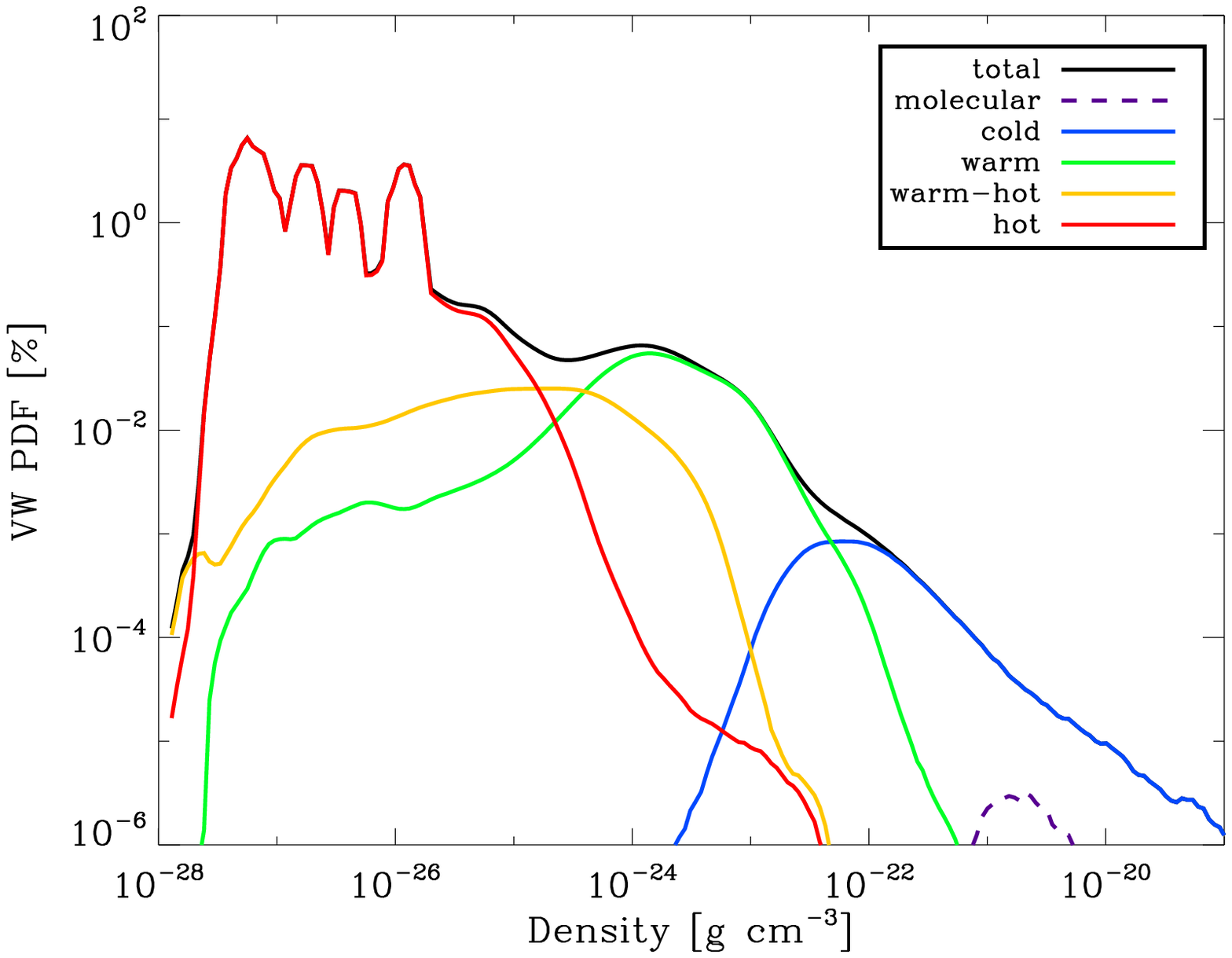} \\
 \end{center}
  \caption{Mass-weighted ({\it 1$^{st}$ row}) and volume-weighted ({\it 2$^{nd}$ row}) density PDFs around $t=50\pm 2.0$ Myr for run {\it S10-KS-clus-mag3}. We take the mean of ten simulation snapshots, which are separated by 0.5 Myrs. The {\it left column} shows the contributions according to the chemical composition and the {\it right column} shows the contributions according to the temperature distribution of the gas. We find that the composition derived from the temperature classification generally does not reflect the gas chemistry.\label{FIG_DENSPDF} } 
\end{minipage}
\end{figure*}

\begin{figure*}    
   \begin{minipage}[b]{1.0\linewidth}
   \begin{center}
  \includegraphics[width=80mm]{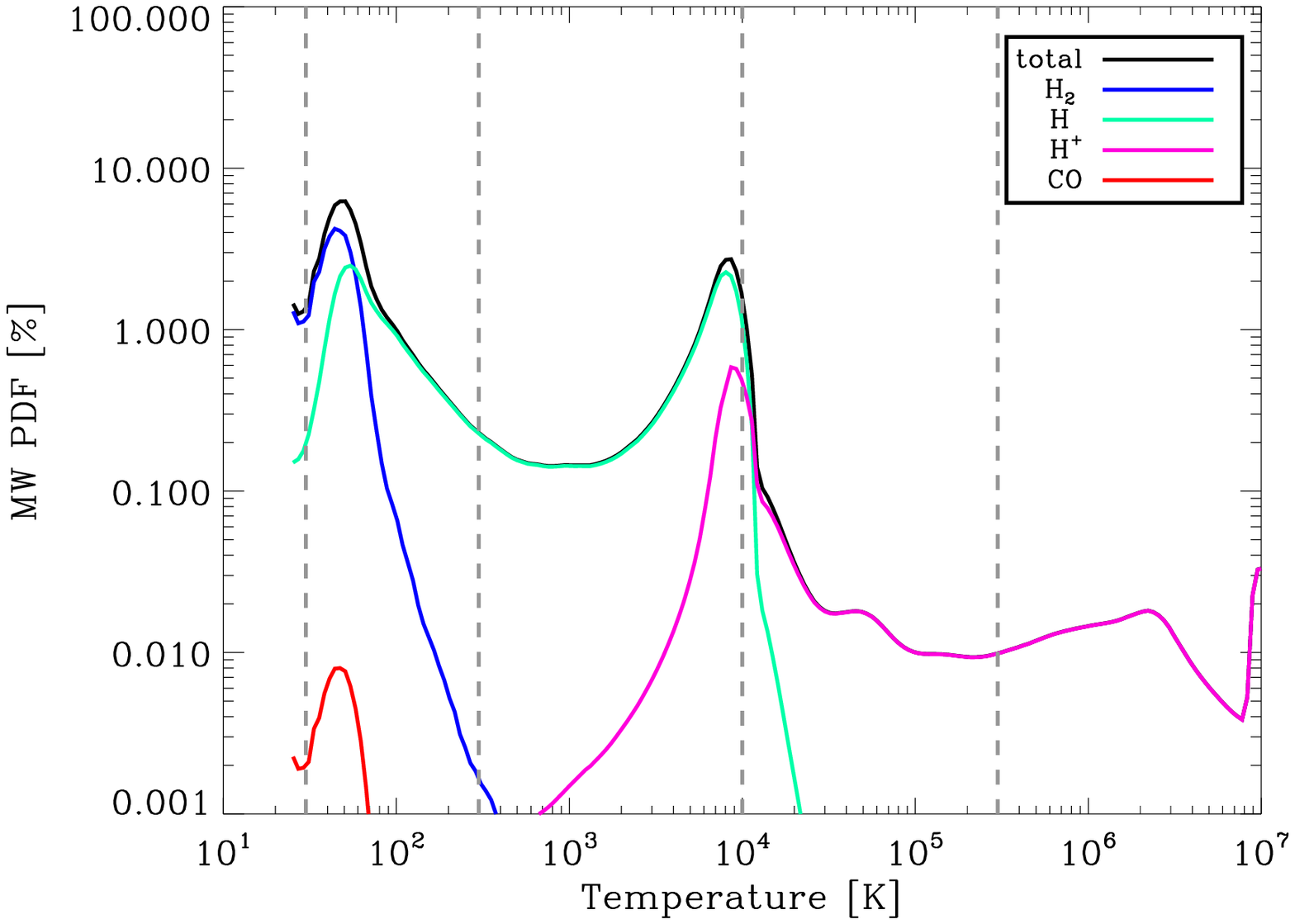}
    \includegraphics[width=80mm]{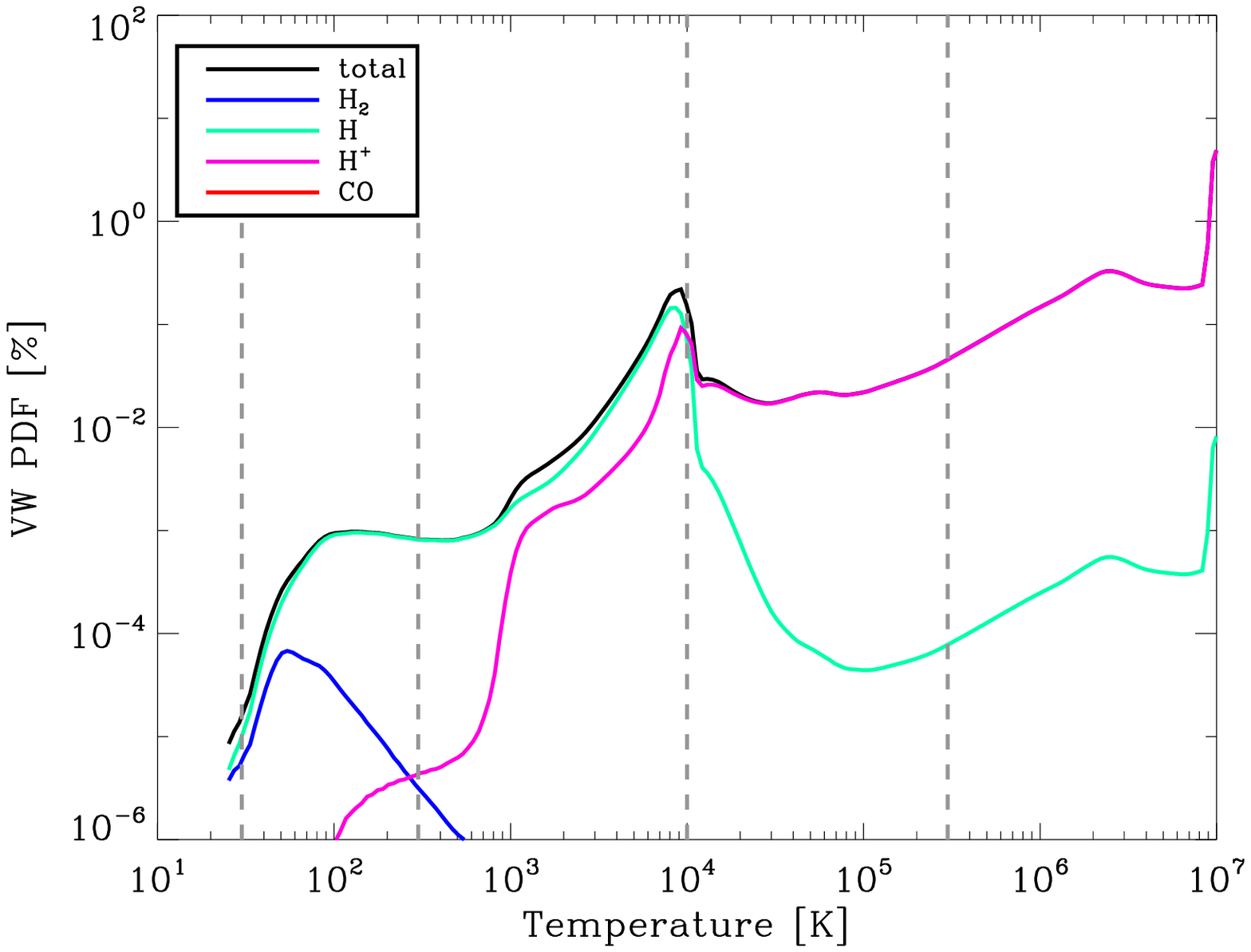}
 \end{center}
  \caption{Mass-weighted ({\it left}) and volume-weighted ({\it right}) temperature PDFs at $t=50\pm 2.0 $ Myr for run {\it S10-KS-clus-mag3}. The grey dashed lines indicate the 5 different temperature regimes we consider in the temperature classification. Each chemical component is spread out over two or more temperature regimes. \label{FIG_TEMPPDF}} 
\end{minipage}
\end{figure*}


\subsection{Chemical composition vs. temperature/density phases}\label{SEC_CHEM}

In Fig. \ref{FIG_DENSPDF} we plot the mass-weighted and volume-weighted density PDFs of run {\it S10-KS-clus-mag3}. The PDFs are split up into individual contributions according to the actual chemical composition (left panels), or gas temperature phase (right panels), respectively. The mass-weighted density PDF appears different when split up by temperature phase rather than chemical composition. The distributions of H$^+$ and H are very broad and overlap significantly, in particular in the thermally unstable (warm-hot) regime. At high densities (above $n\sim 100\;{\rm cm}^{-3}$; dashed purple line), there is still a significant amount of H, which represents the thermally stable, cold neutral medium (CNM). Only above $n\sim 10^4\;{\rm cm}^{-3}$, the medium is predominantly molecular.
The amount of molecular gas predicted from using simple density and temperature cuts ($T \le 30$ K and $n \ge 100\;{\rm cm}^{-3}$) is inconsistent with the actual mass and distribution of molecular gas (H$_2$ and CO). Using the simple estimate, the total mass in molecular gas is under-predicted by a factor of 3 in this run at $t\approx 50$ Myr because there is a warm H$_2$ component.

The volume-weighted density PDFs are in better agreement, at least for the hot and warm gas and H$^+$ and H, respectively. The thermally stable, cold gas again consists of a combination of atomic hydrogen (mostly CNM) and molecular medium. The main difference is found in the distributions of the molecular gas, where the one derived from the simple estimate is more narrow but steeper than the distribution extracted from the simulation, which dynamically follows the formation and dissociation of H$_2$ and CO.

In Fig. \ref{FIG_TEMPPDF} we show the mass-weighted and volume-weighted temperature PDFs for the different chemical components. The grey dashed lines indicate the temperature thresholds used in the simple phase classification. All the chemical components are spread out over two or more temperature regimes, which explains the major differences we see in the density PDFs.\\

In general, none of the total PDFs has a simple lognormal shape or consists of a combination of lognormal distributions as e.g. found in \citet{Hennebelle2014}, who use a simpler cooling description. Concerning the individual components, only the volume-weighted distribution of H$_2$ is consistent with a broad lognormal shape -- although the high-density tail is under resolved and a possible power law tail, which is expected to develop due to local gravitational collapse \citep{Klessen2000, Slyz2005, Girichidis2014PDF, Kainulainen2009, Schneider2011}, cannot be captured. We note that, since the real ISM is neither isothermal nor stirred via coherent large-scale forcing, deriving a turbulent Mach number from fitting a lognormal function to an observed density PDF can easily lead to unreliable results and should be used with caution.
\begin{figure*}  
   \begin{minipage}[b]{1.0\linewidth}
   \begin{center}
  \includegraphics[trim = 5mm 1mm 5mm 10mm, clip,width=85mm, height=53mm]{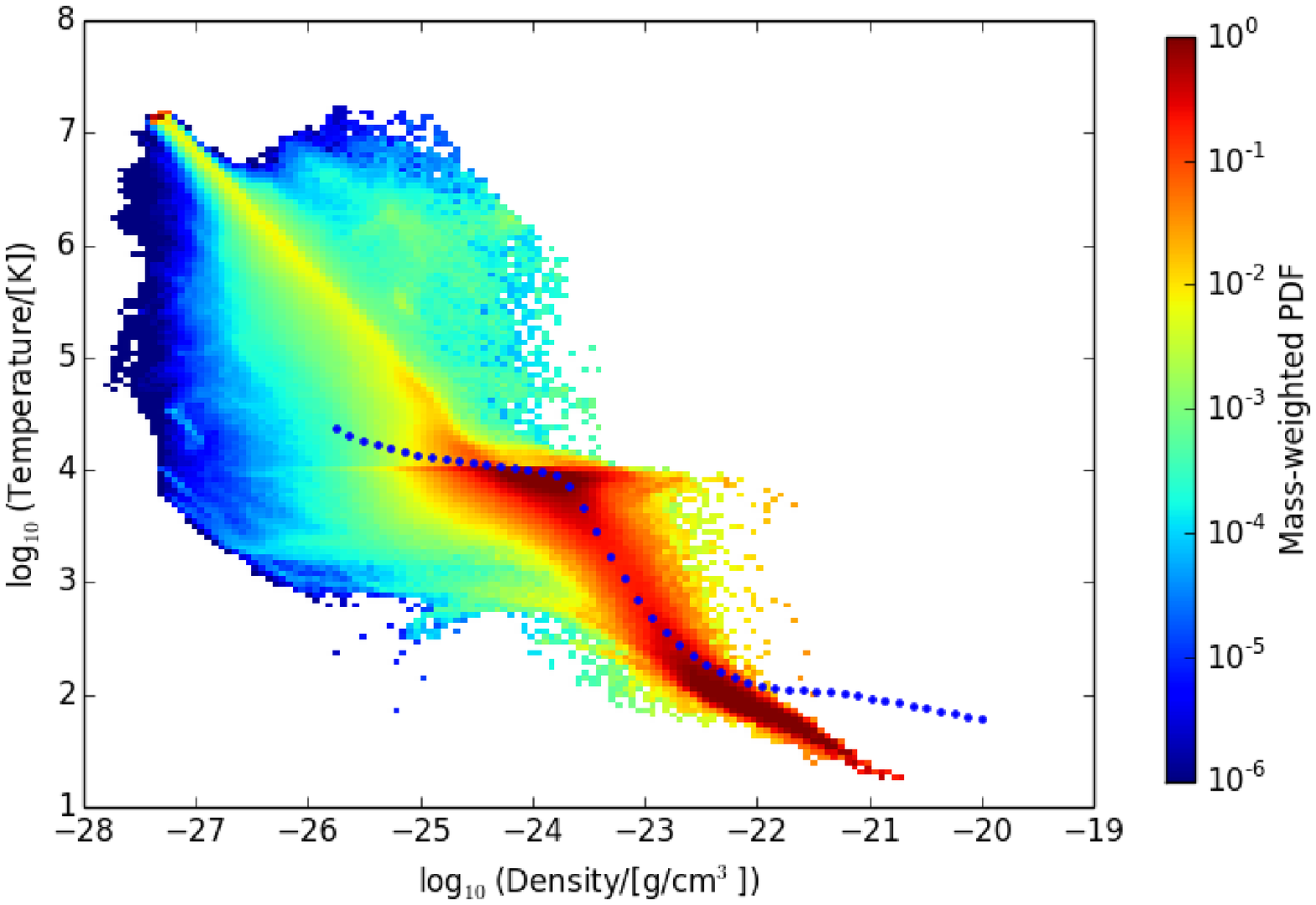}
\includegraphics[trim = 5mm 21mm 5mm 15mm, clip,width=85mm, height=55mm]{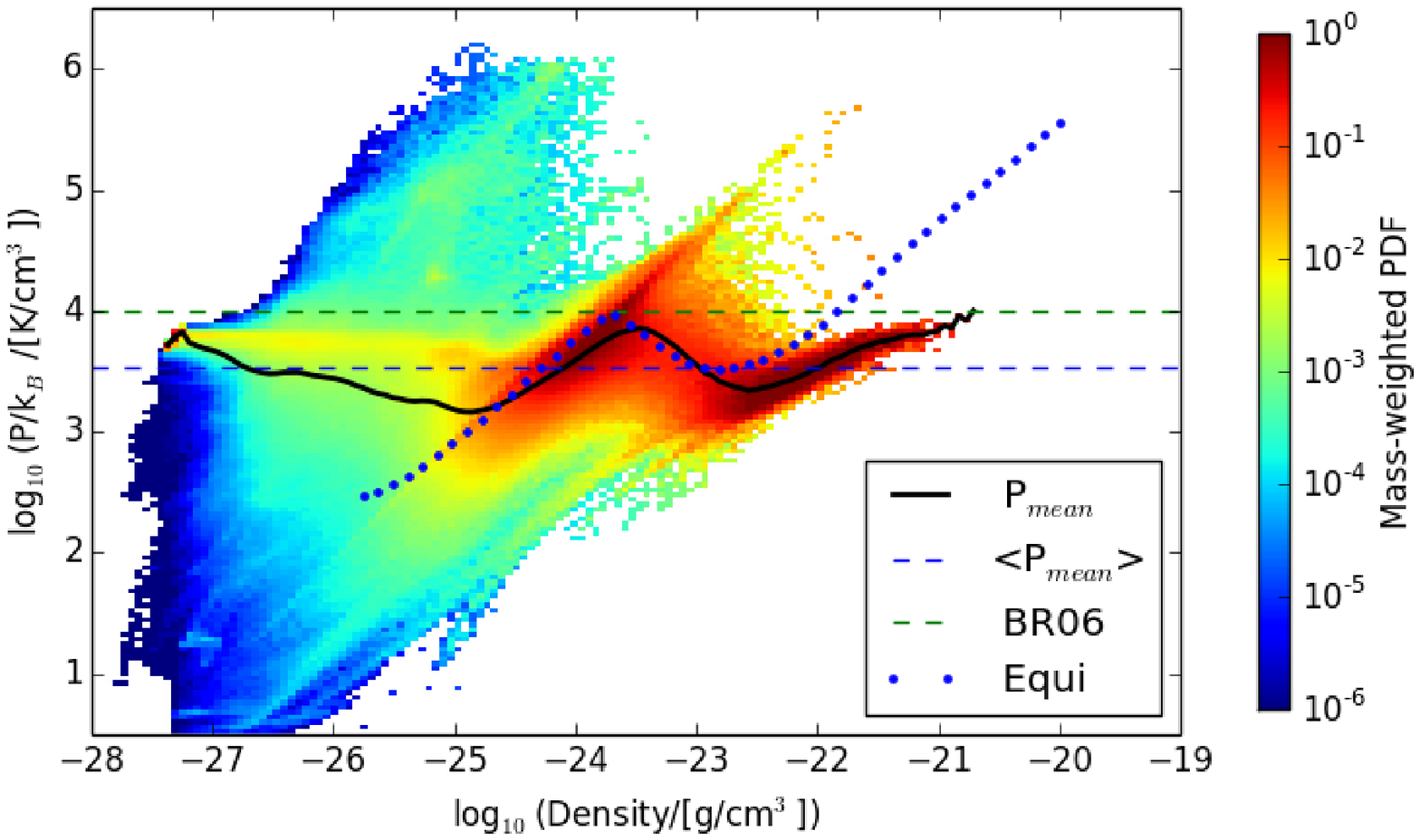}\\
  \includegraphics[trim = 5mm 1mm 5mm 10mm, clip,width=85mm, height=53mm]{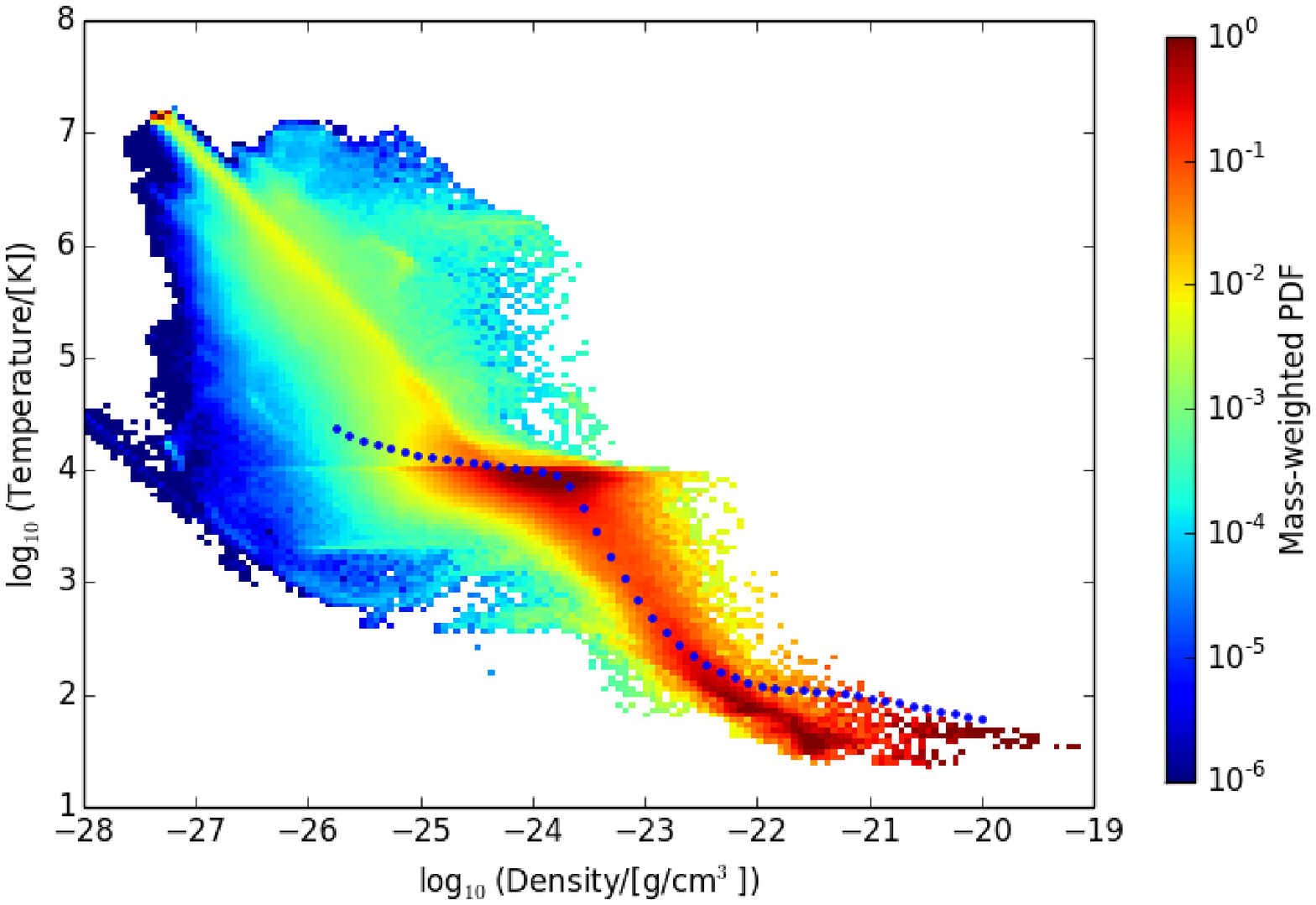}
\includegraphics[trim = 5mm 21mm 5mm 15mm, clip,width=85mm, height=55mm]{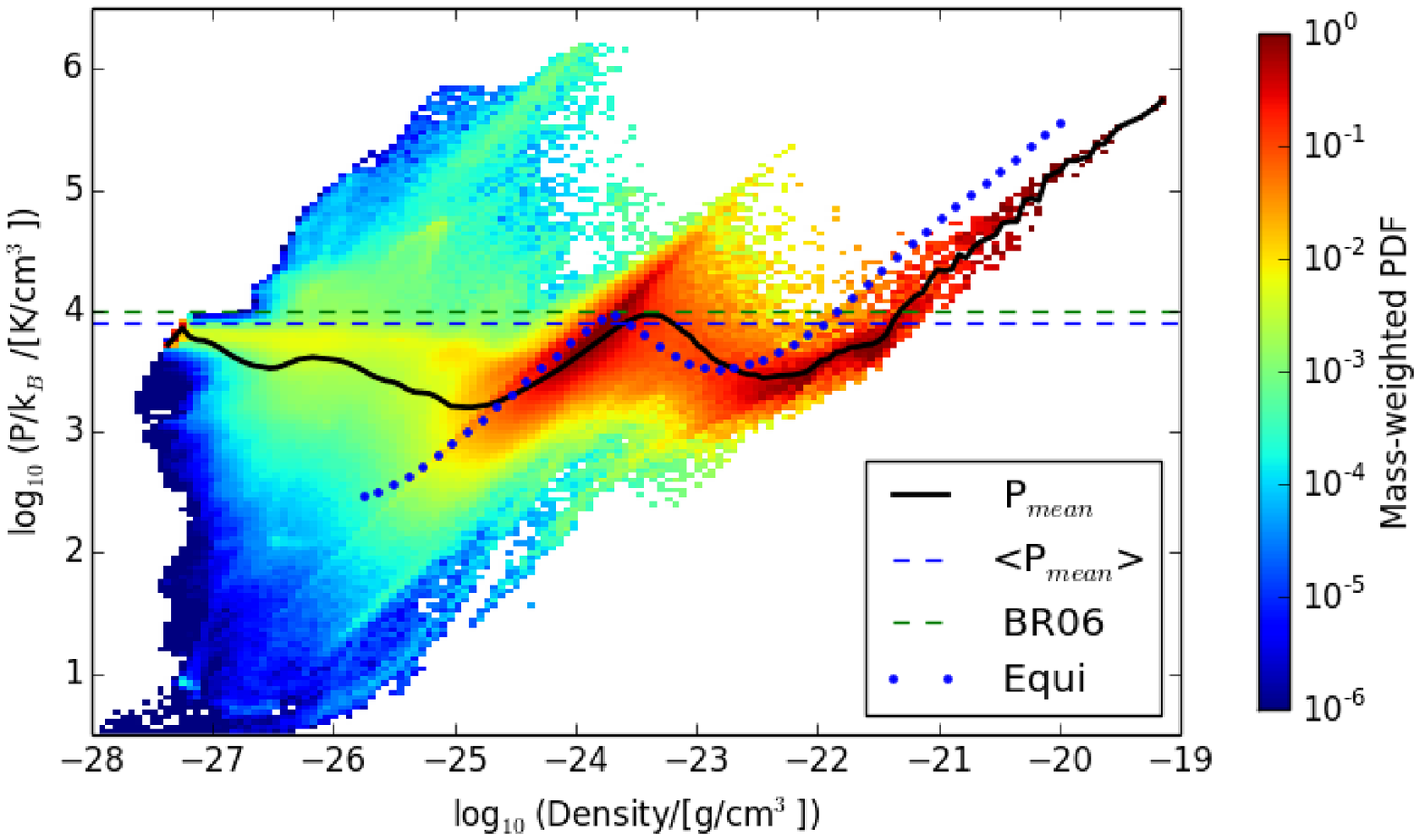}\\  
\includegraphics[trim = 5mm 1mm 5mm 10mm, clip,width=85mm, height=53mm]{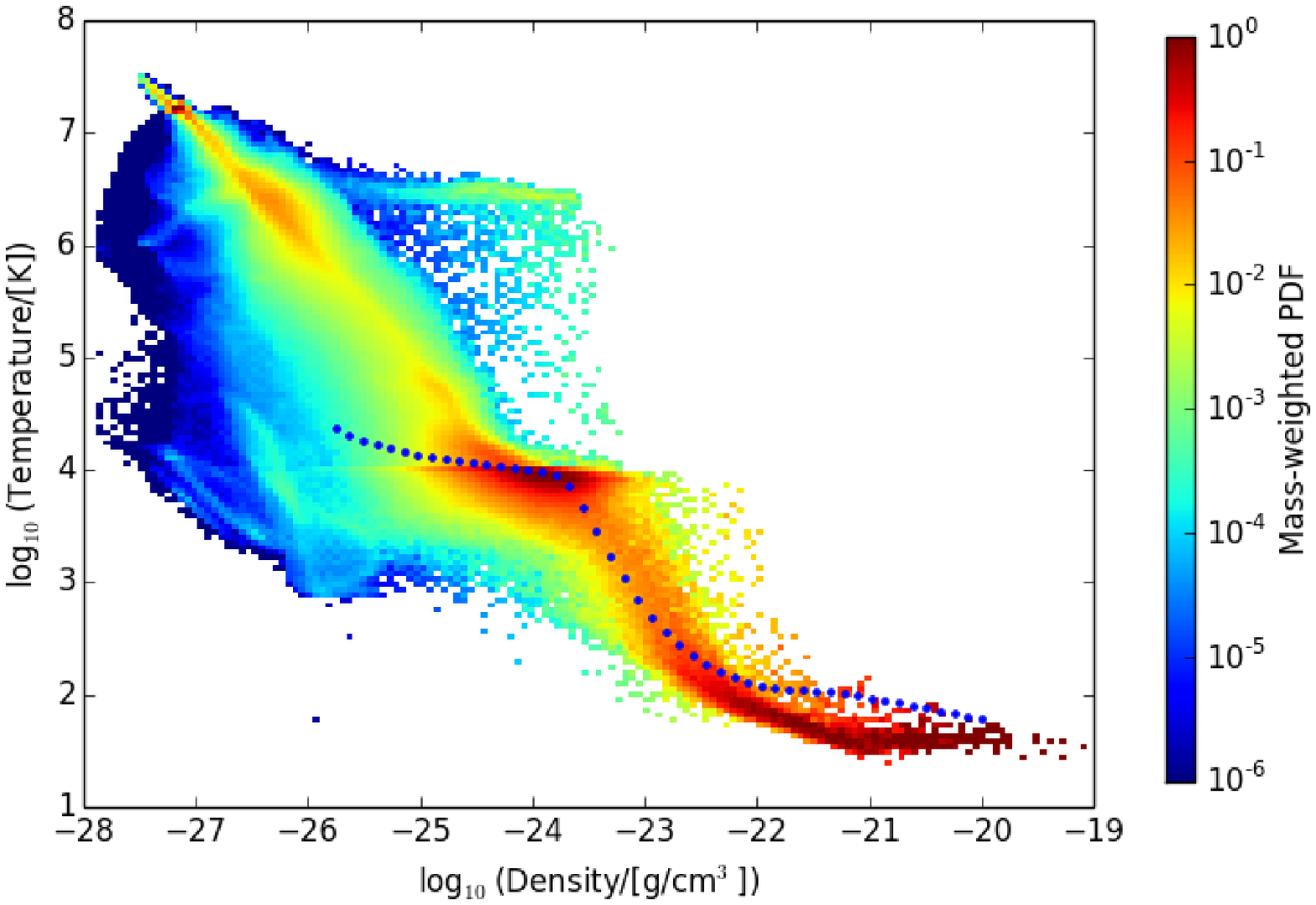}
\includegraphics[trim = 5mm 21mm 5mm 15mm, clip,width=85mm, height=55mm]{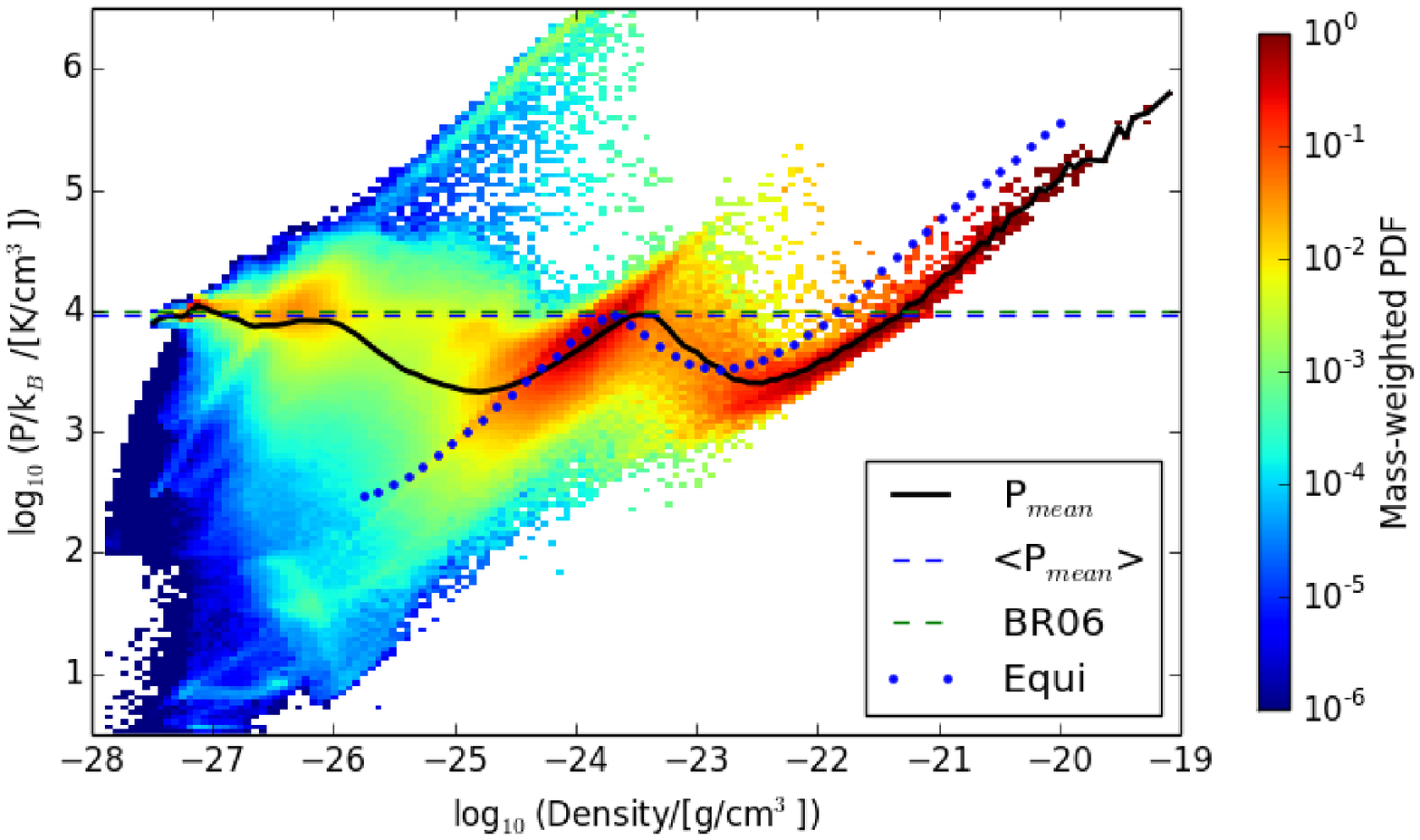}\\
  \includegraphics[trim = 5mm 1mm 5mm 10mm, clip,width=85mm, height=53mm]{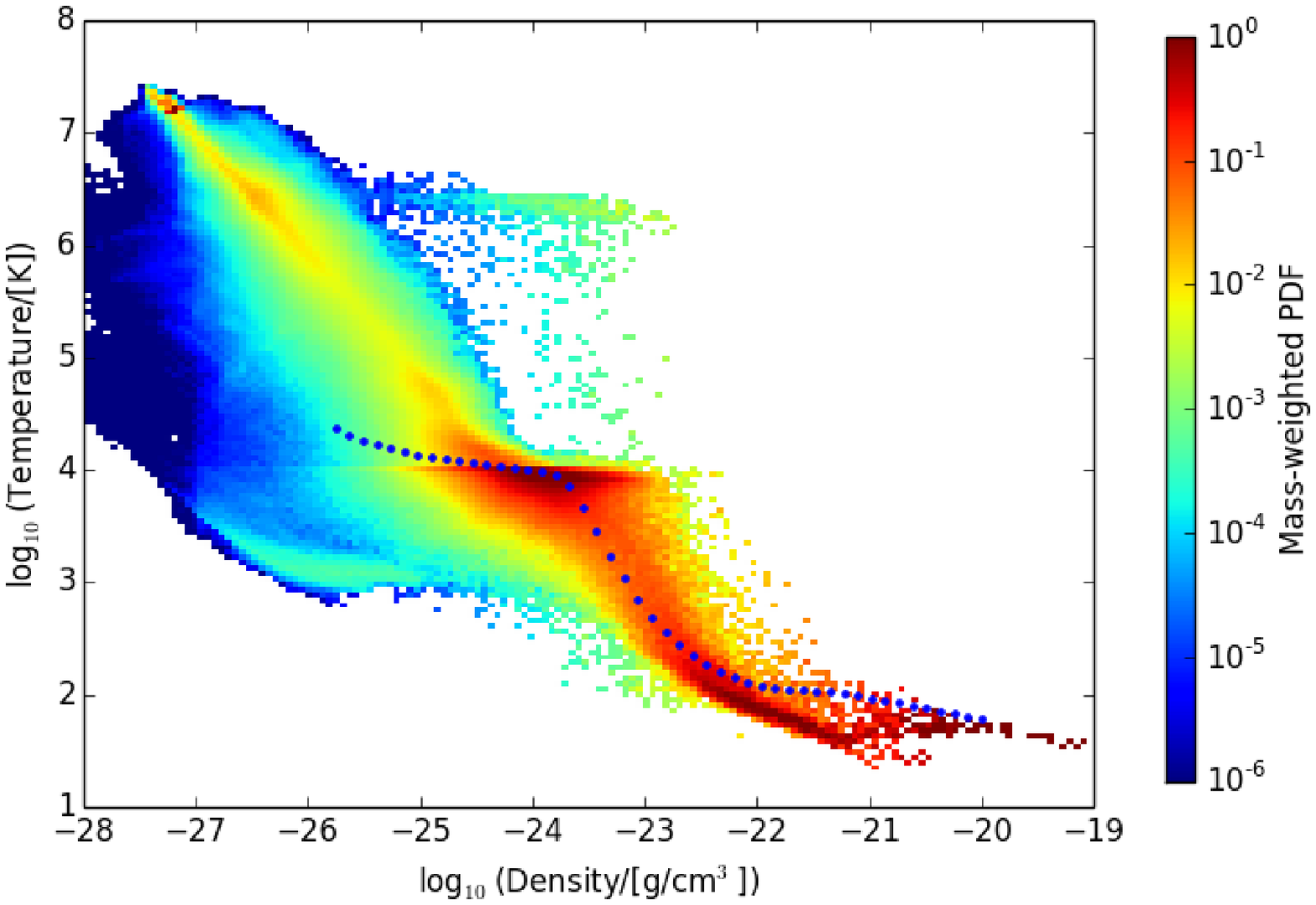}
\includegraphics[trim = 5mm 21mm 5mm 15mm, clip,width=85mm, height=55mm]{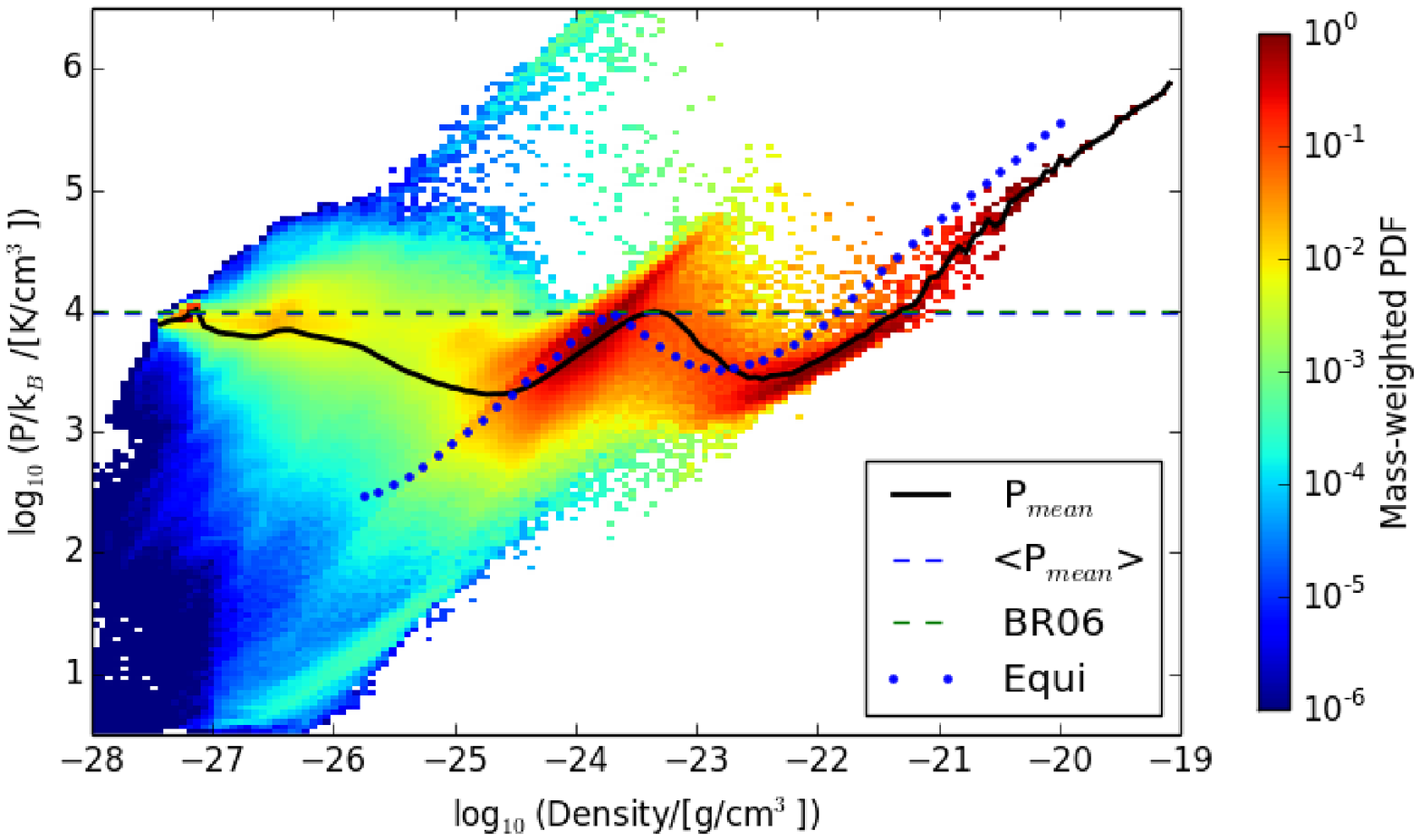}
 \end{center}
  \caption{Mass-weighted phase space probability distributions for runs {\it S10-KS-rand-nsg} (top row), {\it S10-KS-rand} (2$^{\rm nd}$ row), {\it S10-KS-clus} (3$^{\rm rd}$ row), and {\it S10-KS-clus-mag3} (bottom row) at $t=50$ Myr. We over plot the mean pressure for every density bin, $P_\mathrm{mean}$ (black lines), the average pressure computed as $\langle P_\mathrm{mean}\rangle$ (blue dashed lines), and the mid plane pressure derived from \citet{Blitz2006}, $P_\mathrm{BR06}/k_\mathrm{B}=9965\; [{\rm K/cm}^{3}]$ (see Eq. \ref{BR06}; green dashed lines), as well as the equilibrium curve (blue dotted line), which was computed with a constant hydrogen column density of $N_\mathrm{H,tot}=10^{19}\;{\rm cm^{-2}}$ (see Eq. 18) and without considering self-shielding of H$_2$ and CO. \label{FIG_PHASE} }
\end{minipage}
\end{figure*}

\subsection{Phase diagrams}
In the traditional view, the different phases of the ISM, in particular the warm and the cold phase, can co-exist in approximate pressure equilibrium \citep{Field1965, McKee1977}. This picture has been revised since models of turbulent flows under the influence of heating and cooling have shown that the phase space (temperature-density or pressure-density) distributions of the gas are broad due to turbulent heating in dissipative shocks on the one hand, and cooling by expansion on the other hand \citep{Vazquez2009, Seifried2011,Walch2011a, Micic2013, Saury2014}. 

In Fig. \ref{FIG_PHASE}, we plot the temperature-density (left column) and pressure-density (right column) distributions of runs {\it S10-KS-rand-nsg} (top row), {\it S10-KS-rand} (2$^{\rm nd}$ row), {\it S10-KS-clus} (3$^{\rm rd}$ row), and {\it S10-KS-clus-mag3} (bottom row) at $t=50$ Myr using 150 bins in log density and log temperature. We over-plot the mean pressure for each density bin, $P_\mathrm{mean}$ (solid black line), as well as the average pressure of the different phases $\langle P_\mathrm{mean}\rangle$ (blue dashed horizontal line). For comparison, we also show the mid plane pressure as estimated by \citet{Blitz2006}:
\begin{equation}\label{BR06}
P_\mathrm{BR06}/k_\mathrm{B} = \Sigma_{_{\rm GAS}} v \sqrt{2 G \rho_\star}/k_\mathrm{B} = 9965\;[{\rm K\;cm}^{-3}]
\end{equation}
where $v$ is the vertical velocity dispersion of the gas, which we set to $v=8 \;{\rm km\;s}^{-1}$, following \citet{Koyama2009} (green dashed horizontal line). In addition, we over-plot the equilibrium curve (blue dotted line) computed using the chemical network with a constant external hydrogen column density of $N_\mathrm{H,tot}=10^{19}\;{\rm cm^{-2}}$ (see Eq.\ \ref{AV_Ntot}) and without considering the self-shielding of H$_2$ and CO. The warm unstable gas traces the equilibrium curve, but at high densities the gas is colder than predicted since our adoption of a fixed value of $N_\mathrm{H,tot}$ for computing the equilibrium curve dramatically underestimates the actual amount of dust shielding in the high density gas. At a given density we find a range of temperatures for three reasons. First, turbulence broadens the phase space distribution as kinetic energy is thermalised quickly and may heat gas to temperatures above the equilibrium curve. Second, the optical depth is not constant throughout the volume and thus, cells which are highly shielded are cooled to temperatures below the equilibrium curve, whereas cells which sit at lower optical depths have higher temperatures than the ones estimated with $N_\mathrm{H,tot}=10^{19}\;{\rm cm^{-2}}$. A factor of 10 in $N_\mathrm{H,tot}$ can easily change the temperature by a factor of 5 in the thermally unstable regime. Third, cooling by expansion causes low density, cool gas below the equilibrium curve. 

All four simulations have two distinct phases (warm and cold) in approximate pressure equilibrium with each other.  Also, the pressure of the low density, hot phase, which consists of SN shock heated gas, seems to be roughly in pressure equilibrium with the warm phase. However, the pressure distribution at low densities is very broad and spans about five orders of magnitude where SN shock heated and adiabatically cooled gas is both present.

Only for the run without self-gravity, there is a clear difference between $\langle P_\mathrm{mean}\rangle$ and $P_\mathrm{BR06}$ of a factor of 2. In this case $P_\mathrm{mean}$ traces the equilibrium curve slightly better than in all other runs with self-gravity, where $\langle P_\mathrm{mean}\rangle$ adjusts to the local mean pressure of the ISM expected for this galactic disc. All runs with self-gravity also have a much more extended cold branch, which extends to high densities and contains a significant fraction of the total mass. The cold gas stretches towards high pressures ($P_\mathrm{mean}/k_\mathrm{B} \approx 10^5$--$10^6 \: {\rm K \: cm^{-3}}$), thereby following a roughly isothermal behaviour with $T \sim 30$ K. The offset to lower temperatures with respect to the equilibrium curve is caused by the (self-)shielding of the gas. Moreover, the fact that all gas with $\rho \gtrsim 5\times 10^{-22}\;{\rm g\;cm}^{-3}$ is over-pressured with respect to the warm and hot phases indicates that self-gravity dominates above this density threshold. Most of the mass in cold gas has been eaten away from the warm component at $T \approx 10^{4}$ K, which itself spans about two to three orders of magnitude in pressure. The hot gas component is more pronounced in both runs with clustered SN driving, in particular in the run without magnetic fields. The general estimate that the hot gas pressure adjusts to the local pressure that is expected for a given galaxy \citep{Ostriker2010} is met very well in our simulations. A more thorough investigation of all pressure components (thermal, kinetic, and magnetic) is given in the companion paper (SILCC2).


\section{Chemical composition and VFFs}\label{SEC_SUMMARY}
Recently, high-resolution data has greatly enhanced our understanding of the gaseous ISM in different galaxies.
In the previous section, we have shown that the mean thermal pressures derived for our models with random or clustered driving are in good agreement with observed estimates. Here, we will shortly summarise the findings on the mass fractions of H, H$_2$, and CO, and on the volume filling fractions of the different gas phases.

\subsection{Mass fractions}
\begin{figure}  
   \begin{minipage}[b]{1.0\linewidth}
   \begin{center}
  \includegraphics[width=85mm]{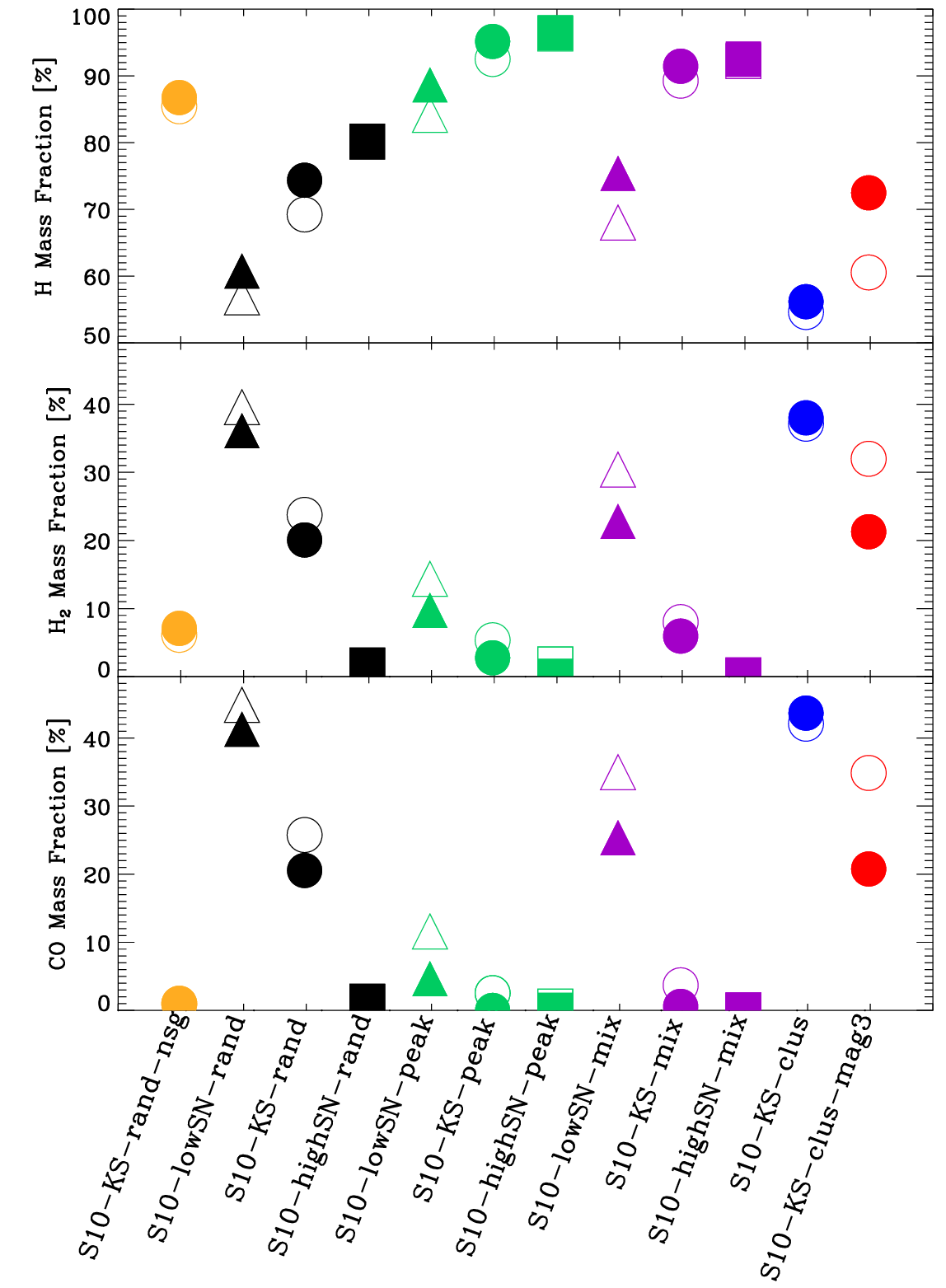}

 \end{center}
  \caption{Time-averaged mass fractions of atomic hydrogen ({\it top}), H$_2$ ({\it middle}), and CO ({\it bottom}) for all simulations presented in this paper. H and H$_2$ mass fractions are computed with respect to the total hydrogen mass and CO is computed with respect to the total mass in carbon. {\it Filled circles} show the average values within 30 Myr--50 Myr, and {\it open circles} show the averages from 30 Myr onward until the end of the simulation. Without self-gravity, the molecular gas fraction is significantly reduced. Compared with peak and mixed driving, random driving results in the highes molecular gas fractions, but the SN rate is as important as the SN positioning. For random driving, a lower SN rate (5 Myr$^{-1}$) gives very similar results as the run with clustered SNe, but the presence of a magnetic field delays the formation of H$_2$ to values similar to those in the fiducial run {\it S10-KS-rand}.\label{FIG_MASS_AV} }
\end{minipage}
\end{figure}

To give a final overview of the mean H, H$_2$, and CO mass fractions that develop in our simulations, we plot the average values between 30 Myr - 50 Myr (filled symbols), as well as the average between 30 Myr and the end of the respective simulation (open symbols) in Fig.~\ref{FIG_MASS_AV}. We find that the run without self-gravity leads to very low H$_2$ fractions of the order of $\sim$5\%. With self-gravity and at a given SN rate, peak (green symbols) and mixed driving (purple symbols) hinder the formation of H$_2$ as long as the dense gas can be dispersed by SN explosions. Therefore, the H$_2$ mass fraction is below 10\% in most of these cases (with the exception of mixed driving at a low SN rate, where the SN heating is not efficient enough to dissociate most of the H$_2$). For random driving at a low SN rate ({\it S10-lowSN-rand}) we find H$_2$ mass fractions of $\sim$ 40\%, comparable with clustered driving ({\it S10-KS-clus}) and, at late times, also clustered driving with magnetic fields ({\it S10-KS-clus-mag3}). In general, these three runs result in the highest H$_2$ mass fractions of our sample. With increasing SN rate, the molecular gas fractions decrease. Overall, we might underestimate the H$_2$ fraction in lower density gas due to our limited resolution, with which we cannot resolve small-scale clumping, which enables and accelerates the formation of molecular hydrogen in this regime \citep{Glover2007}.

In the Milky Way the molecular gas mass fraction drops from $\sim$ 50\% at a Galactocentric radius of 6.5 kpc to $\sim$ 2\% at a distance of 9.5 kpc \citep{Honma1995}. The so-called 'molecular front' is caused by a roughly exponential decrease of the molecular gas mass with Galactocentric radius, while the atomic mass is distributed approximately evenly. 

\citet{Schruba2011} use the IRAM HERACLES survey \citep{Leroy2009} to determine the molecular to atomic gas mass fractions as a function of radius for 33 spiral galaxies. For disc annuli with the given total gas surface density of $10\;{\rm M}_{\odot}{\rm pc}^{-2}$, they find that the H$_2$-to-H surface density ratio varies between 0.1 and $\sim$ 15, with most of the galaxies having ratios between 0.1 and 1. The large variation at this particular total gas surface density arises as a result of a transition from a molecular dominated regime to gas which consists mostly of atomic hydrogen \citep[see also the theoretical work of][]{Krumholz2009}. Therefore, we may say that our findings are consistent with the observations, although the scatter is so large that even the run without self-gravity can provide a reasonable fit (M$_{{\rm H}_2}$/M$_{\rm H} \sim$ 0.065 in this case). However, we certainly do not find ratios of molecular to atomic mass which are greater than 1.

\subsection{Volume filling fractions}
\begin{figure}  
   \begin{minipage}[b]{1.0\linewidth}
   \begin{center}
    $\;\;\;\;\;\;\;\;\;\;\;\;$Time-averaged VFFs within $z=\pm 2$ kpc\\ 
  \includegraphics[width=85mm]{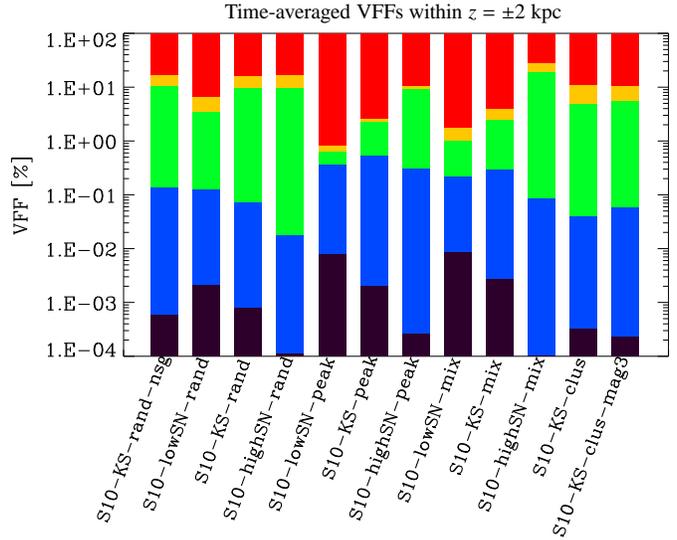}
	\vspace{0.3cm}
 \end{center}
  \caption{Time-averaged volume filling fractions of ISM phases in different temperature regimes for all simulations. We only count gas within $\pm 2$ kpc in the vertical direction. To minimise the effect of fluctuations, we average from 30 Myr onward until the end of each simulation. The colour-coding is the following: {\it Purple:} cold, molecular gas with $T<30$ K; {\it blue} cold, atomic gas with $T < 300$ K; {\it green:} warm gas with $300\;{\rm K} \le T < 12,000 \;{\rm K}$; {\it yellow:} warm, ionised medium with $12,000\;{\rm K} \le T < 300,000 \;{\rm K}$; {\it red:} hot, ionised medium with $T\ge 300,000$ K.   \label{FIG_VFF_AV} }
\end{minipage}
\end{figure}

\begin{figure}  
   \begin{minipage}[b]{1.0\linewidth}
   \begin{center}
   $\;\;\;\;\;\;\;\;\;\;\;\;$Time-averaged VFFs within $z=\pm 150$ pc\\ 
  \includegraphics[width=85mm]{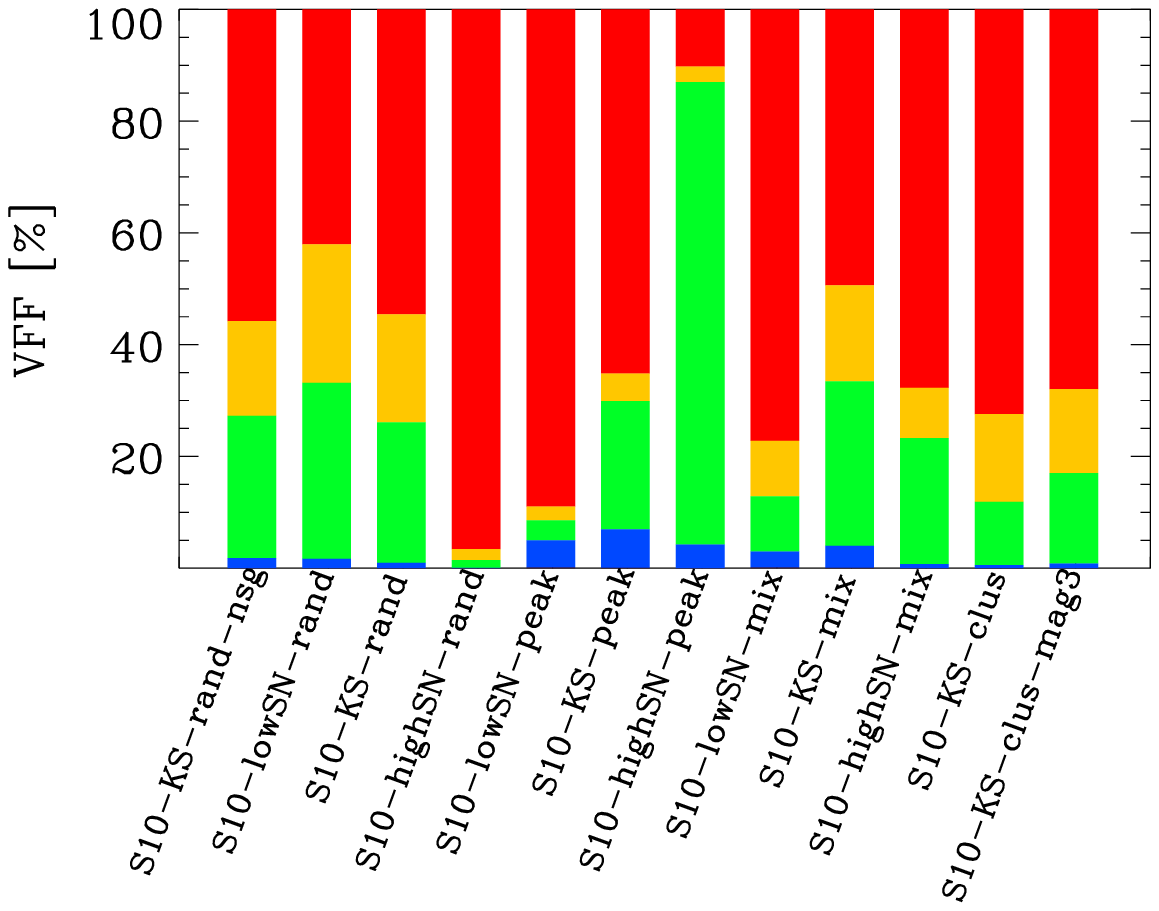}
	\vspace{0.3cm}
 \end{center}
  \caption{Time-averaged VFFs of different temperature regimes, similar to Fig. \ref{FIG_VFF_AV}, for gas within $\pm 150$ pc. Even close to the disc midplane most of the volume  (40\%--55\% for our fiducial runs) is filled with hot gas (red), and a non-negligible fraction is filled with warm-hot gas (yellow) in the thermally unstable regime (10\% - 20\%). Warm gas (green) makes up for another $\sim$15\% - 30\%, and cold, atomic gas (blue) contributes $\lesssim$ 2\%. \label{FIG_VFF_AV2} }
\end{minipage}
\end{figure}

An analysis of the VFFs for different temperature phases is often used to establish an overview of the structure of the multi-phase ISM and attempt to disentangle what drives its evolution.
We show the VFFs for all runs in five different temperature regimes (see section \ref{SEC_PHASES}) within $z=\pm 2$ kpc around the disc mid plane in Fig. \ref{FIG_VFF_AV} and within $z=\pm 150$ pc in Fig. \ref{FIG_VFF_AV2}. Like for the total mass fractions, we average the VFFs from 30 Myr onward until the end of the simulations. 

In Fig. \ref{FIG_VFF_AV} we find that the VFFs are always dominated by hot gas, whereas the warm and cold phases contribute less than 10\% and less than 0.5\%, respectively. Also, the VFFs are very robust if the SNe preferentially explode in low density environments. In these cases, neither the absence of self-gravity, nor clustering or magnetic fields change the mean VFFs significantly. There is a clear trend that a higher SN rate fills the volume with less molecular gas but more cold and warm gas, while, interestingly, the hot gas VFF is rather reduced with increasing SN rate. The reason for the latter is the onset of a larger-scale outflow, which fills a significant volume within the inner 2 kpc around the disc midplane. Overall, a clear difference in the VFF distributions can only be seen for random, peak, and mixed driving.

From Fig. \ref{FIG_VFF_AV2}, we find that even close to the disc mid plane the volume is mostly filled with hot gas (40\% - 55\% for our fiducial runs). In addition, a significant fraction (10\% - 20\%) is filled with warm-hot gas in the thermally unstable regime, while the warm gas only contributes $\sim$ 15\% - 30\%. It seems that our models are in agreement with the classical three-phase ISM model \citet{McKee1977}, where hot gas fills the inter-cloud volume. This is in contradiction to simulations by \citet{deAvillez2004}, \citet{Joung2006} and \citet{Hill2012}, who find that most of the volume is filled with a pervasive, warm medium (VFF ~ 50\%) while the VFF of the hot gas component is smaller and where the hot gas is mostly found in embedded SN bubbles. 

For different SN positioning and SN rates, we find the following trends: For random driving a higher SN rate increases the hot gas VFF. For peak driving the opposite behaviour is found with a much lower hot gas VFF of $\sim$ 10\% for the run with the highest SN rate (45/Myr, run {\it S10-highSN-peak}). Therefore, a high hot gas VFF can also be caused by a vertically more concentrated disc morphology.








\section{Conclusions}\label{SEC_CONCLUSIONS}
We present a self-consistent study of molecular gas formation in a stratified galactic disc, which develops a turbulent, multi-phase ISM under the influence of self-gravity, supernova explosions, and (in one simulation) magnetic fields. We choose conditions resembling the Milky Way disc near the solar neighbourhood. The fully three-dimensional simulations have been carried out with the MHD code {\sc Flash}, which has been significantly extended to include the necessary additional physics.

We find that self-gravity is most important in dense gas. Without self-gravity it is impossible to obtain reliable estimates of the amount of dense, molecular gas in the simulations. In addition, models without self-gravity converge to an arbitrary ISM structure with roughly constant mass fractions within the different chemical components, which can easily be misinterpreted as the establishment of a dynamical equilibrium in these models (after $t \sim 50$ Myr). With self-gravity, the simulations do not evolve towards a dynamical equilibrium state since dense molecular clouds form and merge, but are generally not destroyed again. This long-term evolution of the dense molecular clouds would be changed if star formation and stellar feedback, in particular ionising radiation and stellar winds, would be included self-consistently.

Globally, the external gravitational potential dominates the initial evolution of the gas as it initiates a collapse towards the mid plane. Furthermore, the interplay of the external potential and the turbulent and thermal pressure provided by the supernova feedback sets the vertical distribution of the warm neutral medium seen in atomic hydrogen. 

We explore the impact of different descriptions used for supernova driving by changing the positions relative to the dense clouds or the rate of explosions in different simulations. We investigate four distinct cases: (1) random SN driving; (2) peak SN driving, where the explosion is always placed at the global density maximum; (3) mixed SN driving with a fraction of 50\% peak and 50\% random driving; and (4) clustered SN driving that also includes a population of Type Ia's (20\%), which have a broader vertical distribution. Moreover, we vary the SN rate. As a fiducial value we take the typical value derived from the Kennicutt-Schmidt (KS) relation at $\Sigma_{_{\rm GAS}} =10\;{\rm M}_\odot/{\rm pc}^2$, which results in a SN rate of 15/Myr in the simulated volume. Since this value is quite uncertain, we scale it up by a factor of 3 (to 45/Myr) and down by a factor of 3 (to 5/Myr). 

We find that both the SN rate and position have a major impact on the amount of molecular gas formed as well as on the vertical distributions of the gas. For the KS SN rate, we do recover a disc structure which is in good agreement with the Milky Way. Compared to the impact of the different SN rates,  and peak or mixed vs. random driving, the additional clustering of SNe has a minor effect on the mass distributions, the volume filling fractions, and the phases of the ISM in general. 

Also magnetic fields change the amount of dense and cold gas formed. With moderate initial magnetic fields ($3\;\mu$G), we find that the additional magnetic pressure is significant in dense gas and thus delays the formation of dense and cold, molecular gas. Magnetic fields therefore seem to be important for determining the onset and time-scale of molecular cloud formation. However, they do not change the overall evolution of the disc on longer time scales. We will explore their impact in more detail in a subsequent paper. In any case, magnetic fields and self-gravity have to be included at the same time as the amount of cold and dense gas will be crucially underestimated in MHD runs without self-gravity.

Close to the disc mid plane (within $z=\pm 150$ pc), the hot gas VFF is $\sim$ 50\% for most runs, which is consistent with the three-phase model of \citet{McKee1977}. The hot gas pressure is in approximate equilibrium with the warm phase and the pressures are consistent with the estimates of \citet{Blitz2006}.

\section*{Acknowledgments}%
The SILCC team thanks the Gauss Center for Supercomputing (http://www.gauss-centre.eu/) and the Leibniz-Rechenzentrum Garching (www.lrz.de) for the significant amount of computer time for this project and their user support. Furthermore, we thank the Deutsche Forschungsgemeinschaft (DFG) for funding through the SPP 1573 `The physics of the interstellar medium'. SKW acknowledges the support of the Bonn-Cologne Graduate School for physics and astronomy as well as the SFB 956 on the ``Conditions and impact of star formation".
RSK, SCOG, and CB thank the DFG for funding via the SFB 881 ``The Milky Way System'' (sub-projects B1, B2, and B8). RSK furthermore acknowledges support from the European Research Council under the European Community's Seventh Framework Programme (FP7/2007-2013) via the ERC Advanced Grant STARLIGHT (project number 339177).
T.N. acknowledges support from the Cluster of Excellence "Origin and structure of the Universe".
R.W. acknowledges support by the Czech Science Foundation grant 209/12/1795 and by the project RVO:67985815 of the Academy of Sciences of the Czech Republic.
T.P. acknowledges financial support through a Forschungskredit of the University of Z\"{u}rich, grant no. FK-13-112.
The software used in this work was developed in part by the DOE NNSA ASC- and DOE Office of Science ASCR-supported {\sc Flash} Center for Computational Science at the University of Chicago. To create some of the figures, we have used the free visualisation software {\sc yt} (yt-project.org).
\bibliographystyle{mn2e}
\bibliography{references.tex}%

\clearpage

\end{document}